\documentclass[12pt,preprint]{aastex}

\newcommand{\kms}{\mbox{km\,s$^{-1}$}}
\newcommand{\ctwohtwo}{\mbox{C$_{2}$H$_{2}$}}
\newcommand{\htwo}{\mbox{\ion{H}{2}}}
\newcommand{\hcn}{\mbox{HCN}}
\newcommand{\mgs}{\mbox{MgS}}
\newcommand{\lsun}{\mbox{L$_{\odot}$}}
\newcommand{\msun}{\mbox{M$_{\odot}$}}
\newcommand{\zsun}{\mbox{Z$_{\odot}$}}
\newcommand{\lir}{\mbox{L$_{\rm IR}$}}
\newcommand{\halpha}{\mbox{H$\alpha$}}

\newcommand{\egant}{\mbox{EVP01}}
\newcommand{\eganp}{\mbox{(EVP01)}}
\newcommand{\spitzer}{\emph{Spitzer}}
\newcommand{\iras}{\emph{IRAS}}
\newcommand{\iso}{\emph{ISO}}
\newcommand{\tmass}{\emph{2MASS}}
\newcommand{\msx}{\emph{MSX}}
\newcommand{\irac}{\emph{IRAC}}
\newcommand{\mips}{\emph{MIPS}}
\newcommand{\irs}{\emph{IRS}}

\shorttitle{\emph{Spitzer} IRS spectral atlas of LMC sources}
\shortauthors{Buchanan et al.}

\begin{document}

\title{A \emph{Spitzer} IRS Spectral Atlas of Luminous 8~\micron\ Sources in
  the Large Magellanic Cloud}

\author{Catherine L.\ Buchanan\altaffilmark{1}, Joel H.\
Kastner\altaffilmark{1}, William J.\ Forrest\altaffilmark{2}, Bruce
J.\ Hrivnak\altaffilmark{3}, Raghvendra Sahai\altaffilmark{4}, Michael
Egan\altaffilmark{5}, Adam
Frank\altaffilmark{2}, \& Cecilia Barnbaum\altaffilmark{6}}

\altaffiltext{1}{Center for Imaging Science, Rochester Institute of
Technology, 54 Lomb Memorial Drive, Rochester NY 14623. Email:
clbsps,jhk@cis.rit.edu}
\altaffiltext{2}{Department of Physics \& Astronomy, University of
Rochester, Bausch \& Lomb Hall, P.O. Box 270171, Rochester, NY
14627-0171}
\altaffiltext{3}{Dept. of Physics and Astronomy, Valparaiso
University, Valparaiso, IN 46383}
\altaffiltext{4}{NASA/JPL, 4800 Oak Grove Drive, Pasadena, CA 91109}
\altaffiltext{5}{Air Force Research Laboratory; OASD (NII) Space
Programs, Suite 7000, 1851 S. Bell St., Arlington, VA 22202}
\altaffiltext{6}{Valdosta University, 1500 N Patterson Street,
Valdosta, GA 31698}

\clearpage

\begin{abstract}
We present an atlas of \spitzer\ Space Telescope Infrared Spectrograph (\irs)
spectra of highly luminous, compact mid-infrared sources in the Large
Magellanic Cloud.  Sources were selected on the basis of infrared colors and
8~\micron\ (\msx) fluxes indicative of highly evolved, intermediate- to
high-mass stars with current or recent mass loss at large rates.  We determine
the chemistry of the circumstellar envelope from the mid-IR continuum and
spectral features and classify the spectral types of the stars. In the sample
of 60 sources, we find 21 Red Supergiants (RSGs), 16 C-rich Asymptotic Giant
Branch (AGB) stars, 11 \htwo\ regions, 4 likely O-rich AGB stars, 4 Galactic
O-rich AGB stars, 2 OH/IR stars, and 2 B[e] supergiants with peculiar IR
spectra.  We find that the overwhelming majority of the sample AGB stars (with
typical IR luminosities $\sim 10^4$~\lsun) have C-rich envelopes, while the
O-rich objects are predominantly luminous RSGs with \lir\,$\sim 10^5$~\lsun.
For both classes of evolved star (C-rich AGB stars and RSGs), we use the near-
to mid-infrared spectral energy distributions to determine mean bolometric
corrections to the stellar K-band flux densities. For carbon stars, the
bolometric corrections depend on the infrared color, whereas for RSGs, the
bolometric correction is independent of IR color.  Our results reveal that
objects previously classified as PNe on the basis of IR colors are in fact
compact \protect\htwo\ regions with very red \irs\ spectra that include strong
atomic recombination lines and polycyclic aromatic hydrocarbon (PAH) emission
features.  We demonstrate that the \irs\ spectral classes in our sample
separate clearly in infrared color-color diagrams that use combinations of
\tmass\ data and synthetic \irac/\mips\ fluxes derived from the \irs\
spectra. On this basis, we suggest diagnostics to identify and classify, with
high confidence levels, IR-luminous evolved stars and compact \protect\htwo\
regions in nearby galaxies using \spitzer\ and near-infrared photometry.
\end{abstract}

\keywords{atlases --- stars: AGB and post-AGB --- Magellanic Clouds ---
infrared: stars ---  stars: mass loss --- circumstellar matter}

\section{INTRODUCTION} \label{sec:intro}

While studies in visible light, particularly spectroscopy, have been key to
our understanding the nature and evolution of stars during most of their life
cycles, the early and late stages of stellar evolution are best studied in the
infrared.  This is because, during these stages, the stars are embedded in
regions of dust and gas.  The late stages of stars of intermediate initial
mass (1 -- 8~\msun) are characterized by very high luminosities and
obscuration by dusty, expanding circumstellar envelopes (see review by
\citealt{van03}).  Circumstellar dust envelopes absorb photospheric optical
and ultraviolet emission and re-radiate it in the mid- to far-infrared
(MFIR). Thus evolved intermediate-mass stars are luminous in the infrared (IR)
but may be highly obscured at optical wavelengths.  Current theories of the
later stages of the evolution of mass-losing stars therefore have been
informed by \iras, \iso, and \tmass\ studies of intermediate-mass evolved
stars in the Milky Way Galaxy and Magellanic Clouds.

Understanding the envelope chemistry of optically-obscured post-main sequence
(MS) stars through IR studies is important for our overall view of galaxy
evolution, since these stars dominate the rate of return of stellar processed
material to the ISM \citep{jur89,bar91,leb01}.  Optical surveys (e.g.,
\citealt{gro99} and refs.\ therein) can miss these objects because they are
too faint to be detected. Dusty circumstellar shells in evolved stars are
produced as the star loses mass via a combination of pulsations and radiation
pressure on dust grains newly formed via the pulsation process. Once
accelerated, these grains strip the (predominantly molecular) gas away from
the stellar envelope.  While on the Asymptotic Giant Branch (AGB),
intermediate mass stars undergo dredge-up and surface enrichment in products
of nucleosynthesis.  The enriched material is thus returned to the ISM via the
molecule-rich circumstellar dust shell.

More massive ($>$8~\msun) stars in the red supergiant (RSG) stage may resemble
high mass-loss rate AGB stars, in that they are also surrounded by
optically-thick circumstellar dust shells, though such high mass-loss rate
RSGs are rare.  Highly-obscured, evolved intermediate-mass (AGB) stars with
large mass-loss rates likely dominate the rate of return of nuclear processed
material to the interstellar medium. However, in our own Galaxy, it can be
difficult to distinguish between high mass-loss AGB and RSG stars, due to
their highly uncertain distances.

IR spectroscopy is particularly useful in determining the properties of
circumstellar dust.  The emission and absorption properties of different dust
grains have been studied in the laboratory and modeled (e.g.,
\citealt{dra01,li01}).  IR-bright stars in the Galaxy have been observed with
the \iras\ LRS and, subsequently, with the spectrometers aboard \iso\ (see
reviews by \citealt{wat99a, wat99b, bar99, bar97} and references therein) and
classified using the dust features revealed in these spectra (e.g.,
\citealt{kra02}).  The sensitivity of the \spitzer\ Infrared
Spectrograph\footnote{This work is based on observations made with the
\spitzer\ Space Telescope, which is operated by the Jet Propulsion Laboratory,
California Institute of Technology under a contract with NASA. The IRS was a
collaborative venture between Cornell University and Ball Aerospace
Corporation funded by NASA through the Jet Propulsion Laboratory and Ames
Research Center.} (\irs, \citealt{hou04}) on the \spitzer\ Space Telescope
\citep{wer04} now makes it possible to study the IR spectra of IR-luminous,
evolved stars in nearby galaxies.

In the absence of spectroscopy, photometric tools such as color-color diagrams
can be used to identify and classify evolved stars.  New studies of stellar
populations in the Galaxy as well as in nearby external galaxies will rely in
particular on photometry obtained with the \spitzer\ Infrared Array Camera
(\irac, \citealt{faz04}) and Multiband Imaging Photometer for \spitzer\
(\mips, \citealt{rie04}).  Indeed, the point sources detected in \irac\ and
\mips\ imaging of nearby galaxies likely will be dominated by dust-enshrouded
post-MS stars. As strong silicate and carbonaceous dust features at
$\sim$10~\micron\ distinguish C-rich and O-rich envelope chemistries,
respectively, the use of \irac\ and \mips\ colors can, in principal, avoid the
ambiguity in assigning stellar classes based on near-infrared (NIR) colors
(e.g., \citealt{ega01}, hereafter \egant).  However, such photometric
diagnostics must first be reliably associated with IR spectral properties,
which are best determined from \irs\ spectroscopy.

As previous infrared surveys have demonstrated (e.g., \citealt{gro95, gro98,
zij96, lou97, van97, van98, tra99}), the Large Magellanic Cloud (LMC) is an
ideal nearby galaxy to study the evolution of mass-losing post-MS stars.
Firstly, it contains a large population of IR-luminous, mass-losing objects
found at essentially the same distance, thereby alleviating the distance
ambiguities that haunt \iras\ and \iso\ samples of mass-losing stars in the
Galaxy. This allows the spectral properties of the LMC evolved star population
to be directly related to their luminosities, providing important constraints
on stellar models. Indeed, the luminosity distribution of carbon stars in the
LMC has long been a key test of models of the late stages of stellar evolution
in low metallicity environments (e.g., \citealt{ibe83}).  Secondly, the low
metallicities and high star formation rates of the LMC mimic those of
high-redshift galaxies.  In addition, spectra of LMC objects can be used to
establish ``ground truth'' for IR photometric diagnostics.

Due to the proximity of the LMC, individual objects have been detected by
previous IR satellites: \iras\ and \msx\ each detected $\sim$1800 IR sources
in or towards the LMC (\citealt{sch89}; \egant).  The most sensitive band on
\msx, A band ($\lambda_C$ = 8.3~\micron), was approximately four times more
sensitive than the \iras\ 12~\micron\ band, and permitted the detection of
evolved stars with less extreme IR luminosities than those detected by \iras\
\eganp.  \egant\ combined the \msx\ A band flux (in magnitudes) with near-IR
(JHK) measurements from the \tmass\ survey to form various diagnostic
color-color and color-magnitude diagrams for the LMC objects.  To classify
their nature based upon these IR fluxes, they compared these color-color and
color-magnitude diagrams to analogous diagrams constructed for various classes
of stars based upon the \citet{wai92} ``SKY'' model of the Galaxy.  On this
basis, they characterized the stars in specific categories, including red
supergiants, early (low mass loss) O-rich and C-rich AGB stars (\egant\
category 5), late (high mass loss) O-rich and C-rich AGB stars (category 6),
and PNs and H~II regions (category 7).  Within these categories, the different
types of objects were distinguished somewhat on the basis of colors, although
only for the late C-rich and O-rich AGB stars were the objects well
distinguished within a category.  Note that \egant\ classified $\sim$500 of
the $\sim$1800 objects in their study as foreground Galactic objects.  To
check the accuracy of this classification scheme for the LMC objects, \egant\
investigated the spectral types of the LMC objects using data in SIMBAD.
However, for the bona fide LMC objects, only a small percentage had
classifications; these were especially uncommon for categories 6 and 7.
Although the limited data do not contradict their classifications, these
classifications suffer from limited spectroscopic support.  This is not
surprising, as the IR-luminous objects detected by \msx\ do not commend
themselves to visible spectroscopy, and the requisite mid-IR spectroscopy has
been impractical until only recently.  To date, the study by \egant\ stands as
the most comprehensive classification of luminous 8~\micron\ sources in the
LMC. Therefore it is important to verify and/or correct the \egant\ color
classifications.

We have conducted a spectroscopic study with \spitzer\ \irs\ (PI: J. Kastner;
PID: 3426) of a representative sample of objects selected on the basis of
their IR colors and magnitudes to be the most IR-luminous post-MS stars in the
LMC.  Here we present the atlas of spectra. We then use these spectra as the
basis both for understanding the high-luminosity end of the LMC evolved star
population and for establishing photometric diagnostics for classifying IR
stars.  In \S\ref{sec:obs} we describe the sample selection, observations, and
data reduction. In \S\ref{sec:res} we present the spectra and classify them
according to continuum and spectral features and bolometric luminosities.  In
\S\ref{sec:dis} we describe the classes of objects identified and compare the
classifications with the types assigned by \egant. We then use our library of
\irs\ spectra to show how \spitzer/\tmass\ colors can be used to classify
IR-luminous sources in external galaxies reliably and unambiguously. On the
basis of our new classifications, we discuss the implications for the late
stages of stellar evolution. Detailed analysis of individual spectra and of
the spectral classes established in this study will be presented in future
papers.

\section{SAMPLE AND OBSERVATIONS} \label{sec:obs}

The complete sample observed by \irs\ comprises 62 IR sources, selected on the
basis of large 8.3~\micron\ fluxes from the comprehensive catalog of
\tmass/\msx\ sources compiled by \egant. The \egant\ sources were classified
based on IR colors.  These sources have coordinates accurate to $<1''$. We
selected those \msx\ 8.3~\micron\ (A-band) sources with $F_{8.3} > 150$~mJy
(A$ \le 6.5$) that were classified by \egant, on the basis of \tmass/\msx\ IR
colors, as ``late'' (i.e. very red) O-rich AGB (`O'or `OH/IR'), ``late''
C-rich AGB (`C IR'), red supergiants (`RSG'), or planetary nebulae
(`PN'). This resulted in a preliminary sample of $\sim170$ sources. Of these,
$\sim50$\% were classified as OH/IR, $\sim20$\% as PN, $\sim12$\% as RSG,
$\sim12$\% as O AGB, and $\sim5$\% as C IR. We then randomly deselected
sources so as to ``thin'' this preliminary list to 59 targets, distributed so
as to include 10 -- 15 sources from each of these five evolved star spectral
classes defined in the \egant\ \tmass/\msx\ study.  To ensure representation
in the sample of the most rapidly mass-losing, and hence highly obscured,
objects --- which may have not been detected by \tmass\ and therefore may have
been excluded from the \egant\ sample --- we include seven additional \iras\
sources.  These were sources selected from known or candidate evolved stars
\citep{tra99}, that satisfied our 8~\micron\ flux criterion, for which \tmass\
positions were made available subsequent to publication of the \egant\ MSX LMC
catalog, and which were not listed on the \spitzer\ Reserved Observations
Catalog.  Of this final sample of 66 sources, four targets (1 \egant\ source
and 3 \iras\ sources) were not observed, due to conflicts with other \spitzer\
programs.  The 62 observed sample objects are listed in Table
\ref{tab:sample}, where we list \tmass\ (J, H, and K) and \msx\ (A)
magnitudes, IR types determined from the IR colors by \egant, and object types
and other star names we obtained from SIMBAD.

The targets were observed by \spitzer\ using the \irs\ in staring mode. The
short-low (SL) and long-low (LL) modules were used; these modules cover the
wavelength ranges 5.2 -- 14~\micron\ and 14 -- 38~\micron, respectively, with
resolving powers of 64 -- 128. Two nod positions along the slit were observed
to facilitate rejection of artifacts in the data. For most objects, four
exposures of 30 and 14 seconds for the SL and LL modules, respectively, were
obtained at each nod position. A few unusually bright objects were observed
with shorter exposure times to avoid saturating the detector. To accurately
position each object in the slit, a peak-up procedure was used (\irs\ or
PCRS). Two observations (MSX~LMC~769 and 773) were unsuccessful due to peak-up
failure; the fields were crowded and the peak-up centered on the wrong
object. Thus, this atlas presents spectra of 60 objects.

The raw data were processed through the \spitzer\ pipeline version S11.0.
After pipeline processing, the basic calibrated data (BCDs) were processed
using the SMART software\footnote{SMART was developed by the IRS Team at
Cornell University and is available through the \spitzer\ Science Center at
Caltech.} \citep{hig04}.  The two-dimensional spectral images (BCDs) of the
multiple exposures were median-combined and, in most cases, the off-source
images were subtracted to remove the sky background. In the case of four
objects, the off order was contaminated and could not be used for sky
subtraction. In these cases, the sky was estimated locally, on either side of
the extracted aperture.  

Spectra were extracted and the fluxes were calibrated using the default point
source extraction apertures in SMART.  The modules at each nod position were
merged and the edges and overlapping regions of the modules were trimmed.  The
spectra from the two nod positions were then averaged to produce the final
spectrum. The \spitzer\ pipeline produces uncertainty images; however the
uncertainty images produced by the S11 pipeline are unreliable.  Therefore, we
estimated the flux uncertainty in each spectrum based upon the differences
between the two nod position spectra at each wavelength. The final uncertainty
for each spectrum was assumed to be the median value of the uncertainties at
all wavelengths; these generally ranged from 2 -- 20~mJy, though 4 objects had
median uncertainties $>$30~mJy.  We found that the relative uncertainties (the
median value of the uncertainty at each wavelength divided by the flux at that
wavelength) were typically $\sim$2\%, with a maximum of 5\%.  We note that the
wavelength calibration in the S11 pipeline is incorrect (too red) by
0.04~\micron\ for the SL module and also uncertain (by $\le$0.14~\micron) for
the LL module. These wavelength inaccuracies are not expected to have any
effect on the analysis or conclusions of this paper.

\section{RESULTS: THE SPECTRA} \label{sec:res}

\subsection{SPECTRAL FEATURES} \label{subsec:res_feat}

Mass-losing evolved stars are expected to show distinctive IR features,
depending on the chemistry of the circumstellar dust (e.g., \citealt{dra84}).
Oxygen-rich dust shows silicate features at 9.7 and 18~\micron\ which appear
in emission at low optical depths and absorption at high optical depths
(i.e. high mass-loss rates). Carbon-rich stars show a broad SiC dust emission
feature at 11.5~\micron\ and a narrow absorption feature at 13.7~\micron, due
to \ctwohtwo\ gas in the circumstellar envelope.

The IRS spectra are presented in Figure \ref{fig:spec}.  The spectra are
grouped according to the dominant spectral features: features indicative of a
predominance of silicate dust (Fig. \ref{fig:spec}{\it a}); features
indicative of a predominance of carbonaceous dust and/or molecules
(Fig. \ref{fig:spec}{\it b}); or polycyclic aromatic hydrocarbon (PAH)
features and narrow emission lines (Fig. \ref{fig:spec}{\it c}).  The O-rich
objects are further divided into subgroups (see \S\ref{subsec:res_clas}).
Within each group, the objects are then placed in order of increasing
luminosity, which we have calculated (see \S\ref{subsec:res_lum}).  Figure
\ref{fig:spec} also shows the infrared spectral energy distributions (SEDs) of
the sample objects, including the \irs\ spectra, \msx\ 8.3~\micron\ fluxes,
and \tmass\ and \iras\ fluxes (obtained from SIMBAD) where available.  Two
objects (MSX~LMC~1072 and 1524) were observed using only the LL modules, as
data were already being obtained for \spitzer\ project PID:1094. A third
object (MSX~LMC~1280) was observed using only the SL modules, as data were
being obtained for program PID:3505.

The largest of the three groups of spectra is that of the oxygen-rich
objects. These mostly show the expected strong silicate emission features and
have continua that are dominated by photospheric emission at short wavelengths
and dust emission at longer wavelengths. The silicate features are generally
quite smooth, indicating an amorphous composition for the silicate dust
(MSX~LMC~1326 is a notable exception, showing peaks indicative of crystalline
silicate dust; \citealt{kas06}).  Many of the spectra of the evolved O-rich
objects (most of which are RSGs; see \S\ref{subsec:dis_class_o}) also show a
narrow 6.2~\micron\ emission feature. This is unexpected, since this feature
is generally attributed to PAHs, carbon-based molecules which are not expected
to be found in the circumstellar environment of an O-rich evolved star.  The
2-dimensional spectra show that this feature is commonly seen in emission in
the sky (i.e., in the LMC) in the vicinity of these O-rich objects, and that
there is some spectral contamination that is not completely canceled out in
the sky subtraction process. However, this sky emission is at very low flux
density levels, not enough to explain the bright features observed.
Extraction of off-source spectra indicates that the 6.2~\micron\ feature can
is only detectable on the source, and is not apparent in off-source spectra,
with or without sky subtraction. The presence of this 6.2~\micron\ carbon
feature, and the suggestion of the 11.3~\micron\ PAH feature, in the spectra
of some O-rich objects is a puzzle.  The second group of objects, the C-rich
sources, show the expected SiC emission and \ctwohtwo\ absorption features and
have continua dominated by dust emission.  The broad ``30 micron'' feature
seen in C-rich evolved sources and attributed to MgS is also present in most.
There are also a significant number of very red objects that show forbidden
emission lines and strong features attributed to PAHs; these make up the third
group of sources in our sample.  In Figure \ref{fig:egspec} we present a
typical spectrum from each of the three groups, with the primary spectral
features indicated.

Spurious features appear in several of the spectra, at wavelengths 15.6, 18.7,
33.5, 34.8, and 36.0~\micron. Examination of the 2-dimensional spectral images
shows that these are due to background ``sky'' emission lines originating in
diffuse gas in the LMC; the lines appear in off-source as well as on-source
spectral images and are unrelated to the program sources. The wavelengths
correspond to known [\ion{S}{3}], [\ion{Si}{2}], and [\ion{Ne}{3}] emission
lines.  These spurious features appear in our spectra of C-rich and O-rich
sources in emission or absorption; because this diffuse emission is not
uniform across the region covered by the spectrograph slit, it is not always
removed properly by our sky subtraction. These features are seen at much
stronger levels in emission in the very red emission-line objects, and in
these cases they clearly arise from the sources themselves.

\subsection{LUMINOSITIES} \label{subsec:res_lum}

Infrared luminosities were calculated by integrating the flux over the
wavelength range 1 -- 100~\micron. The \irs\ spectra were used to determine
the flux densities over the wavelength range 5.0 -- 36.0~\micron. The
integrated fluxes over the range 1 -- 5~\micron\ were determined using a
blackbody scaled to match the \tmass\ flux densities, and the flux densities
from 36 -- 100~\micron\ were estimated using a blackbody weighted by
$\lambda^{-1.5}$ (to approximate optically-thin emission) and scaled to match
the \irs\ spectrum and available \iras\ data. An example of this procedure for
each of the three groups is shown in Figure \ref{fig:lumcalc}.  A distance to
the LMC of 50.1~kpc (distance modulus 18.50~mag; \citealt{fre01}) was assumed.

The uncertainties in the luminosities are dominated by the uncertainty in the
shapes of the spectral energy distribution in the wavelength ranges outside
the IRS spectrum.  The temperatures of the blackbodies were varied to estimate
the uncertainties in the measurement of the luminosities, which were found to
be 5 -- 10\% for the O-rich and C-rich evolved stars. The assumption that the
shape of the SED between 2 and 5~\micron\ can be approximated by a blackbody
curve was investigated using ISO spectra of Galactic objects and might
contribute an additional uncertainty to the luminosities of up to a further
19\% (typically 4\%). In calculating the luminosities of the mass-losing
stars, we note that we have neglected the effects of variability. Long period
variables may vary by up to 1 magnitude in the NIR, which results in up to a
factor 2.5 uncertainty in the derived luminosities of these objects, assuming
that the shape of the 1 -- 100~\micron\ SED remains the same as the flux
varies. Although not affected by variability, the uncertainties in the
measurement of the luminosities of the red emission-line objects are large
because most of the emission is from wavelengths longer than the IRS spectral
range, where the shape of the spectral energy distribution is poorly
constrained. The uncertainties in the luminosities of these sources may be up
to a factor of a few.  The bolometric luminosity will be dominated by the IR
luminosity for most sources. The RSGs are the only stars in the sample for
which optical emission might be expected to contribute significantly to the
SED; however, we find that the derived bolometric correction for the RSGs has
very little dispersion between objects (\S\ref{subsec:dis_cols}), so there can
be little contribution shortward of JHK.  Luminosities are listed in Table
\ref{tab:data}.  Histograms of the luminosities of the stars in each group are
shown in Figure \ref{fig:lumhist}.

\subsection{CLASSIFICATION OF THE SPECTRA} \label{subsec:res_clas}

Several IR spectral classification schemes have been established since the
advent of satellite mid-IR spectroscopy.  \iras\ had a Low-Resolution
Spectrometer (LRS) that obtained spectra in the range 8$-$23 with a resolution
that varied from $\sim$40 at the short end to $\sim$20 at the long end.  The
original \iras\ LRS atlas contained spectra of 5425 sources \citep{oln86}.
These were classified based upon the continuum shape (blue or red) and the
presence or absence of certain dominant spectral features (i.e., silicate
emission or absorption or C-rich features).  \citet{vol89}, \citet{vol91}, and
\citet{kwo97} extracted spectra of additional objects from the \iras\ database
to arrive at a total of 11,224 sources.  These were classified into nine
groups, with a one-letter designation based on the dominant spectral
feature(s) and/or the shape of the continuum.

\citet{kra02} developed a new classification system (hereafter the KSPW
system) that they applied to the over 900 objects for which complete spectra
were obtained by \iso\ with the Short Wavelength Spectrometer (SWS).  The SWS
had a wavelength range of 2.4$-$45 $\mu$m with a resolution of 1000 -- 2000.
Since our IRS spectra extend to 38~\micron\ and have a resolution of 64 --
120, they are better matched to the \iso\ spectra, and we will consequently
classify our spectra according to the KSPW system.

We compared our \irs\ spectra to typical \iso\ spectra in the KSPW spectral
classes\footnote{\protect\iso\ spectra of template objects were obtained from
http://isc.astro.cornell.edu/~sloan/library/swsatlas/aot1.html.}.  In the KSPW
system, stars are classified according to their continua, with subclasses
assigned according to the appearance of specific spectral features. The
continuum is characterized by the temperature of the dominant emitter, in
particular whether it is photospheric or dust emission.  Spectral features
include broad features associated with silicate or carbon rich dust emission,
complex features thought to be due to PAH emission, and narrow atomic and
molecular features. We classified each of our sample objects according to the
KSPW system.

We also assigned an ``\irs\ type'' to each star in our sample based on the
observed spectral features and luminosities (as well as optical information
for the two peculiar objects).  The various classifications of the sample
objects so derived from the \irs\ spectra are listed in Table \ref{tab:data},
along with the original \egant\ type. The new classifications are discussed in
the next section.

\section{DISCUSSION} \label{sec:dis}

\subsection{O-RICH STARS} \label{subsec:dis_class_o}

O-rich objects represent the largest group in the sample, with 33/60 objects
(55\%) having IR features indicative of O-rich dust. The majority of these
O-rich spectra (25/33) are dominated by photospheric emission at shorter
wavelengths and by dust emission at longer wavelengths (group 2 in the KSPW
system).  Most of these (22 objects) have strong silicate dust emission
features, and hence we classify these objects as belonging to KSPW subgroup
2SEc, while the remaining three are best classified as 2SEa.  These 3 objects
show weaker, broader 10~\micron\ emission and weaker, more sharply peaked
18-20~\micron\ emission than the 2SEc stars, as well as a 13~\micron\ feature
in emission, possibly due to aluminum-rich silicate grains \citep{slo03}.  All
except one (MSX~LMC~1677, discussed below) of the 22 O-rich objects with
strong dust features and significant photospheric emission have IR
luminosities $> 5\times 10^{4}$~\lsun, and so we identify them as RSGs.  The
previous classifications of these objects \eganp\ were either O-rich AGB stars
or RSGs (see Table \ref{tab:data}).

Examination of the luminosities of the three O-rich stars in subgroup 2SEa and
MSX~LMC~1677 (subgroup 2SEc) reveals that, if these stars were at the distance
of the LMC, they would have luminosities $\ge$5.8$\times 10^5$~\lsun. These
luminosities would place them significantly above the distribution of
luminosities for the RSGs (Figure \ref{fig:lumhist}; these four objects are
not included in the histogram). We conclude that these four stars are in fact
in the halo of the Milky Way Galaxy.  One of the stars, MSX~LMC~412 (RS Men),
is a known Galactic Cap Mira variable with a heliocentric radial velocity of
$>$140~\kms\ and period of 304 days \citep{whi94}. These authors derive a
distance to this star of 4.75~kpc using the period-luminosity relation to
determine the luminosity and inferring the distance from the measured flux.
We estimate luminosities for the other three Galactic stars using their K
magnitudes and color-period-luminosity relationships determined by
\citet{whi94}.  Distances of 4.2, 3.3, and 5.1~kpc, for MSX~LMC~1150, 1686,
and 1677, respectively, were inferred for the three sources using luminosities
so derived.  Three of the stars are identified by \egant\
as O~AGB stars from their blue NIR colors, while one star, MSX~LMC~1686, had
been misidentified as a C AGB star. The NIR colors of MSX~LMC~1686 are
consistent with the categories of both O AGB and C AGB \eganp\ and \egant\ 
note that there is some confusion and overlap between these categories. These
Galactic Cap stars will be discussed in detail in a forthcoming paper.

Three, possibly four, of the O-rich stars, MSX~LMC~642, 1072, 1524, and 1280,
have spectra dominated by warm dust (group 3). We classify these as KSPW
subgroup 3SE, i.e., similar to the majority of the O-rich stars (subgroup
2SEc) with strong silicate emission at 10 and 18~\micron\ from optically thin
dust emission, but with thermal continuum from an optically thick dust shell.
The SED, $F_{\nu}$, peaks around 10 -- 15~\micron, indicating a strong warm
dust contribution, although $\nu F_{\nu}$ peaks closer to 2~\micron, due to
the contribution from the reddened stellar photosphere.  The silicate emission
features are stronger than the Galactic O-rich AGB stars discussed above.  The
combined SL+LL spectra of MSX~LMC~1072 and 1524, kindly provided by
Marckwick-Kemper, confirm the classifications we made based on our LL spectra.
The classification of MSX~LMC~1280 is based on our SL spectrum, which covers
only the 5 -- 14~\micron\ wavelength range, and is therefore somewhat
tentative. The luminosity of this source was derived by assuming the same
spectral shape as MSX~LMC~1189, and is highly uncertain.  These four O-rich
stars with dust-dominated spectra have IR luminosities of $\sim$2 -- 7 $\times
10^{4}$~\lsun, placing them very high on the AGB, so we tentatively identify
these as luminous O-rich AGB stars.  Although the theoretical limit for the
luminosity of AGB stars is $6 \times 10^{4}$~\lsun, luminosities slightly
higher than this may be possible from higher mass ($\sim$10~\msun) progenitors
at low metallicity (i.e., super-AGB stars \citealt{her05}).  Our
classifications of these objects based upon the mid-IR spectra differ
significantly from those of \egant: One of the 3SE stars (MSX~LMC~1072) has
\egant\ type PN, but its \irs\ spectrum clearly lacks the red continuum. It is
likely to be a O AGB star, as it only narrowly falls outside of the criteria
for identifying O AGB stars in \egant: H-K = 0.76, 0.01 magnitudes too
red. The other stars classified as 3SE have \egant\ type C IR. MSX~LMC~1280,
1524, and 642 have \tmass/\msx\ colors inconsistent with the \egant\ criteria
for O AGB, OH/IR stars and RSGs but their \irs\ spectra clearly indicate
O-rich envelopes.

We find no population of less luminous O-rich AGB stars in the sample. It is
possible lower luminosity O-rich objects were excluded from the sample due to
our A-band selection criterion ($F_{8.3} > 150$~mJy; \S\ref{sec:obs}). Unlike
the C-rich AGB stars, the 8~\micron\ flux is a small fraction of the
bolometric luminosity for the O-rich AGB stars, therefore lower luminosity
O-rich AGB stars are relatively fainter at 8~\micron.  The 4 Galactic O-rich
AGB stars in our sample, for example, would have $F_{8.3} < 40$~mJy if they
were at the distance of the LMC, and so would have been excluded from our
sample. Identifications of O-rich AGB stars fainter than those studied in
this work are necessary to confirm the possibility that the lack of lower
luminosity O-rich AGB stars in our sample is a selection effect.

Two of the O-rich stars show self-absorbed silicate emission features at
10~\micron, indicative of optically thick dust shells.  One of the objects,
MSX~LMC~1182, shows broad silicate emission with strong, narrower central
absorption, and has an IR luminosity of 4.7$\times 10^5$~\lsun, confirming its
status as an OH/IR supergiant \citep{woo86, roc93}. The spectrum of
MSX~LMC~1182 is dominated by very cool dust so we classify it as group 5 in
the KSPW system, though its exact subclassification is ambiguous.  This star
is one of the \iras\ sources included in our sample to ensure representation
of the most highly obscured objects (\S\ref{sec:obs}). It was not selected on
the basis of a classification by \egant; it is in the \egant\ sample, but does
not have an \egant\ type.  The other object, MSX~LMC~807, shows less strongly
self-absorbed 10~\micron\ emission, and has luminosity
4$\times$10$^{4}$~\lsun. We identify this object as a likely OH/IR star on the
AGB, in agreement with its \egant\ type, although we note that OH maser
detection is required to confirm its formal OH/IR star classification. This
spectrum falls into KSPW subclass 3SB, which is associated with AGB and OH/IR
stars.

The remaining two O-rich objects, MSX~LMC~890 and MSX~LMC~1326, show
relatively flat continua with strong broad silicate emission at 10~\micron\
and 18 -- 20~\micron. We assign these objects \irs\ type
``Peculiar''. MSX~LMC~890 and MSX~LMC~1326 are optically identified as
hypergiant B stars R~126 and R~66, respectively, and the IR spectra are best
matched by the class 5SE template spectrum, which are generally young
stars. The \irs\ spectra of these stars are likely to be indicative of
circumstellar, perhaps circumbinary, disks, and are modeled and discussed in
detail by \citet{kas06}, who classify them as B[e] hypergiants.  MSX~LMC~890
and MSX~LMC~1326 have \egant\ type PN based on their NIR colors. \egant\
separate PNe from early type stars with surrounding dust (such as these B
stars) using the H-K color, as PNe are redder than the younger objects in the
\citet{wai92} model. They note that their classification may be confused for
objects with H-K colors near the boundary, i.e. H-K = 0.75, and other authors
have shown that Galactic PNe can have NIR colors considerably bluer than this
\citep{whi85, ram05}.  MSX~LMC~890 and MSX~LMC~1326 are slightly redward of
this boundary (H-K = 0.94 and 0.81, respectively), and so were misclassified
by \egant\ as PNe when they are instead dusty B stars.  It has been noted by
\citet{whi85} that the H-K colors of PNe are in fact not dissimilar from hot
stars surrounded by dust, though a J-band excess in PNe due to He line
emission produces redder J-K colors which may distinguish these objects from
younger stars.

\subsection{C-RICH STARS} \label{subsec:dis_class_c}

Sixteen of the 60 sample stars (27\%) have spectra typical of warm carbon-rich
dust envelopes. The IR luminosities of the C-rich stars are 5$\times 10^3$ --
2$\times 10^4$~\lsun, so we identify all 16 C-rich stars as AGB stars. The
large number of C-rich AGB stars in the sample stands in stark contrast to the
lack of O-rich AGB stars found (\S\ref{subsec:dis_class_o}) and has
implications for the late stages of stellar evolution (see
\S\ref{subsec:dis_evol}).  In addition to the silicon carbide emission feature
and the narrow \ctwohtwo\ absorption feature at 13.7~\micron, there is a
\ctwohtwo + \hcn\ absorption feature at 7.5~\micron\ seen in almost all these
spectra.  Another broad absorption feature, due to \ctwohtwo\ and \hcn, may be
present between 14 -- 16~\micron\ \citep{slo98,aok99}, particularly in
MSX~LMC~1384. A broad emission bump, peaking at wavelengths $\sim$26~\micron,
is present in 11 objects.  This feature, known as the ``30 micron'' emission
feature, is usually attributed to \mgs\ \citep{goe85}. \citet{for81} detected
it initially in the most optically thick Galactic carbon stars and in C-rich
PNe. It has subsequently been seen in several carbon-rich PNe and
approximately a dozen carbon-rich proto-PNe \citep{vol02}. There is evidence
from \iso\ that this broad feature is composed of two features, peaking at
$\sim$26~\micron\ and $\sim$33~\micron, and with varying strengths depending
upon the evolutionary state of the star \citep{hri00,vol02}.  \citet{hon02}
have examined this broad feature in a variety of evolved carbon-rich objects
and are able to fit it by using MgS grains of varying temperatures.

Most of the carbon-rich stars (12/16) have spectra that are dominated by warm
dust. These are classified as subgroup 3CE.  Only three of these 12 have
\egant\ type C IR, while one has no \egant\ type (no \tmass\ data available)
and 8 have \egant\ type OH/IR. These eight stars have very red NIR colors for
carbon-rich stars, compared with the \citet{wai92} model predictions and
\egant\ classifications.  Only one star (MSX~LMC~218) has H-K and K-A colors
close to the boundary for C IR stars, while the other seven have colors too
red to be categorized as \egant\ type C IR, C AGB or RSG.  Two carbon-rich
stars, MSX~LMC~1652 and 1592, have spectra dominated by cooler dust; they fall
into subgroup 3CR. MSX~LMC~1652 has \egant\ type OH/IR, with extremely red NIR
colors. MSX~LMC~1592 has no \tmass\ data so has no \egant\ type.  One star,
MSX~LMC~1400, shows clear SiC features, but matches equally well the 3CE and
3CR templates, so has been classified as '3C?'. It also has very red NIR
colors for a C-rich star and has \egant\ type OH/IR.

The final object classified as a carbon-rich star, MSX~LMC~775, shows a nearly
featureless spectrum with a narrow peak at 5 -- 6~\micron\ and relatively blue
continuum.  The 13.7~\micron\ \ctwohtwo\ absorption feature apparent in all
other carbon-rich stars in our sample is notably absent.  This spectrum is not
well matched by any of the naked star (group 1) or stellar photosphere + dust
(group 2) \iso\ template spectra. The spectrum of MCX~LMC~775 is very similar
in slope to that of MSX~LMC~1492, but without the clear 13.7~\micron\
feature. The lack of strong O-rich or C-rich dust features suggests that this
object may be an S star, with approximately equal carbon and oxygen abundances
(e.g., \citealt{jur88}). However, the \irs\ spectrum of MSX~LMC~775 does not
resemble the \iso\ spectrum of the S star W~Aql and shows a hint of a SiC
feature at 11.5~\micron, which is rare for S stars \citep{che93}, suggesting
the emission may be due to optically thick carbon-rich dust.  This unusual
object may be similar to R~Corona Borealis, which typifies a class of stars
that have hydrogen-deficient and carbon-rich atmospheres and are highly
variable due to periods of dust formation (see review by \citealt{cla96}).
These objects show nearly featureless mid-IR spectra, that are dominated by
blackbody continuum \citep{cla96, kra05}.  Two such R~CrB stars have recently
been identified in the Small Magellanic Cloud \citep{kra05}; these objects
show similar \spitzer\ spectra to MSX~LMC~775, although MSX~LMC~775 is
well-described by a modified blackbody (multiplied by $\lambda^{-1}$), rather
than a simple blackbody continuum as seen in these two.  RCB stars are
identified mainly by the absence of hydrogen features in their optical spectra
and by variability characterized by dramatic declines in the optical
brightness at irregular intervals \citep{cla96}.  Further data, such as
optical spectroscopy and an optical light curve, are necessary to confirm if
this candidate is indeed an R~CrB source or an S star.  MSX~LMC~775 has
\egant\ type C IR.

\subsection{RED, EMISSION-LINE OBJECTS} \label{subsec:dis_class_pn}

Approximately one fifth of the sample (11 objects) is comprised of very red
objects which have a peak in their SED longer than the wavelengths observed by
\irs\ (group 5). These objects are dominated by cool dust and show strong PAH
emission features and narrow emission lines.  These red, emission-line objects
were classified by \egant\ as type PN, which they distinguish from \htwo\
regions associated with massive star formation by H-K colors (H-K $>$ 0.75).
Indeed, the spectra of these objects resemble the \iso\ spectra of PNe that
are classified as 4PU in the KSPW scheme.  However, various lines of evidence,
discussed below, indicate that these objects are \htwo\ regions rather than
PNe.

The luminosities of the red objects are rather high, $2 \times 10^{4}$ -- $5
\times 10^{5}$~\lsun. Most of these are too luminous to be the immediate
descendents of evolved, intermediate-mass stars (Figure \ref{fig:lumhist}),
but rather they agree with the luminosities of \htwo\ regions powered by
massive young O stars. The IR luminosities of many of these objects are
somewhat uncertain, as much of the flux is emitted beyond 40~\micron, where
the shape is the spectral energy distribution is poorly, if at all,
constrained (\S\ref{subsec:res_lum}). The infrared flux densities of the
\htwo\ regions are, if anything, more likely to be underestimated, due to
extended emission falling out of the slit. Thus the luminosities argue against
most of these objects being PNe.

Examination of the optical (Digital Sky Survey) and IR (\msx\ A band) images
reveals widespread nebulosity around these objects, suggesting they are the
bright, compact cores of \htwo\ regions associated with massive star formation
regions, rather than PNe. This is shown in Figure \ref{fig:hiimontage}.  The
\msx\ 8.3~\micron\ images in particular probe the dust and reveal extended
nebulous emission, probably dominated by PAH's. Several of the DSS images also
show diffuse emission, possibly dominated by \halpha, while, in contrast, the
\tmass\ NIR images, which we do not display, generally show only the bright
point sources.

The spectra of the red, PAH-dominated objects match KSPW subgroup 5UE
(Fig. \ref{fig:pncomp}), and show strong atomic fine-structure lines and PAH
features. Galactic objects in this class are mostly young objects.  In
addition, these \irs\ spectra are lacking many of the high-ionization narrow
emission lines common in Galactic PNe \citep{kra02} and LMC PNe \citep{ber04},
such as [\ion{Ne}{5}]~24.3~\micron\ and [\ion{O}{4}]~25.9~\micron, as well as
[\ion{Ne}{6}]~7.65~\micron. This is shown in Figure \ref{fig:pncomp}. The
\irs\ spectra of the PAH-dominated objects show strong [\ion{S}{3}] emission
peaks at 18.7 and 33.5~\micron\ and [\ion{Ne}{2}]~12.8~\micron, emission lines
found in both PNe and \htwo\ regions \citep{giv02}. The similarity of the
continuum shapes of the PN and \htwo\ region in Figure \ref{fig:pncomp}
suggests that these types of objects will have similar mid-IR (5 --
35~\micron) colors and so will be difficult to distinguish photometrically in
this wavelength range.

The \irs\ spectra provide further evidence that the PAH-dominated sources are
extended \htwo\ regions and not PNe.  The discontinuity seen at
$\sim$14~\micron\ between the spectra formed by the short and long wavelength
modules (see Fig. \ref{fig:egspec}{\it c}) is attributed to the extended
nature of these sources (a similar phenomenon also affects certain \iso\
spectra of extended sources; see, e.g., \citealt{lam96}).  This ``jump''
indicates that the spatial extents of the PAH-dominated sources are greater
than the width of the SL slit (3.6\arcsec), which is smaller than the LL slit
(10.5\arcsec).  A spatial extent of 5\arcsec\ at 50.1~kpc corresponds to a
physical size of 25000~AU (1.2~pc), confirming that these sources are unlikely
to be PNe.  The red, emission-line sources also appear extended in 2-D
spectral images and this is confirmed by fitting Gaussian functions along the
spectrum of each source and comparing the FWHM to a similar fit to one of the
O-rich point sources (MSX~LMC~1150).

In summary, we find that all of the program objects selected as PNe based on
their NIR/\msx\ colors are, in fact, dusty \htwo\ regions associated with
luminous young O/B stars.  Thus the color distinction between these two groups
made by \egant\ is not supported. This in fact confirms their caveat that many
of their objects classified as PN may be \htwo\ regions with enhanced
reddening. Revised criteria will need to be developed to distinguish PN from
\htwo\ regions photometrically, and these will probably need to rely upon
wavelengths other than just the near- and mid-IR.

\subsection{INFRARED COLOR-COLOR AND COLOR-MAGNITUDE CLASSIFICATIONS OF LMC
  SOURCES} \label{subsec:dis_cols}

Figure \ref{fig:nircols} shows IR color-color diagrams for the stars in this
study, based upon the \tmass\ and \msx\ photometry.  The stars of most of the
various IRS spectral types separate quite cleanly into different regions of
this diagram, confirming that IR color criteria can be used to classify the
spectral types, although this may not be the case for distinguishing PNe and
\htwo\ regions, and PNe may occupy the region between the AGB stars and the
\htwo\ regions and make the separation less clean. As previously noted,
however, our IRS spectral classifications frequently differ from the original
\egant\ classes assigned to these objects. \egant\ indeed mention that the
O-rich and C-rich AGB star classifications are rather uncertain due to the
similar regions in color space that these objects occupy. Other authors have
also shown, both observationally \citep{van98, zij06} and theoretically
\citep{gro06}, the difficulty of distinguishing O-rich and C-rich evolved
stars based on their IR colors.  We nevertheless propose modified IR color
criteria, based on the locations of objects in Figure \ref{fig:nircols},
subject to the uncertainty due to our sample incompleteness. The isolated
locations of the two OH/IR stars in our sample in the color-color diagrams
suggests that other \egant\ objects with \tmass/\msx\ colors K-A$>$4 and
J-K$>$2, but outside the C-rich AGB area, are candidate OH/IR stars. The four
objects we classify as O-rich AGB stars lie fairly close to each other on the
color-color and color-magnitude diagrams, indicating that objects with K-A = 1
-- 3 and J-K = 2 -- 3 are candidate O-rich AGB stars in the LMC.  Further
observations are required to probe the borderline regions where
classifications overlap or are uncertain and to identify where PNe in the LMC
lie on these diagrams.

Color-magnitude diagrams are also extremely useful in discriminating between
spectral types of stars (Figure \ref{fig:colmag}). The Galactic and LMC
objects separate clearly, with Galactic objects having brighter apparent
magnitudes due to their relatively modest distances. The AGB star luminosity
sequence is also evident, with the C-rich objects in the majority and the most
IR-luminous AGB stars being O-rich.  The most IR-luminous LMC objects in our
sample are the RSGs and the lone OH/IR supergiant (MSX~LMC~1182), the latter
of which is redder due to the thick dust shell presumably surrounding this
source.  The lone OH/IR AGB star (MSX~LMC~807) has a lower IR luminosity,
comparable to the C-rich AGB stars.  The \htwo\ regions, which have J-band
magnitudes similar to, but K magnitudes fainter than, the C-rich AGB stars
(due to their bluer J-K colors; Fig.\ \ref{fig:nircols}), display the reddest
K-A colors (Fig.\ \ref{fig:nircols}).

One of the primary goals of this program was to determine reliable relations
between \spitzer\ \irac\ and \mips\ photometry and \irs\ spectral types for
compact, infrared-luminous objects, especially evolved stars.  This will
enable the classification and study of evolved stars in other nearby galaxies
using \spitzer\ photometry. Therefore, we also derived \irac\ and \mips\ flux
densities for the objects using the \irs\ spectra and the spectral response
functions of the \irac\ 5.8 and 8.0~\micron\ bands and the \mips\ 24~\micron\
band.  Figure \ref{fig:sstcols} shows the simulated \irac/\mips\ colors of our
stars as derived from their \irs\ spectra.  Also plotted on Figure
\ref{fig:sstcols} are Galactic stars in each spectral type, as well as PNe,
for comparison. The three Galactic PNe plotted overlap with the \htwo\ regions
and appear to fill the gap between the evolved stars and \htwo\
regions. Clearly, observations of LMC PNe are required to determine whether
the proposed photometric classifications of evolved stars and \htwo\ regions
are ``contaminated'' by PNe.  The comparison objects were selected at random
from the database of \citet{kra02} --- subject to the constraint that the
\iso\ spectrum had reasonable signal-to-noise ratio --- and the synthetic
\irac\ and \mips\ colors were derived from the \iso\ spectra. The Galactic
objects fall in the same regions of the diagram as the LMC stars, indicating
that the \irac/\mips\ color-color diagrams are not very sensitive to
metallicity. Therefore these diagnostics can be applied to external galaxies
without being adversely affected by metallicity differences, at least in the
range $\sim$0.5 -- 1~\zsun.  It is possible to construct a large number of
color-color diagrams using the observed \tmass/\msx\ photometry and synthetic
\spitzer\ \irac/\mips\ colors. In Figure \ref{fig:sstcrit} we present two such
color-color diagrams that appear to show the best separation of the object
types. Based on these two diagrams, we propose color criteria to classify
luminous 8~\micron\ sources in external galaxies (shown as boxes in
Fig. \ref{fig:sstcrit} and listed in Table \ref{tab:diag}). Figure
\ref{fig:sstcolmag} shows color-magnitude diagrams derived solely from the
synthetic \spitzer\ flux densities of our sample objects. The figure indicates
that such \irac/\mips\ color-magnitude diagrams represent an effective means
to distinguish between evolved stars and \htwo\ regions in external galaxies,
and should easily identify the most luminous, rapidly mass-losing stars (i.e.,
OH/IR supergiants) in such galaxies.

IR colors may also be useful to infer the luminosity of evolved stars. Using
the measured luminosities of the C-rich AGB stars and the RSGs, for which we
have a sufficiently large number of objects, we investigated the relation
between the luminosity and the NIR (K-band) magnitude. We defined a
quasi-``bolometric correction'' to be the ratio of the bolometric luminosity
to the K-band flux density. For the C-rich AGB stars, this ratio shows a clear
trend with the K-A color (Figure \ref{fig:cbolcor}).  This trend can be fitted
with an exponential function, of the form $L_{bol}/F_{K} = a({\rm K-A})^b +
c$, where $L_{bol}$ is the bolometric (IR) luminosity in units of $L_{sun}$,
$F_{K}$ is the \tmass\ K-band flux density in Jy, and (K-A) is the 2 --
8~\micron\ color in magnitudes, derived from the \tmass\ and \msx\ K and A
magnitudes. The best fit parameters are $a = 0.0015$, $b = 6.26$, and $c =
14.9$.  Using this fit to the points shown in Figure \ref{fig:cbolcor}, the
bolometric correction for a source of a given K-A color can be estimated and
hence the luminosity derived from the K-band magnitude. The largest error in
the luminosity predicted in this way is less than a factor of 2, with most of
predicted luminosities within 20\% of the measured luminosity.

\notetoeditor{Figure 12 approximately here.}

For the RSGs, there is a unique value of the ``bolometric correction'' for all
K magnitudes (Figure \ref{fig:rsgbolcor}). The mean value for all sources is
$L_{bol}/F_{K} = 1.46 \stackrel{^{+0.21}}{_{-0.17}}\, \times\,
10^{5}$~\lsun/Jy.  Thus the luminosities of RSGs can be predicted from their K
magnitudes with an uncertainty of less than 15\%.

\notetoeditor{Figure 13 approximately here.}

These results are potentially very useful in the study of Galactic carbon
stars and RSGs, assuming that metallicity does not affect the bolometric
correction. The bolometric corrections derived here will enable estimates of
luminosities, and hence distances, of stars to be made simply from their IR
colors and fluxes.

\subsection{IMPLICATIONS FOR LATE STAGES OF STELLAR EVOLUTION}
\label{subsec:dis_evol}  

The predominance of C-rich stellar envelopes among the lower-luminosity
members of our sample, and a relatively sharp dividing line between C-rich and
O-rich infrared-luminous, evolved stars, is apparent in Table \ref{tab:data}
and Fig.\ \ref{fig:lumhist}. These results suggest constraints on models of
the processes of nuclear burning and dredge-up in intermediate- to high-mass
stars and, at the same time, may indicate the importance of the metallicity of
the star-forming environment on the ultimate evolutionary endpoints of such
stars.

We caution that our IR-luminous LMC sample was selected based on a
classification scheme that has proven somewhat unreliable, especially with
regard to the distinctions between O-rich and C-rich AGB stars and between PNe
and \htwo\ regions. This sample therefore may not well represent the
IR-luminous, evolved stellar population of the LMC.  The flux limit imposed
may also have biased the sample against O-rich AGB stars.  Nevertheless, the
fact that we find $\sim80$\% of the AGB stars in our sample are C-rich argues
that the low metallicity of the LMC serves to facilitiate carbon star
formation, due to the relative ease of inverting the C/O ratio at the stellar
surface.  At the same time, the apparent upper cutoff in the luminosity of
IR-luminous carbon stars in the LMC, $\sim2\times10^4$~\lsun\ (Table
\ref{tab:data}), implies an upper limit of $\sim$3.7~\msun\ for the initial
mass of carbon star progenitors in the LMC \citep{mar99}, though we note that
this depends on assumptions about the mass loss, dredge-up efficiency and
envelope burning. The apparent presence of such a carbon star luminosity
cutoff, and the fact that the only three AGB stars brighter than this are
O-rich, suggests that ``hot bottom burning'' (see \citealt{her05} and
references therein) inhibits carbon star formation in the LMC for initial
masses larger than $\sim$3.7~\msun.  Furthermore, the fact that all RSGs in
the sample are O-rich suggests that the He-burning phase for such massive
($\gtrsim 8$~\msun) stars is short, that the onset of C burning in such stars
is rapid, and/or that the envelopes of such stars are sufficiently massive
that it is difficult to invert the C/O ratio even given low initial
metallicity.

We also note that, for the O-rich evolved stars, all except one (MSX~LMC~1072)
with \lir$> 6 \times 10^{4}$~\lsun\ have spectral shape 2SEc, while all except
one (MSX~LMC~815) with \lir$< 6 \times 10^{4}$~\lsun\ have spectral class 3SE.
This transition coincides with the theoretical upper limit on AGB luminosity
\citep{ibe83}, although luminous O-rich stars in a hot bottom burning phase
may fall above the classical AGB limit (e.g., \citealt{blo91}) so the
luminosity separation of AGB stars and RSGs is uncertain.  The two exceptions
are the ones closest to this dividing luminosity and it may be that
variability is playing a role in confusing the classification of the two
objects.  Thus is it possible that the \irs\ spectral shape may provide clues
to an object's luminosity and progenitor mass, which would be particularly
useful in cases where the luminosity is otherwise ambiguous.

We caution, however, that the present LMC \irs\ sample was selected to
represent the types of object in the LMC but not their relative numbers, and
so is not proportionally representative of the luminous mid-IR population of
the LMC.  For example, the most luminous sources listed in \egant\ are
dominated by objects classified as ``OH/IR stars'' --- which, our IRS survey
demonstrates, are predominantly or perhaps exclusively C-rich AGB stars ---
whereas such objects constituted only $\sim20$\% of the sample we have
surveyed thus far with IRS.  Thus, it is important to revisit the complete
sample of luminous 8 $\mu$m sources in the \egant\ lists, to reclassify these
objects according to our revised criteria and, thereby, determine the
distribution of envelope spectral types among these stars. This work will be
the subject of a future paper. In addition, it is crucial to obtain additional
\irs\ spectra of the most highly obscured yet luminous evolved stars. Such
objects would likely appear as \msx\ and \iras\ sources but not as \tmass\
sources and, hence, are not necessarily well represented in the present
sample. The OH/IR supergiant MSC~LMC~1182, along with the two luminous C-rich
AGB stars that are also \iras\ sources (MSX~LMC~1592 and 1652), offer
indications of the types of sources we might expect to find among this group.

\section{CONCLUSIONS} \label{sec:con}

We have obtained \spitzer\ low-resolution spectra for a representative sample
of 60 luminous 8~\micron\ sources in the LMC. The sources observed were
selected to include a variety of evolved, mass-losing stars, based on recently
established \tmass/\msx\ color criteria. The \irs\ spectra covered the
wavelength range 5 -- 35~\micron, allowing determination of the envelope
chemistry from detailed spectral features, particularly silicate emission
features at 10 and 18~\micron\ in O-rich objects and the SiC emission feature
at 11~\micron\ and \ctwohtwo\ absorption feature at 13.7~\micron\ in C-rich
objects. We derived IR luminosities from the combined \tmass\ and \irs\
spectral energy distributions of the objects and used these, along with the
spectra, to classify the object types. We identify 16 C-rich AGB stars, 4
O-rich AGB stars, 21 RSGs, 11 \htwo\ regions, and 2 OH/IR stars (one
supergiant and one AGB star). We also find 2 B supergiants that have peculiar
IR spectra and 4 O-rich AGB stars that are in fact foreground Galactic Mira
variables.  We find the vast majority of the AGB stars, with IR luminosities
$\sim 10^4$~\lsun, have C-rich envelopes, while the O-rich objects are almost
all luminous RSGs with \lir$\gtrsim 10^5$. The predominance of C-rich stars
along the AGB in the LMC is consistent with the hypothesis that C-rich stars
form more easily than O-rich stars in lower metallicity environments.
However, all C-rich AGB stars in our sample have \lir\ $< 2 \times
10^{4}$~\lsun, while all four O-rich AGB stars have \lir\ $\gtrsim 2 \times
10^{4}$~\lsun, suggesting a relatively firm upper limit ($\sim$3.7~\msun) for
the progenitor mass of LMC Carbon stars.  The fact that all of the supergiant
stars are O-rich, in turn, may indicate that massive supergiants do not have
sufficiently long He-burning lifetimes to produce C-rich surfaces, and hence
C-rich ejecta, despite the low metallicity of the LMC.

Comparison of the \irs\ classifications with the NIR color-based types \eganp\
revealed a number of inaccuracies in the \tmass/\msx\ classification
criteria. All of the sample objects with \egant\ type PN were reclassified,
mostly as \htwo\ regions. Stars classified as OH/IR stars on the basis of NIR
colors are mostly C-rich AGB stars, based on their \irs\ spectra. The O-rich
AGB stars in the sample were reclassified as either RSGs or Galactic O~AGB
stars, based on their \irs\ spectra and luminosities.  On the basis of these
reclassifications, we have developed revised \tmass/\msx\ color-color and
color-magnitude classification criteria for LMC objects and, by extension, for
luminous 8~\micron\ sources in other external galaxies.

For the C-rich AGB and RSG stars, bolometric corrections to the stellar K-band
flux densities were derived using the observed IR SEDs. For C-rich AGB stars,
the bolometric corrections were found to depend on the K-A color, allowing the
bolometric luminosity to be estimated from the K and A-band fluxes with a
typical uncertainty of less than 20\%, and at most a factor of 2.  The
bolometric corrections of the RSGs were found to be independent of IR color;
for these objects the bolometric luminosities can be predicted from the K-band
magnitude with an uncertainty of less than 15\%.  The derived bolometric
corrections could prove very useful for determining the luminosities of carbon
stars and RSGs in the Galaxy and other nearby galaxies.

Infrared (\irac/\mips/\tmass) colors were determined from the \irs\ spectra
and \tmass\ fluxes. We found that the spectral types separate well in various
mid-IR color-color diagrams, indicating that \spitzer\ photometry can be an
effective means to infer the spectral class of dust-enshrouded objects in
external galaxies. We suggest diagnostics for classifying IR-luminous sources
in the Galaxy and nearby galaxies, using \spitzer\ photometry.  These colors
may also prove useful in the analysis of \spitzer\ photometry of unresolved
stellar populations in high-redshift galaxies.

\acknowledgments

This work is based on observations made with the \emph{Spitzer} Space
Telescope, which is operated by the Jet Propulsion Laboratory, California
Institute of Technology under a contract with NASA. Support for this work was
provided by NASA through awards issued by JPL/Caltech to the authors'
institutions. We would like to thank Dan Watson for providing peak-up
stars, without which many of these observations would not have been possible.
We thank the referee, A. Zijlstra, for helpful comments that improved the
paper. 

{\it Facilities:} \facility{\spitzer}

\clearpage

\begin{deluxetable}{lrrlllllllll}
\setlength{\tabcolsep}{0.03in}
\tabletypesize{\tiny}
\tablewidth{0pt}
\tablecolumns{12}
\tablecaption{LMC Spitzer sample\label{tab:sample}}
\tablehead{
\colhead{MSX} &
\colhead{RA\tablenotemark{a}} &
\colhead{Dec\tablenotemark{a}} &
\colhead{J} &
\colhead{H} &
\colhead{K} &
\colhead{A} &
\colhead{\protect\egant} &
\colhead{SIMBAD} &
\colhead{\protect\tmass} &
\colhead{\protect\iras} &
\colhead{Other} \\ 
\colhead{LMC} &
\colhead{J2000} &
\colhead{J2000} &
\colhead{(mag)} &
\colhead{(mag)} &
\colhead{(mag)} &
\colhead{(mag)} &
\colhead{type} &
\colhead{type\tablenotemark{b}} &
\colhead{Name} &
\colhead{Name} &
\colhead{Name\tablenotemark{b}}
\\ 
\colhead{No.} &
& & & & & & & & & & \\
}
\startdata
22   &  05 04 47.14  &  -66 40 30.4  & 14.18 & 13.75 & 12.96 &  6.06 &   PN      & IR source       & J05044715-6640307 & 05047-6644          &  \\ 
87   &  05 10 19.63  &  -69 49 51.2  & 14.60 & 12.73 & 11.07 &  6.42 &   OH/IR   &                 & J05101962-6949514 &   &            \\ 
95   &  05 10 00.00  &  -69 56 09.6  & 13.39 & 11.49 & 10.01 &  6.27 &   C IR    &                 & J05095999-6956097 &   &            \\ 
141  &  05 05 33.48  &  -70 33 46.8  &  8.84 &  7.97 &  7.63 &  6.36 &   RSG     & M               & J05053350-7033469 &          &  [L72] LH 24- 15           \\ 
217  &  05 13 24.67  &  -69 10 48.4  & 14.69 & 14.21 & 13.23 &  5.92 &   PN      & IR source       & J05132465-6910480 & 05137-6914          &  \\ 
218  &  05 13 16.39  &  -68 44 10.0  & 13.39 & 11.47 &  9.93 &  6.15 &   OH/IR   &                 & J05131640-6844099 &   &            \\ 
220  &  05 12 32.06  &  -69 15 40.7  & 12.91 & 11.15 &  9.78 &  6.21 &   C IR    & V*              & J05123206-6915404 &  &DCMC J051232.11-691540.5            \\ 
222  &  05 13 42.48  &  -69 35 21.8  & 14.74 & 15.14 & 13.95 &  6.44 &   PN      &                 & J05134244-6935219 &   &            \\ 
264  &  05 14 49.75  &  -67 27 19.8  &  8.62 &  7.75 &  7.39 &  5.85 &   RSG     & M3Iab: V*       & J05144972-6727197 & 05148-6730                   & HV 916 \\ 
412  &  05 15 41.28  &  -73 47 13.9  &  6.59 &  5.71 &  5.26 &  4.57 &   O AGB   & Me V*           & J05154126-7347137 & 05169-7350                   & RS Men \\ 
438  &  05 25 19.51  &  -71 04 02.6  & 16.06 & 13.23 & 10.96 &  6.09 &   OH/IR   & IR source       & J05251951-7104027 &               &  LI-LMC 1028          \\ 
529  &  05 23 43.63  &  -65 41 59.6  &  9.00 &  8.10 &  7.74 &  6.24 &   RSG     & M3/M4 V*        & J05234361-6541596 & 05235-6544                 & HV 12793 \\ 
549  &  05 26 11.38  &  -66 12 10.8  &  9.32 &  8.26 &  7.71 &  6.04 &   O AGB   & Star in cluster & J05261135-6612111 &           &  NGC 1948 WBT 54          \\ 
559  &  05 25 49.25  &  -66 15 08.3  & 14.38 & 13.58 & 12.49 &  5.98 &   PN      & UV source       & J05254923-6615087 &        & [HCB95] LH 52 4978           \\ 
587  &  05 31 04.15  &  -69 19 03.3  &  9.12 &  8.12 &  7.65 &  5.64 &   O AGB   &                 & J05310418-6919030 &   &            \\ 
589  &  05 26 34.80  &  -68 51 40.0  &  8.50 &  7.65 &  7.29 &  5.90 &   RSG     & M2Iab:+...      & J05263479-6851400 &               & [GMP94] 301           \\ 
593  &  05 28 28.87  &  -68 07 08.0  &  8.48 &  7.66 &  7.34 &  5.95 &   RSG     & M0Ia V*         & J05282886-6807078 &                   & HV 2561           \\ 
597  &  05 29 42.24  &  -68 57 17.3  &  8.00 &  7.24 &  6.96 &  6.21 &   RSG     & M1Ia            & J05294221-6857173 & 05300-6859          &  \\ 
642  &  05 28 48.14  &  -71 02 28.7  & 11.31 &  9.70 &  8.64 &  6.30 &   C IR    & M8              & J05284817-7102289 & 05294-7104          &  \\ 
764  &  05 32 52.68  &  -69 46 22.8  & 13.92 & 13.76 & 12.84 &  4.93 &   PN      & IR source       & J05325272-6946226 & 05333-6948          &  \\ 
769  &  05 31 42.41  &  -68 34 53.7  & 15.08 & 14.55 & 13.45 &  5.91 &   PN      &                 & J05314242-6834539 &   &            \\ 
773  &  05 35 41.14  &  -69 11 59.6  &  9.58 &  8.55 &  8.06 &  6.07 &   O AGB   &                 & J05354110-6911596 &   &            \\ 
775  &  05 32 56.18  &  -68 12 49.0  & 13.24 & 11.32 &  9.85 &  6.44 &   C IR    &                 & J05325618-6812487 &   &            \\ 
807  &  05 32 37.15  &  -67 06 56.5  & 15.50 & 13.63 & 11.93 &  5.66 &   OH/IR   &                 & J05323716-6706564 &   &            \\ 
810  &  05 30 20.66  &  -66 53 01.7  &  8.78 &  7.92 &  7.72 &  6.21 &   RSG     & Double star     & J05302067-6653018 &         & CCDM J05303-6653B           \\ 
815  &  05 35 14.09  &  -67 43 55.6  &  9.49 &  8.68 &  8.14 &  6.37 &   O AGB   & M4 V*           & J05351409-6743558 &                   & HV 1001           \\ 
836  &  05 32 31.99  &  -66 27 15.1  & 15.94 & 15.01 & 14.00 &  5.85 &   PN      & IR source       & J05323195-6627154 & 05325-6629          &  \\ 
839  &  05 31 36.82  &  -66 30 07.9  &  8.44 &  7.66 &  7.36 &  6.27 &   RSG     &                 & J05313681-6630076 &   &            \\ 
870  &  05 35 28.32  &  -66 56 02.4  &  8.45 &  7.65 &  7.30 &  6.20 &   RSG     & M3Iab:          & J05352832-6656024 & 05354-6657           & Dachs LMC 2-16 \\ 
889  &  05 38 31.63  &  -69 02 14.6  & 14.53 & 13.80 & 12.68 &  4.65 &   PN      & Molecular cloud & J05383167-6902146 &         & [JGB98] 30 Dor-06           \\ 
890  &  05 36 25.87  &  -69 22 55.9  & 10.12 &  9.73 &  8.79 &  4.75 &   PN      & B:e Em. object  & J05362586-6922558 & 05368-6924            & HD 37974 \\ 
894  &  05 38 44.71  &  -69 24 39.6  & 13.91 & 14.00 & 12.85 &  6.26 &   PN      & Em. object      & J05384470-6924395 &            & LHA 120-N 158B           \\ 
897  &  05 40 43.73  &  -69 21 57.9  &  8.87 &  7.96 &  7.49 &  6.29 &   O AGB   &                 & J05404375-6921581 &   &            \\ 
934  &  05 39 15.86  &  -69 30 38.5  & 14.84 & 14.42 & 13.48 &  4.95 &   PN      & Molecular cloud & J05391587-6930384 &            & [JGB98] N158-2           \\ 
939  &  05 40 48.48  &  -69 33 36.0  &  9.24 &  8.26 &  7.75 &  5.71 &   O AGB   & Open cluster    & J05404850-6933360 &                & [HS66] 385           \\ 
943  &  05 41 10.63  &  -69 38 03.8  &  8.80 &  8.00 &  7.66 &  6.42 &   RSG     & M0 V*           & J05411066-6938040 &   &            \\ 
1007\tablenotemark{c} &  04 28 30.17  &  -69 30 50.0  & 16.29 & 13.83 & 11.87 &  6.66 &   OH/IR   &                 & \nodata           & 04286-6937          &  \\    
1072 &  04 40 28.51  &  -69 55 13.8  & 10.16 &  8.92 &  8.16 &  5.35 &   PN      & M7.5            & J04402848-6955135 & 04407-7000          &  \\ 
1120 &  04 47 16.08  &  -68 24 25.5  & 14.46 & 12.30 & 10.66 &  6.35 &   OH/IR   & IR source       & J04471609-6824256 &                 & LI-LMC 31           \\ 
1132 &  04 49 22.46  &  -69 24 34.6  &  9.11 &  8.18 &  7.78 &  6.17 &   O AGB   & M V*            & J04492246-6924344 &                   & HV 2236           \\ 
1150 &  04 39 23.66  &  -73 11 02.8  &  6.92 &  5.99 &  5.52 &  4.70 &   O AGB   &                 & J04392369-7311028 &   &            \\ 
1182\tablenotemark{c} &  04 55 10.49  &  -68 20 29.8  &  9.25 &  7.74 &  6.85 &  2.43 &   \nodata & M7.5            & J04551048-6820298 & 04553-6825          &  \\
1189 &  04 55 03.07  &  -69 29 12.8  &  8.67 &  7.68 &  7.20 &  5.32 &   O AGB   & IR source M2    & J04550307-6929127 & 04553-6933          &  \\ 
1204 &  04 55 16.03  &  -69 19 12.0  &  8.55 &  7.74 &  7.37 &  6.35 &   RSG     & Em.-line star M & J04551604-6919120 &           & [BE74] 164           \\ 
1280 &  05 00 19.03  &  -67 07 58.1  & 12.03 & 10.44 &  9.28 &  6.44 &   C IR    & M9              & J05001899-6707580 & 05003-6712          &  \\ 
1282 &  05 01 0.86   &  -67 35 23.6  & 17.24 & 14.51 & 12.18 &  6.43 &   OH/IR   &                 & J05010087-6735236 &   &            \\ 
1306 &  04 52 58.80  &  -68 02 56.8  & 15.85 & 15.45 & 14.15 &  6.34 &   PN      & Em. object      & J04525878-6802569 &               & LHA 120-S 5           \\ 
1326 &  04 56 47.04  &  -69 50 24.7  & 10.05 &  9.65 &  8.84 &  4.85 &   PN      & Em.-line star B8Ia & J04564705-6950247 & 04571-6954        & HD 268835 \\ 
1328 &  04 57 43.30  &  -70 08 50.3  &  8.42 &  7.62 &  7.32 &  5.60 &   RSG     & M4 V*           & J04574331-7008503 & 04581-7013                  & HV 2255 \\ 
1330 &  04 55 21.65  &  -69 47 16.8  &  8.87 &  7.95 &  7.61 &  6.03 &   RSG     & M               & J04552165-6947167 &         & [M2002] LMC 24410           \\ 
1384 &  05 43 36.05  &  -70 10 35.0  & 16.37 & 13.53 & 11.44 &  6.37 &   OH/IR   & V*              & J05433602-7010351 &   &            \\ 
1400 &  05 40 20.62  &  -66 14 44.2  & 16.60 & 13.91 & 11.77 &  6.02 &   OH/IR   &                 & J05402057-6614442 &   &            \\ 
1429 &  05 44 13.70  &  -66 16 44.8  &  9.01 &  7.92 &  7.48 &  6.48 &   O AGB   & M0.5 V*         & J05441373-6616445 &                   & HV 2834           \\ 
1488 &  05 50 6.72   &  -71 46 03.0  & 15.14 & 12.91 & 11.04 &  6.50 &   OH/IR   & IR source       & J05500676-7146026 & 05508-7146          &  \\ 
1492 &  05 49 8.83   &  -71 32 07.1  & 12.29 & 10.66 &  9.49 &  6.46 &   C IR    &                 & J05490888-7132069 &   &            \\ 
1524 &  05 55 21.05  &  -70 00 03.2  & 11.97 & 10.32 &  9.11 &  6.04 &   C IR    &                 & J05552103-7000030 & 05558-7000          &  \\ 
1592\tablenotemark{c} &  05 56 38.76  &  -67 53 34.4  & \nodata & \nodata & \nodata & 5.81 & \nodata & Carbon star  & \nodata & 05568-6753  &  \\
1652 &  06 02 31.03  &  -67 12 46.8  & 17.85 & 15.64 & 12.99 &  6.14 &   OH/IR   & Carbon star     & J06023105-6712469 & 06025-6712          &  \\ 
1677 &  06 01 27.82  &  -65 05 23.3  &  6.81 &  5.88 &  5.15 &  3.05 &   O AGB   & IR source       & J06012780-6505231 & 06013-6505          &  \\ 
1686 &  06 06 47.81  &  -66 48 12.6  &  6.33 &  5.43 &  4.84 &  3.58 &   C AGB   &                 & J06064779-6648125 &   &            \\ 
1794 &  05 40 43.99  &  -69 25 54.5  & 15.89 & 15.36 & 14.23 &  6.44 &   PN      &                 & J05404401-6925546 &   &            \\ 
04374-6831\tablenotemark{c}$^{\rm{,}}$\tablenotemark{d} & 04 37 22.73 &  -68 25 03.3 &  \nodata   &  \nodata  &  \nodata  &  \nodata   & \nodata  &     & \nodata  & 04374-6831 &   \\
\enddata
\tablenotetext{a}{Source positions are from the \protect\tmass\ catalog.}
\tablenotetext{b}{Source types and names from SIMBAD,
simbad.u-strasbg.fr/sim-fid.pl. Spectral types are indicated where known,
blank indicates that the object is a star of unknown spectral type. V*
indicates a variable star, and ``Em.'' indicates emission.}
\tablenotetext{c}{Additional \protect\iras\ source included in sample (see
\protect\S\ref{sec:obs}).}
\tablenotetext{d}{This object was not detected by \protect\msx, so is designated by
its \iras\ name. \iras\ flux densities for this object are $F$(12\,\micron):
0.199~Jy; $F$(25\,\micron): 0.25~Jy.}
\end{deluxetable}

\clearpage

\begin{deluxetable}{llllll}
\tabletypesize{\scriptsize}
\tablewidth{0pt}
\tablecolumns{6}
\tablecaption{Properties of the IRS spectra.\label{tab:data}}
\tablehead{
\colhead{\protect\msx} &
\colhead{Envelope} &
\colhead{IRS} &
\colhead{\protect\lir} &
\colhead{KSPW} &
\colhead{\protect\egant}  \\
\colhead{Name} &
\colhead{chemistry} &
\colhead{type} &
\colhead{\protect\lsun} &
\colhead{class} &
\colhead{type} \\
\colhead{[1]} &
\colhead{[2]} &
\colhead{[3]} &
\colhead{[4]} &
\colhead{[5]} &
\colhead{[6]}  \\
}
\startdata
1280\tablenotemark{a}$^{\rm{,}}$\tablenotemark{b}    & O-rich  & O AGB     & 1.7 $\times 10^{4}$: & 3SE     &  C IR    \\
1524     & O-rich  & O AGB & 4.1 $\times 10^{4}$ &  3SE  &  C IR    \\
642      & O-rich  & O AGB & 4.2 $\times 10^{4}$ &  3SE  &  C IR    \\
1072     & O-rich  & O AGB & 6.5 $\times 10^{4}$ &  3SE  &  PN      \\
815      & O-rich  & RSG  & 5.5 $\times 10^{4}$ &  2SEc &  O AGB   \\
1132     & O-rich  & RSG  & 7.7 $\times 10^{4}$ &  2SEc &  O AGB   \\
549      & O-rich  & RSG  & 7.9 $\times 10^{4}$ &  2SEc &  O AGB   \\
529      & O-rich  & RSG  & 7.9 $\times 10^{4}$ &  2SEc &  RSG     \\
939\tablenotemark{a}     & O-rich  & RSG       & 8.1 $\times 10^{4}$  & 2SEc    &  O AGB   \\
587      & O-rich  & RSG  & 8.5 $\times 10^{4}$ &  2SEc &  O AGB   \\
943      & O-rich  & RSG  & 8.7 $\times 10^{4}$ &  2SEc &  RSG     \\
1429     & O-rich  & RSG  & 9.1 $\times 10^{4}$ &  2SEc &  O AGB   \\
141      & O-rich  & RSG  & 9.2 $\times 10^{4}$ &  2SEc &  RSG     \\
1330     & O-rich  & RSG  & 9.3 $\times 10^{4}$ &  2SEc &  RSG     \\
897\tablenotemark{a}     & O-rich  & RSG       & 9.4 $\times 10^{4}$  & 2SEc    &  O AGB   \\
810\tablenotemark{c}     & O-rich  & RSG       & 9.4 $\times 10^{4}$: & 2SEc    &  RSG     \\
1204     & O-rich  & RSG  & 1.0 $\times 10^{5}$ &  2SEc &  RSG     \\
264      & O-rich  & RSG  & 1.1 $\times 10^{5}$ &  2SEc &  RSG     \\
839      & O-rich  & RSG  & 1.2 $\times 10^{5}$ &  2SEc &  RSG     \\
593      & O-rich  & RSG  & 1.2 $\times 10^{5}$ &  2SEc &  RSG     \\
870      & O-rich  & RSG  & 1.3 $\times 10^{5}$ &  2SEc &  RSG     \\
589\tablenotemark{a}     & O-rich  & RSG       & 1.3 $\times 10^{5}$  & 2SEc    &  RSG     \\
1328     & O-rich  & RSG  & 1.3 $\times 10^{5}$ &  2SEc &  RSG     \\
1189     & O-rich  & RSG  & 1.3 $\times 10^{5}$ &  2SEc &  O AGB   \\
597      & O-rich  & RSG  & 1.6 $\times 10^{5}$ &  2SEc &  RSG     \\
1150\tablenotemark{d}     & O-rich  & MW O AGB  & 4.0 $\times 10^{3}$ &  2SEa &  O AGB   \\
1686\tablenotemark{d}     & O-rich  & MW O AGB  & 4.6 $\times 10^{3}$ &  2SEa &  C AGB   \\
412\tablenotemark{d}     & O-rich  & MW O AGB  & 6.0 $\times 10^{3}$ &  2SEa &  O AGB   \\
1677\tablenotemark{d}     & O-rich  & MW O AGB  & 9.7 $\times 10^{3}$ &  2SEc &  O AGB   \\
807\tablenotemark{f}     & O-rich  & OH/IR     & 4.0 $\times 10^{4}$  & 3SB     &  OH/IR   \\
1182     & O-rich  & OH/IR SG & 4.7 $\times 10^{5}$ &  5?   &  -       \\
1326\tablenotemark{e}     & O-rich  & Peculiar & 5.5 $\times 10^{4}$ &  5?   &  PN      \\
890\tablenotemark{e}     & O-rich  & Peculiar & 7.1 $\times 10^{4}$ &  5?   &  PN      \\
1384     & C-rich  & C AGB  & 5.8 $\times 10^{3}$ &  3CE  &  OH/IR   \\
1400     & C-rich  & C AGB  & 7.1 $\times 10^{3}$ &  3C?  &  OH/IR   \\
1007\tablenotemark{g}    & C-rich  & C AGB     & 8.2 $\times 10^{3}$  & 3CE     &  OH/IR       \\
04374-6831\tablenotemark{h} & C-rich  & C AGB  & 1.0 $\times 10^{4}$ &  3CE  &  -       \\
1488     & C-rich  & C AGB  & 1.1 $\times 10^{4}$ &  3CE  &  OH/IR   \\
438      & C-rich  & C AGB  & 1.1 $\times 10^{4}$ &  3CE  &  OH/IR   \\
87       & C-rich  & C AGB  & 1.1 $\times 10^{4}$ &  3CE  &  OH/IR   \\
1120     & C-rich  & C AGB  & 1.1 $\times 10^{4}$ &  3CE  &  OH/IR   \\
1652     & C-rich  & C AGB  & 1.2 $\times 10^{4}$ &  3CR  &  OH/IR   \\
218      & C-rich  & C AGB  & 1.3 $\times 10^{4}$ &  3CE  &  OH/IR   \\
95       & C-rich  & C AGB  & 1.4 $\times 10^{4}$ &  3CE  &  C IR    \\
1282     & C-rich  & C AGB  & 1.5 $\times 10^{4}$ &  3CE  &  OH/IR   \\
220      & C-rich  & C AGB  & 1.6 $\times 10^{4}$ &  3CE  &  C IR    \\
1592     & C-rich  & C AGB  & 1.6 $\times 10^{4}$ &  3CR  &  -       \\
1492     & C-rich  & C AGB  & 1.9 $\times 10^{4}$ &  3CE  &  C IR    \\
775      & C-rich  & C AGB  & 1.9 $\times 10^{4}$ &  C?   &  C IR    \\
1794     & PAH     & H{\sc ii} & 1.7 $\times 10^{4}$ &  5UE  &  PN      \\
222      & PAH     & H{\sc ii} & 3.8 $\times 10^{4}$ &  5UE  &  PN      \\
1306     & PAH     & H{\sc ii} & 4.3 $\times 10^{4}$ &  5UE  &  PN      \\
894      & PAH     & H{\sc ii} & 5.6 $\times 10^{4}$ &  5UE  &  PN      \\
22       & PAH     & H{\sc ii} & 6.8 $\times 10^{4}$ &  5UE  &  PN      \\
836      & PAH     & H{\sc ii} & 7.7 $\times 10^{4}$ &  5UE  &  PN      \\
217      & PAH     & H{\sc ii} & 8.1 $\times 10^{4}$ &  5UE  &  PN      \\
559      & PAH     & H{\sc ii} & 2.0 $\times 10^{5}$ &  5UE  &  PN      \\
934      & PAH     & H{\sc ii} & 2.5 $\times 10^{5}$ &  5UE  &  PN      \\
764      & PAH     & H{\sc ii} & 2.7 $\times 10^{5}$ &  5UE  &  PN      \\
889      & PAH     & H{\sc ii} & 4.0 $\times 10^{5}$ &  5UE  &  PN      \\
769\tablenotemark{i}     & \nodata & \nodata   & \nodata              & \nodata &  PN      \\
773\tablenotemark{i}     & \nodata & \nodata   & \nodata              & \nodata &  O AGB   \\
\enddata

\tablenotetext{a}{For these objects, local sky subtraction was used.}
\tablenotetext{b}{Only SL module spectra were obtained for this object. The
luminosity was calculated using MSX~LMC~1189 as a template for the spectral
shape, so is highly uncertain.}
\tablenotetext{c}{This object exhibits extended emission due to a nearby or
companion object (resolved in SL but not LL spectral images). Slit losses
result in a large flux jump between the SL and LL modules, so the SL spectra
were scaled by a factor of 1.86 to match the LL2 spectrum; the luminosity is
therefore very uncertain.}
\tablenotetext{d}{The luminosities for these Milky Way O~AGB stars
  were calculated using the following assumed distances: MSX~LMC~1150
  4.2~kpc, MSX~LMC~1686 3.3~kpc, MSX~LMC~412 4.6~kpc, and MSX~LMC~1677
  5.1~kpc.  These distances were estimated using the K-band distance
  modulus with absolute K-band magnitudes derived using the
  color-period-luminosity relations for Mira variables
  \protect\citep{whi94}.}
\tablenotetext{e}{The luminosities for these two objects were taken
from \citet{kas06}. \protect\lir\ obtained from the flux density
integrated over the wavelength range 2.2 -- 35~\micron\ only.}
\tablenotetext{f}{A second source fell partially in the slit for several
modules and nod positions of this observation. The extraction aperture used
for this object was defined manually, using a tapering column of the default
SMART widths to ensure accurate flux calibration.}
\tablenotetext{g}{For the SL1 module of this spectrum, only the second slit
position was used in the final spectrum, as the flux density of the SL1
spectrum in the first nod position did not match the rest of the modules and
nod positions.}
\tablenotetext{h}{This object was not detected by \protect\msx, so is
designated by its \iras\ name.}
\tablenotetext{i}{These two objects were not observed due to peak-up failure.}
\tablecomments{Columns: [1] MSX/IRAS name; [2] the dominant chemistry
of the dust envelope, determined by the IR spectral features; [3] our
classification of the likely type of object, based on the IRS spectral
features, luminosity and all other available information; [4] the IR
(1 -- 100~\micron) luminosity derived as described in
\S\ref{subsec:res_lum}. Uncertain luminosities are indicated by a
colon; [5] the spectral class of the object in the scheme of
\protect\citet{kra02}; [6] the type of the object determined from the
NIR colors by \protect\egant.}
\end{deluxetable}

\clearpage

\begin{deluxetable}{lrrrr}
\tabletypesize{\small}
\tablewidth{0pt}
\tablecolumns{5}
\tablecaption{{\it Spitzer}/{\it 2MASS} color diagnostics for object types.\label{tab:diag}}
\tablehead{
\colhead{Class} &
\multicolumn{4}{c}{Color-color diagnostics}  \\
 & 
\colhead{H-[5.8]} &
\colhead{[8.0]-[24]} &
\colhead{K-[5.8]} &
\colhead{H-K} \\
\colhead{[1]} &
\colhead{[2]} &
\colhead{[3]} &
\colhead{[4]} &
\colhead{[5]} \\
}
\startdata
RSG                    & -2.5 -- -1.0 & [$\frac{H-[5.8]}{3}$\,+3.0] -- [$\frac{H-[5.8]}{3}$\,+4.5] & [$\frac{H-K-1.15}{0.3}$] -- [$\frac{H-K-0.8}{0.3}$]  & 0.2 -- 0.6 \\
O AGB\tablenotemark{a} & -0.5 -- 1.2  & 3.5 -- 4.0                                                 & -1.6 -- 0.0   & 0.7 -- 1.3  \\
MW O AGB               & -2.3 -- 0.0  & [$\frac{H-[5.8]}{3}$\,+2.2] -- [$\frac{H-[5.8]}{3}$\,+2.7] & [$\frac{H-K-1.05}{0.2}$] -- [$\frac{H-K-0.9}{0.2}$]  & 0.4 -- 0.75 \\
C AGB                  &  0.5 -- 6.0  & [$\frac{H-[5.8]}{6}$\,+0.5] -- [$\frac{H-[5.8]}{6}$\,+2.5] & [$\frac{H-K-2.0}{0.3}$] -- [$\frac{H-K-1.2}{0.3}$]   & 1.1 -- 2.7  \\
\htwo\                 & 2.0 -- 4.5   & 4.5 -- 8.0                                                 & 0.2 --3.0 & 0.7 -- 1.5                                             \\
\enddata 
\tablecomments{Columns: [1] Type of IR object; [2]-[3] Color
diagnostics for classifying objects based on \protect\spitzer\ and \protect\tmass\ colors.}
\tablenotetext{a}{These criteria are based on the colors of only four objects and thus are very uncertain.}
\end{deluxetable}

\clearpage

\begin{figure}
\epsscale{0.85}
\plotone{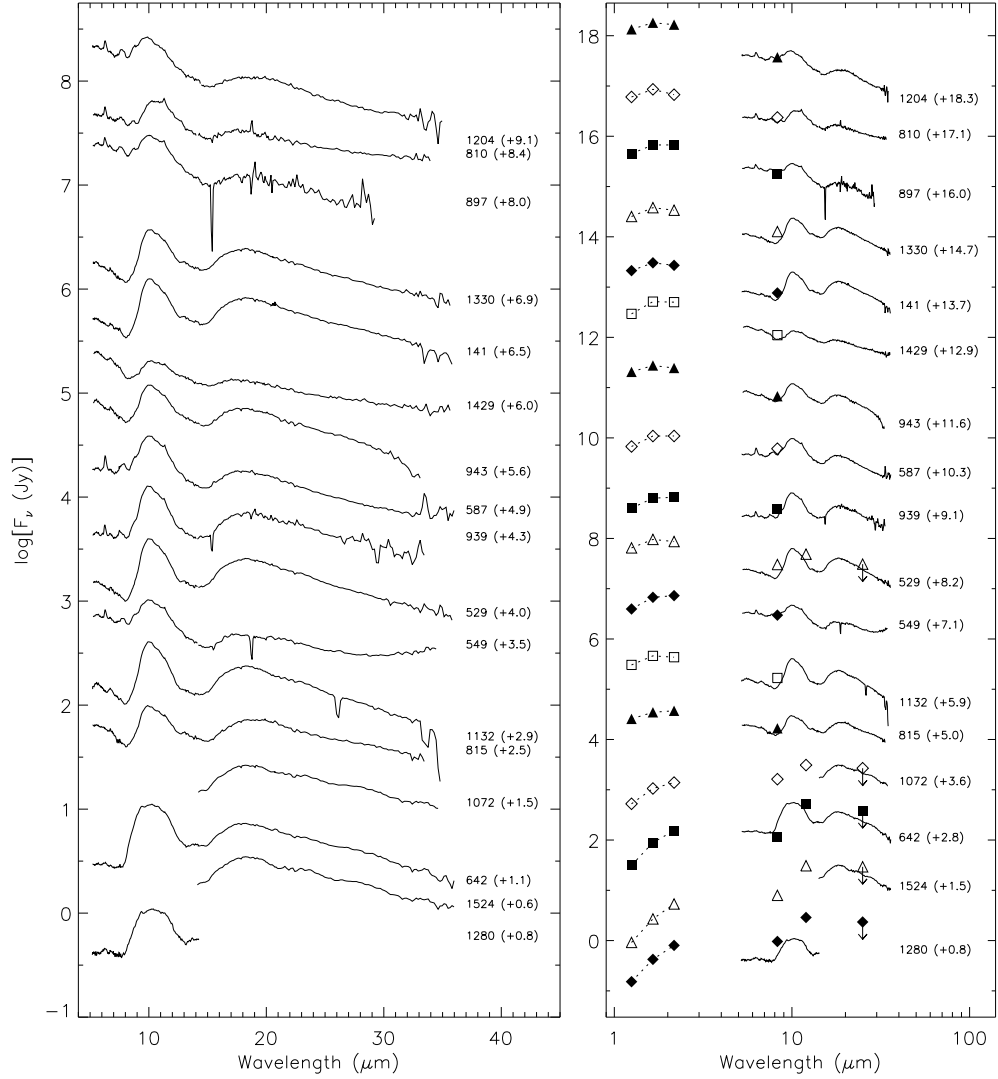}
\caption{{\it (a)} Spitzer IRS spectra of the stars showing silicate
features. The spectra are arranged according to IRS spectral type and then
object bolometric luminosity (Table \protect\ref{tab:data}). For each object,
the left panel shows the IRS spectrum on a linear wavelength scale, and the
right panel shows the spectrum with photometry from \protect\tmass,
\protect\msx, and \protect\iras, on a logarithmic wavelength
scale. Alternating symbols are used to make clear which photometric points are
associated with which IRS spectrum. The flux density is plotted on a
logarithmic scale and the value in parentheses next to the object name
indicates the offset in flux density applied to the
spectrum. \label{fig:spec}} \epsscale{1.0}
\end{figure}

\clearpage

\begin{figure}
\addtocounter{figure}{-1}
\epsscale{0.85}
\plotone{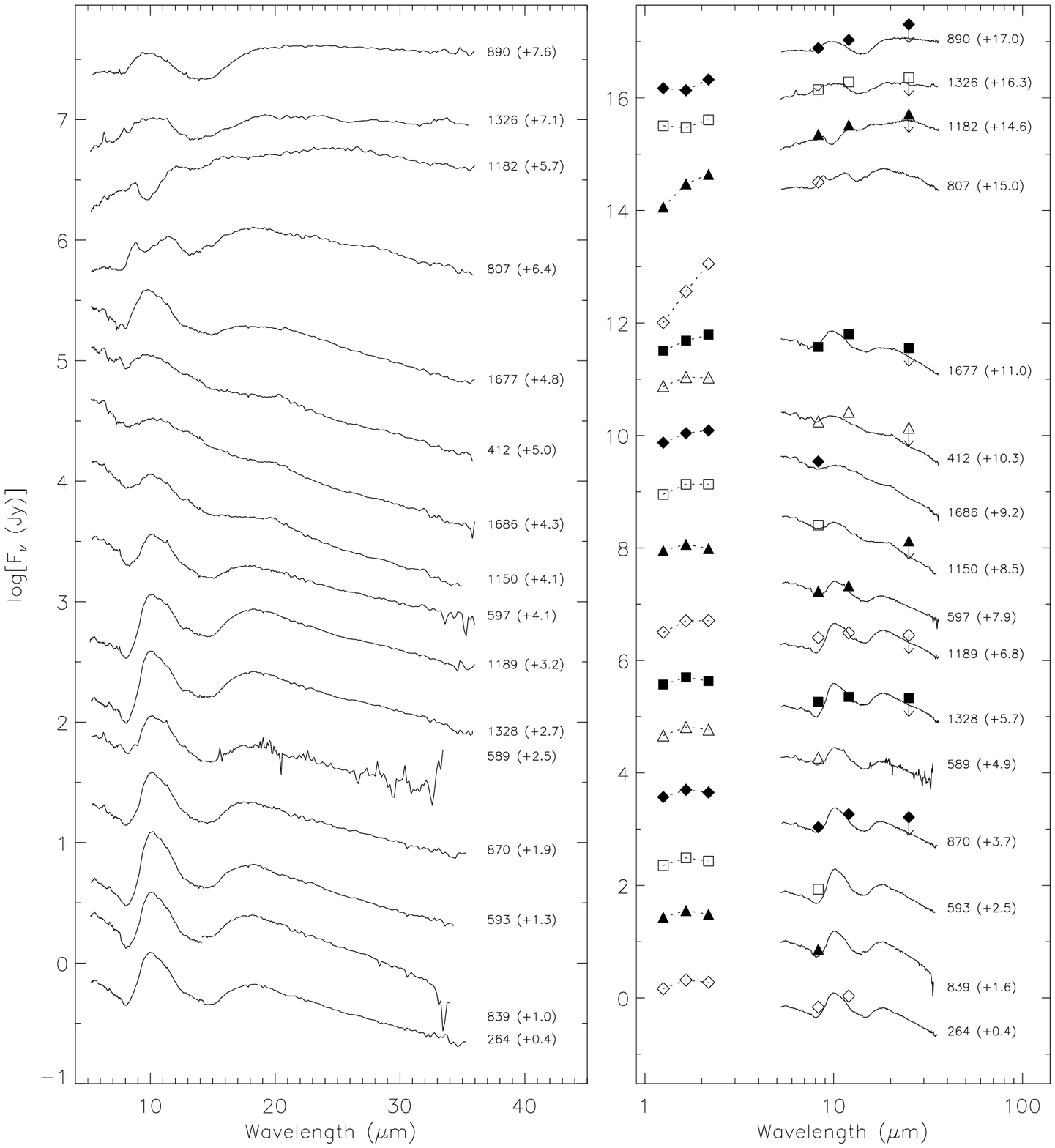}
\caption{\em (a) Continued}
\epsscale{1.0}
\end{figure}

\clearpage

\begin{figure}
\epsscale{0.85}
\addtocounter{figure}{-1}
\plotone{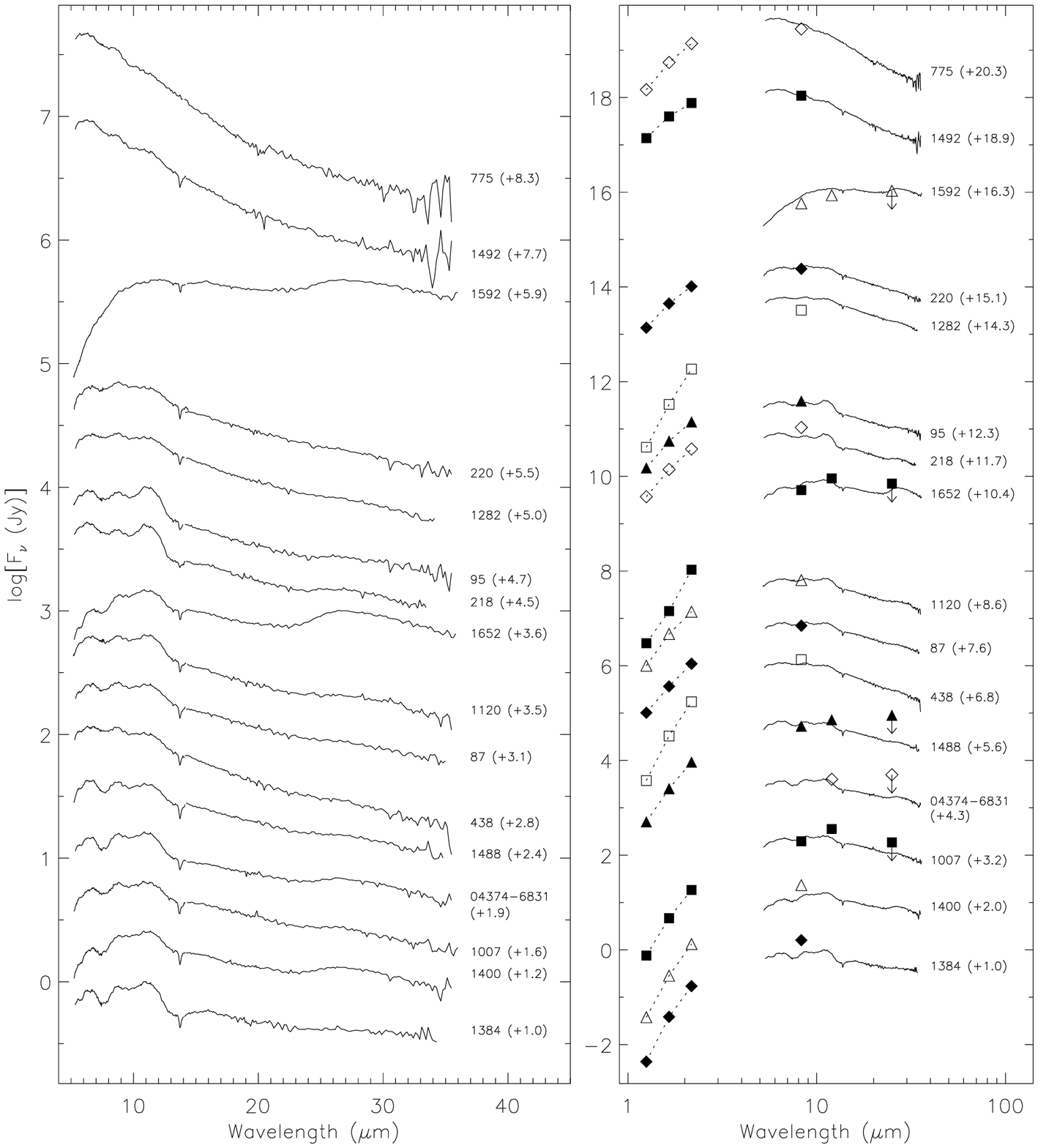}
\caption{{\it (b)} Spitzer IRS spectra of the stars showing SiC
features. In the right panel, \protect\tmass, \protect\msx, and
\protect\iras\ data are also shown where available. }  \epsscale{1.0}
\end{figure}

\clearpage

\begin{figure}
\epsscale{0.85}
\addtocounter{figure}{-1}
\plotone{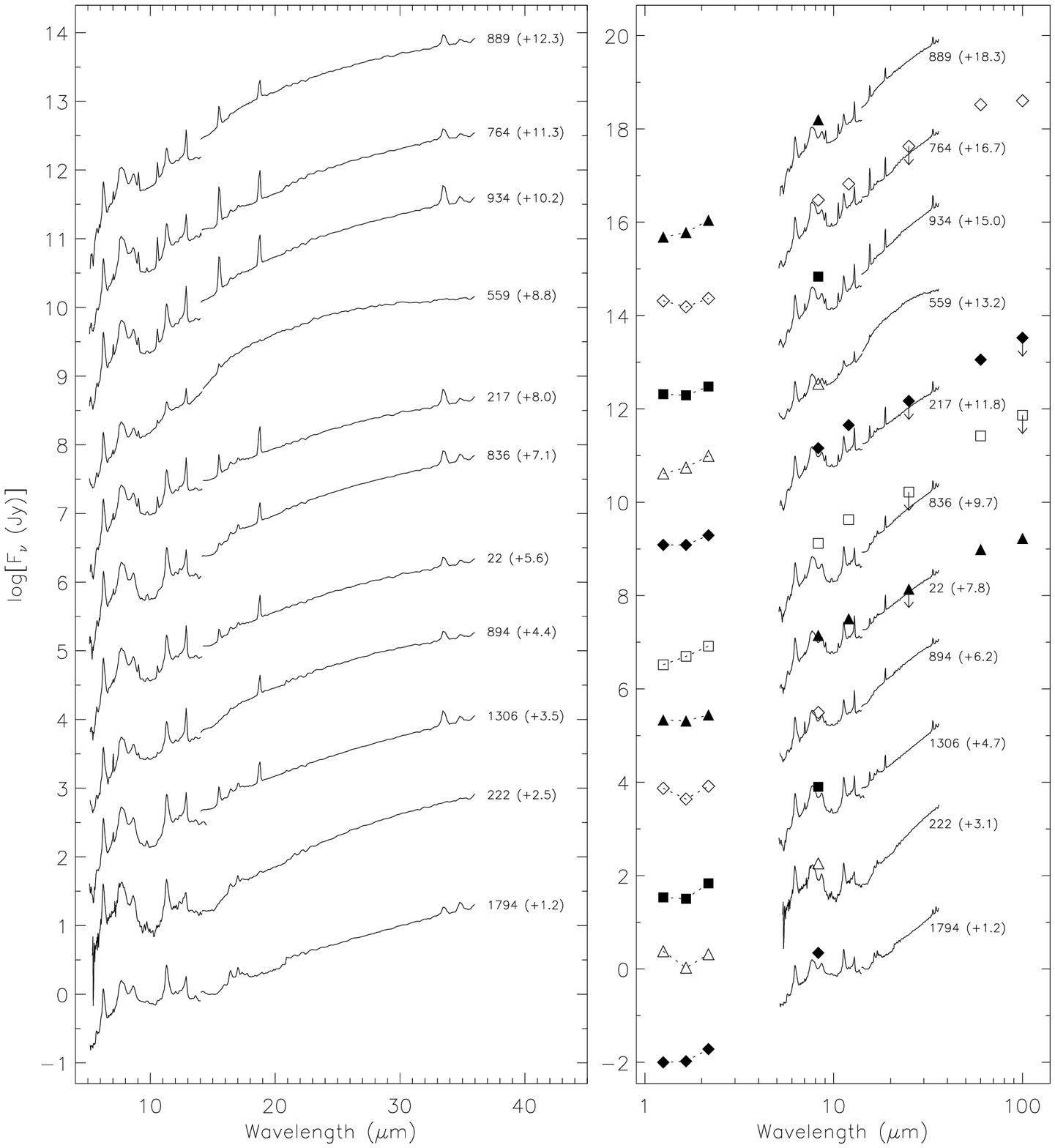}
\caption{{\it (c)} Spitzer IRS spectra of red objects showing emission
features.  In the right panel, \protect\tmass, \protect\msx, and
\protect\iras\ data are also shown where available.}  \epsscale{1.0}
\end{figure}

\clearpage

\begin{figure}
\epsscale{0.5}
\plotone{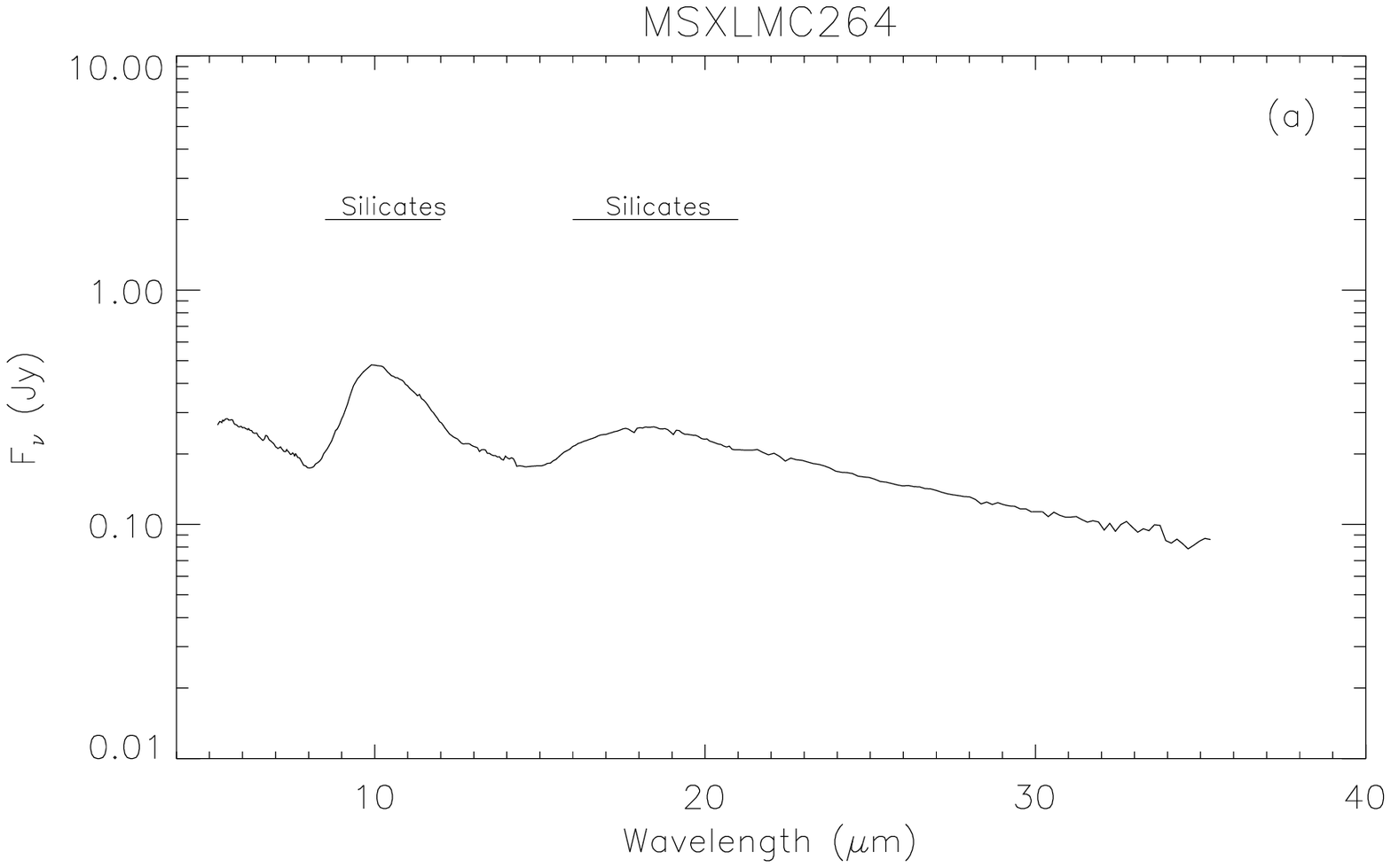}
\plotone{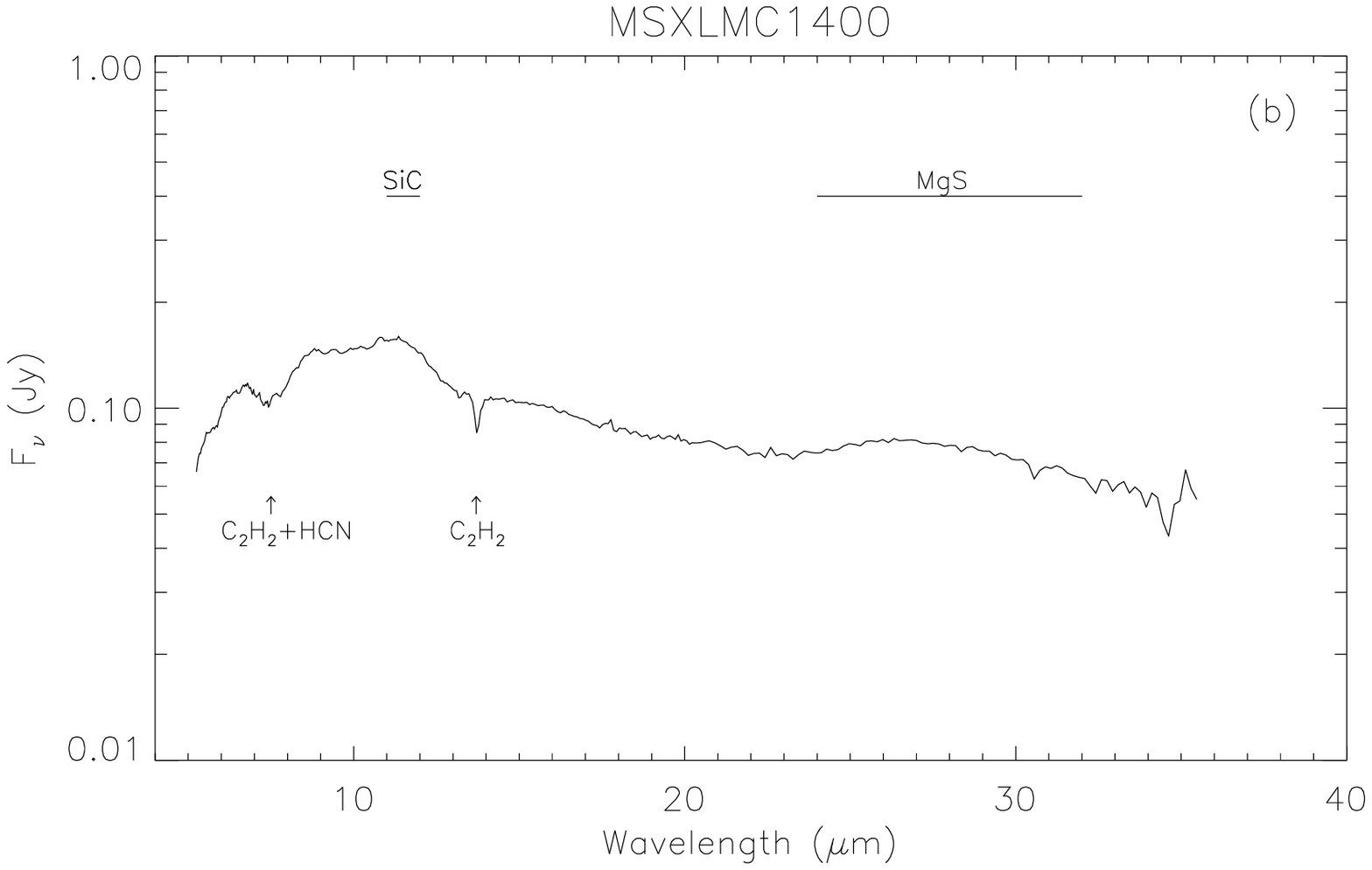}
\plotone{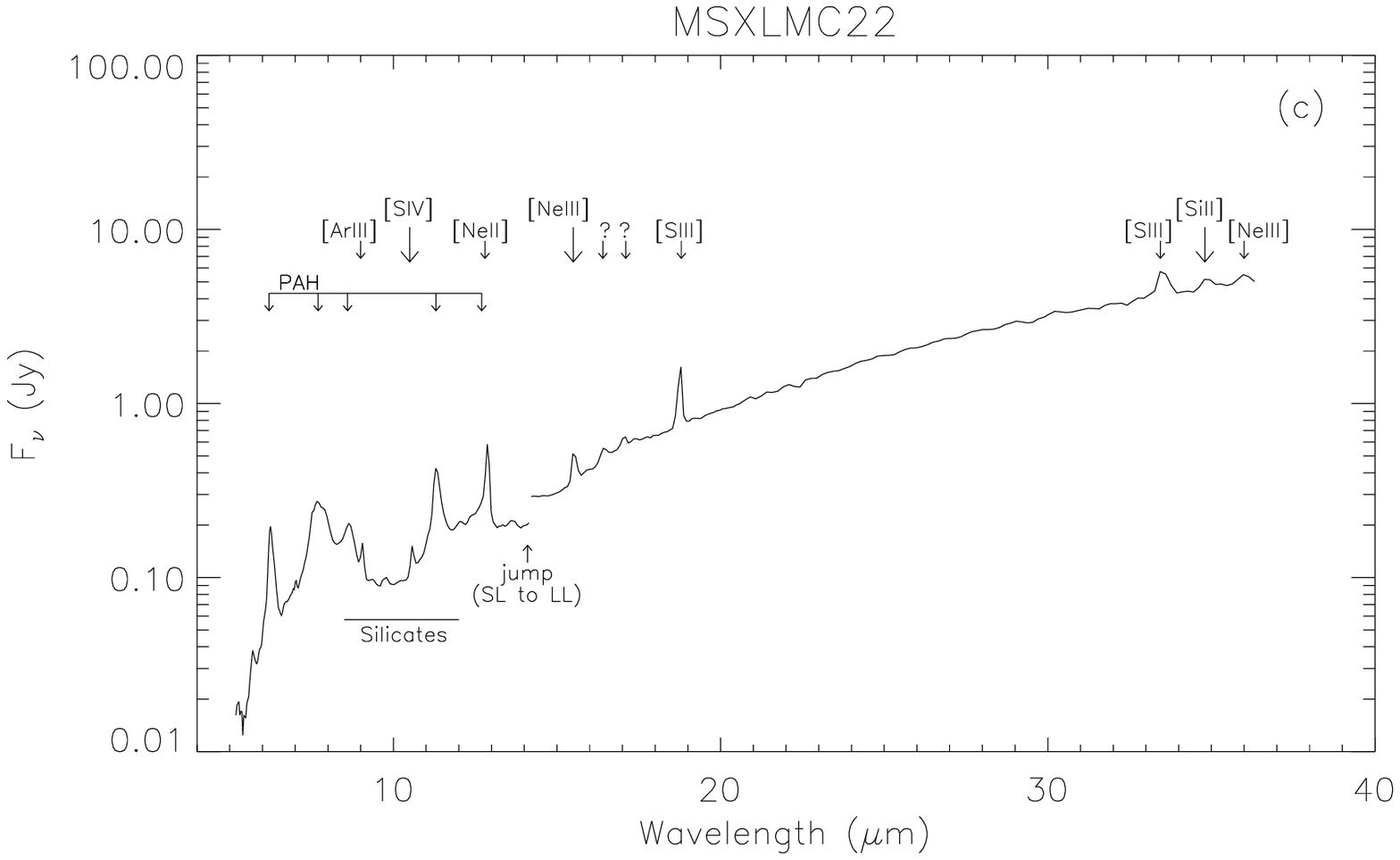}
\caption{A typical spectrum from each of the groups of objects in the
  sample, showing the primary spectral features detected for
  each. {\it (a)} The spectrum of O-rich object MSX~LMC~264. {\it (b)}
  The spectrum of C-rich object MSX~LMC~1400. {\it (c)} The spectrum of
  emission-line object MSX~LMC~22. The ``jump'' at $\sim$14~\micron\ is
  discussed in
  \S\ref{subsec:dis_class_pn}. \label{fig:egspec}}
  \epsscale{1.0}
\end{figure}

\clearpage

\begin{figure}
\epsscale{0.45}
\plotone{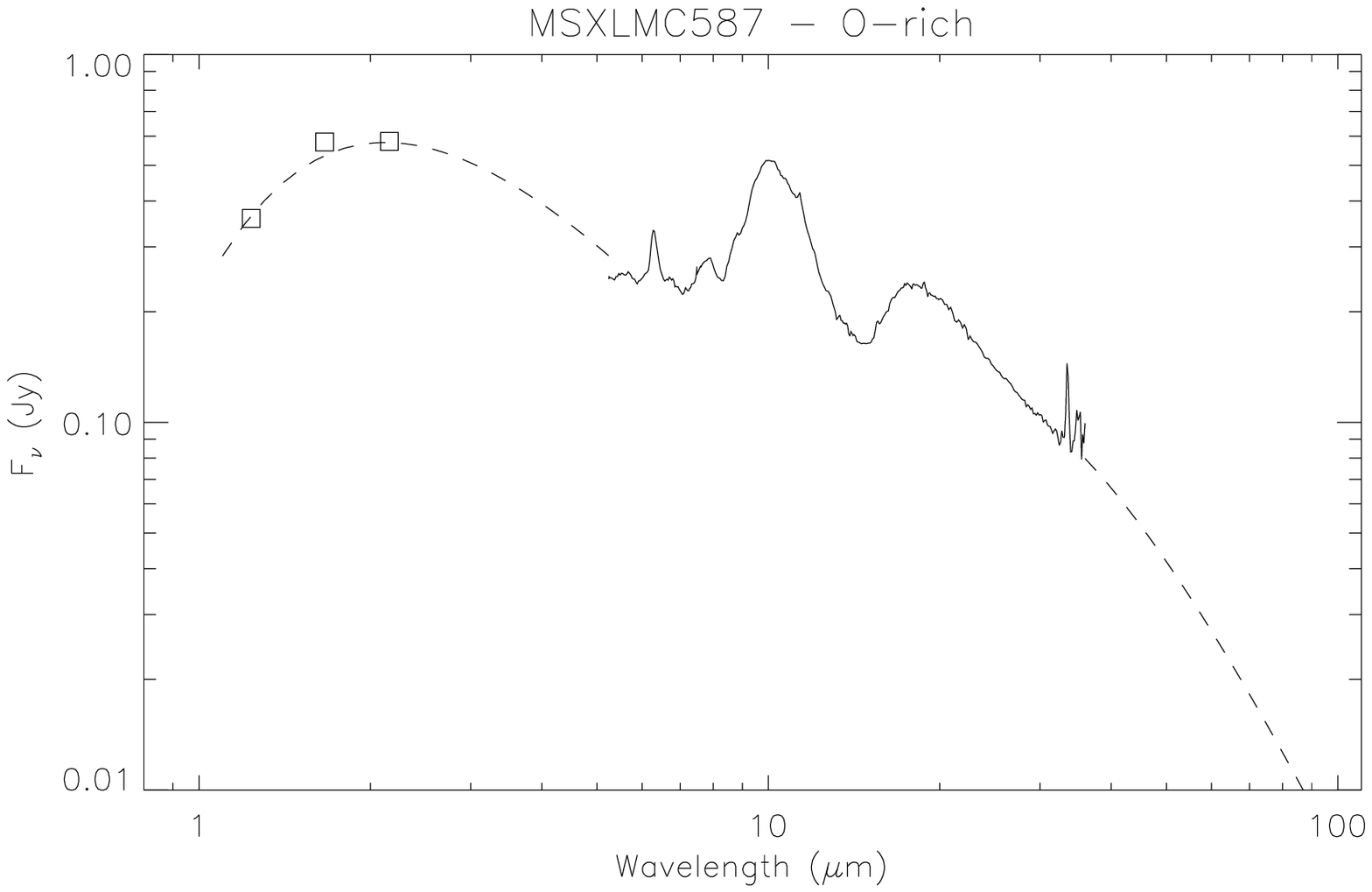}
\plotone{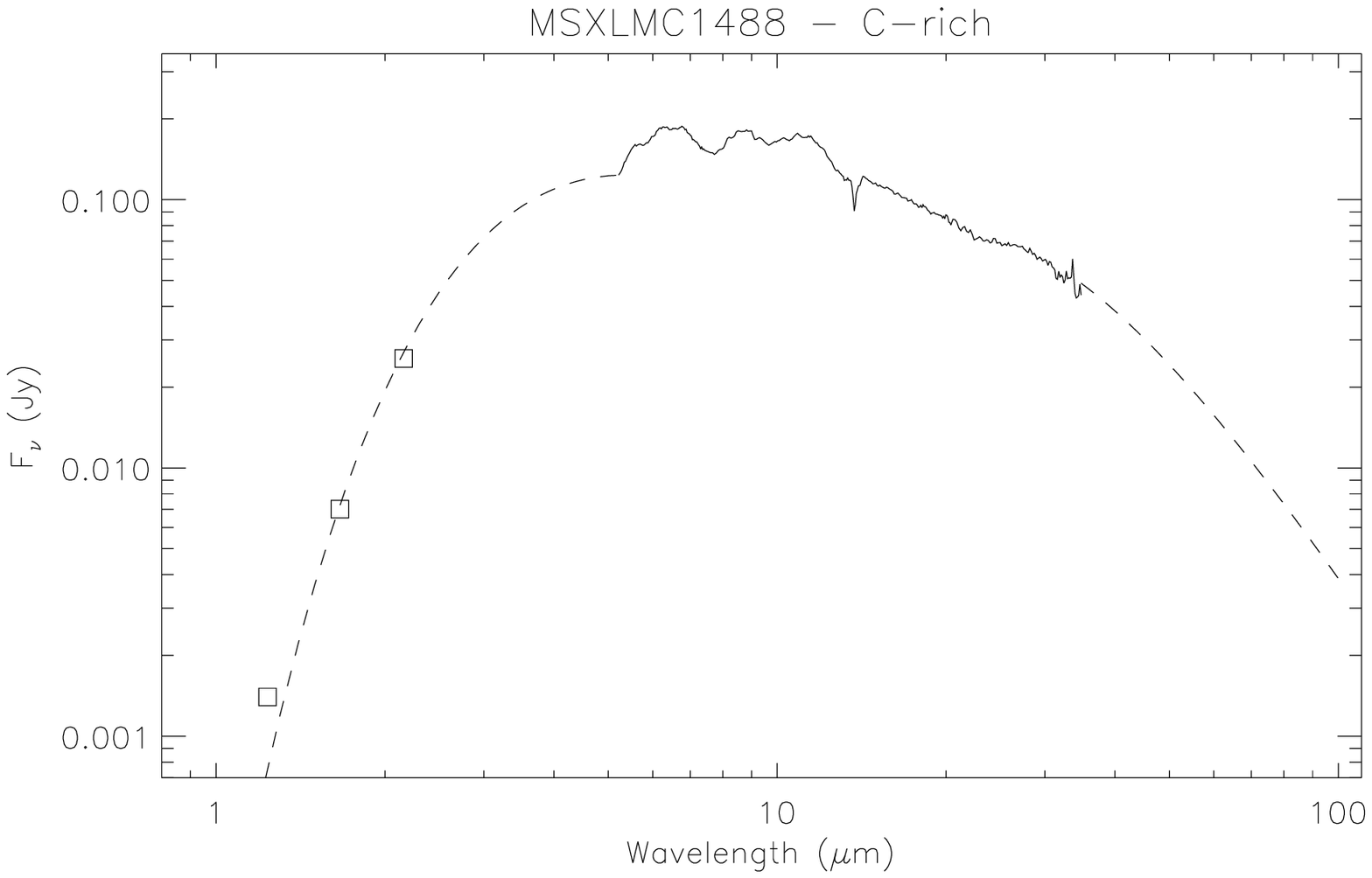}
\plotone{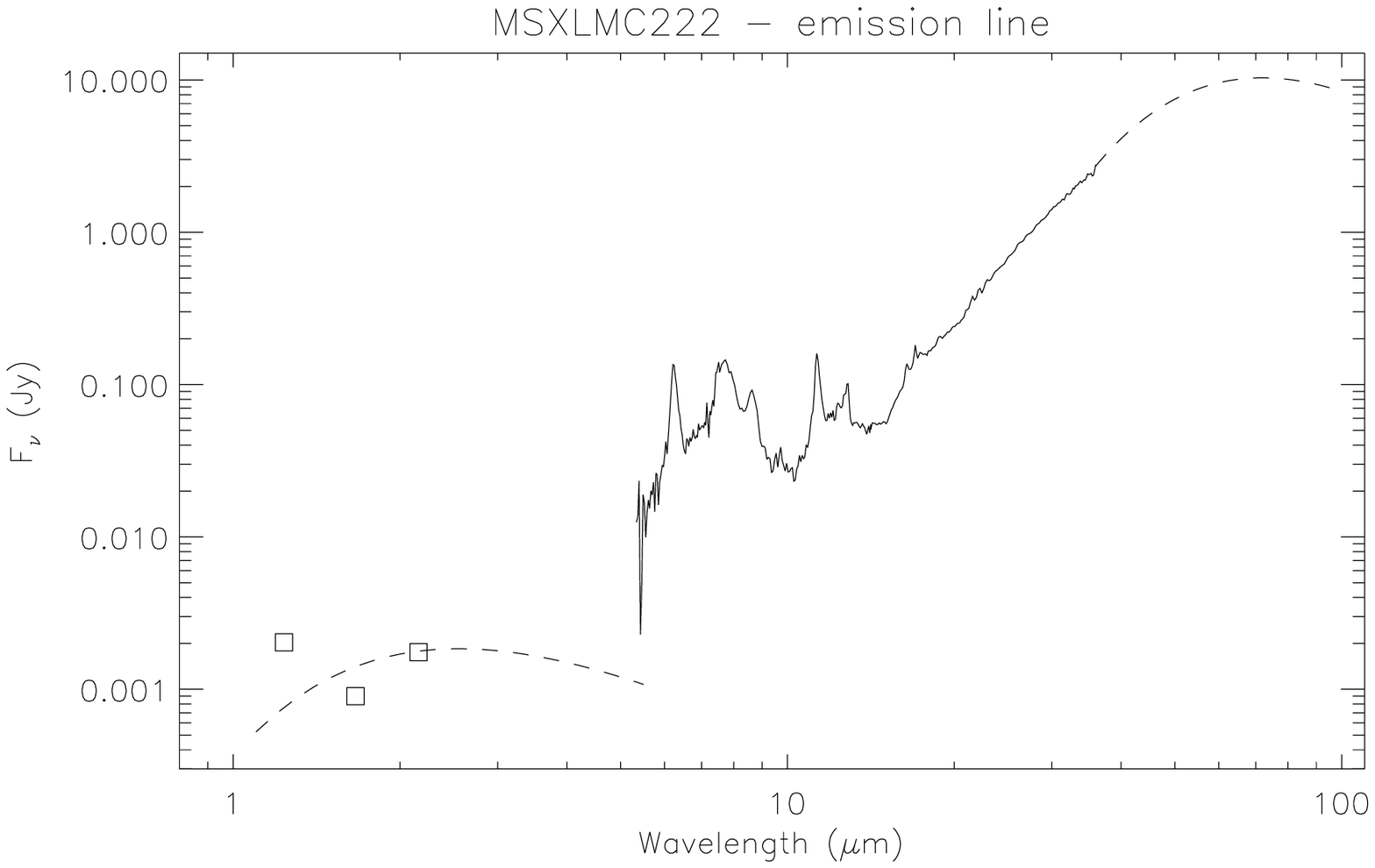}
\caption{Figures illustrating the method of calculating the luminosities for
the O-rich {\it (top)}, C-rich {\it (middle)} and emission-line {\it (bottom)}
objects. The flux density was integrated under the \protect\irs\ spectrum
between 5 and 36~\micron\ {\it (solid line)}; a warm blackbody {\it (dotted
line)}, scaled to match the \protect\tmass\ photometry {\it (open squares)},
was used to estimate the 1 -- 5~\micron\ flux densities; and a cool,
emissivity-weighted blackbody {\it (dashed line)} was used to estimate the 36
-- 100~\micron\ flux densities. \label{fig:lumcalc}} \epsscale{1.0}
\end{figure}

\clearpage

\begin{figure}
\epsscale{0.8} 
\plotone{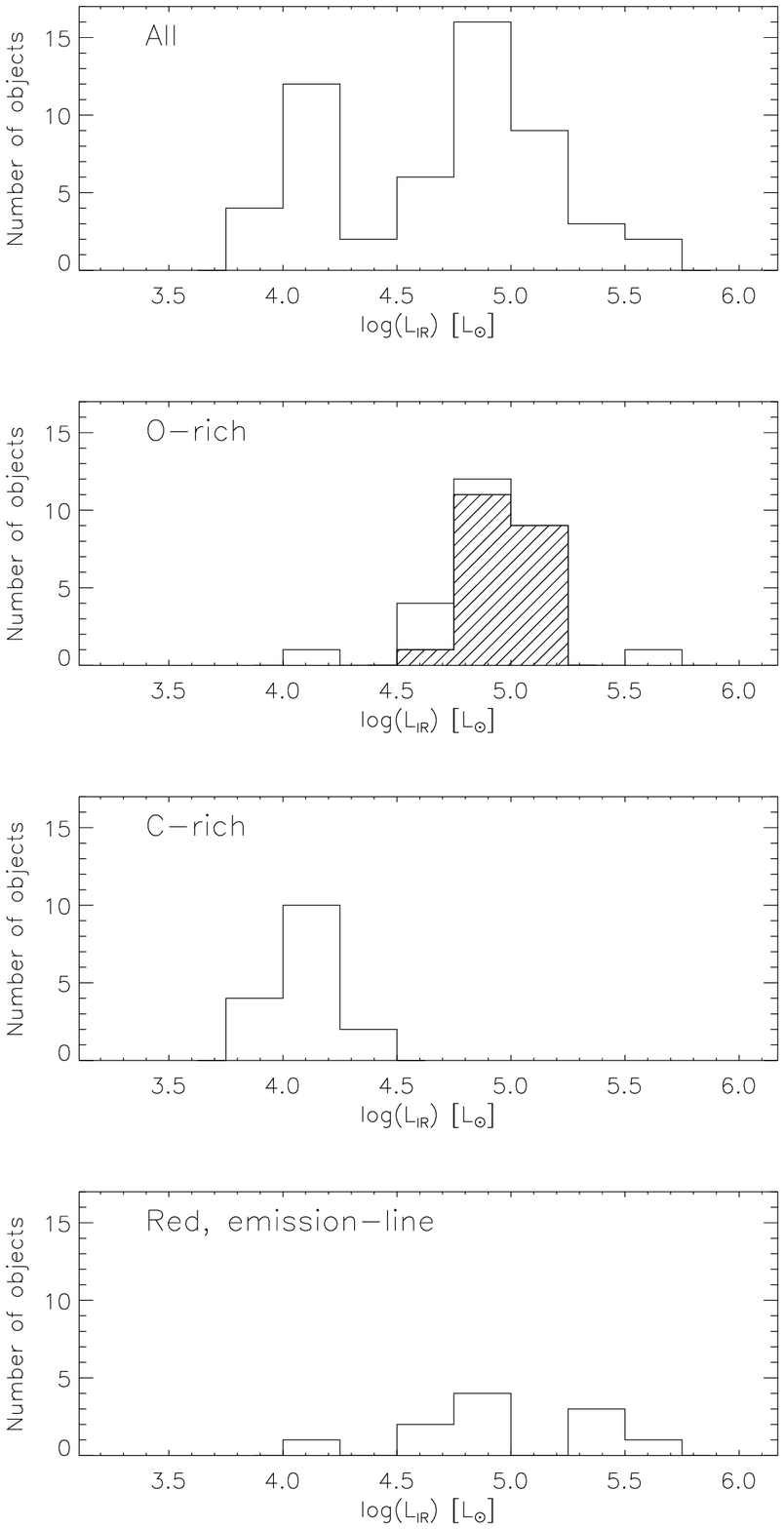}
\caption{Histograms showing the distributions of the IR luminosity for the
  whole sample {\it (upper panel)} and for the O-rich, C-rich and PAH-rich
  stars separately {\it (lower panels)}.  The hatched regions of the histogram
  for O-rich objects indicate stars classified as RSGs based on their IR
  luminosities {\it (diagonal)}. Stars identified as Galactic objects (see
  Table \protect\ref{tab:data}) are not shown. \label{fig:lumhist}}
  \epsscale{1.0}
\end{figure}

\clearpage

\begin{figure}
\epsscale{0.85}
\plottwo{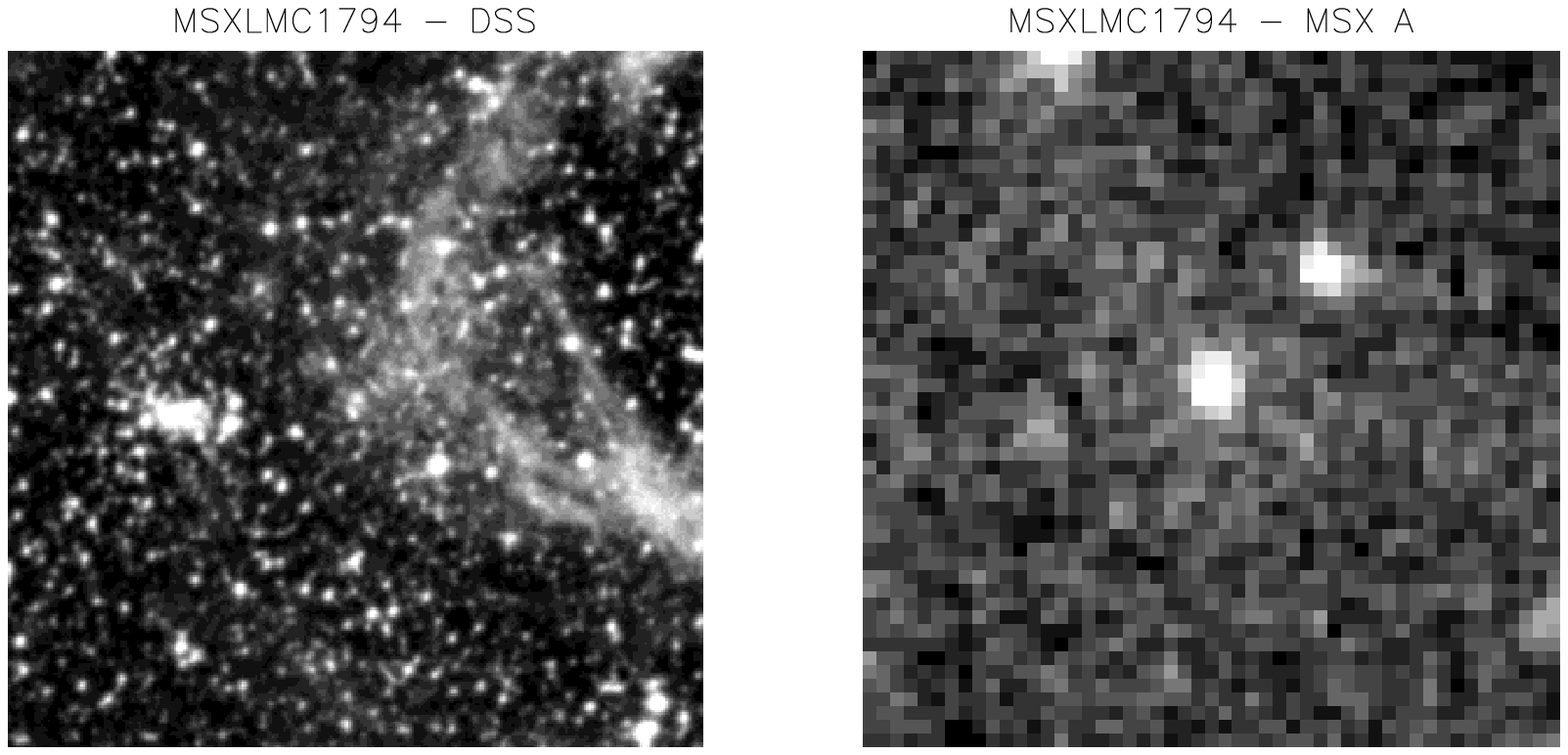}{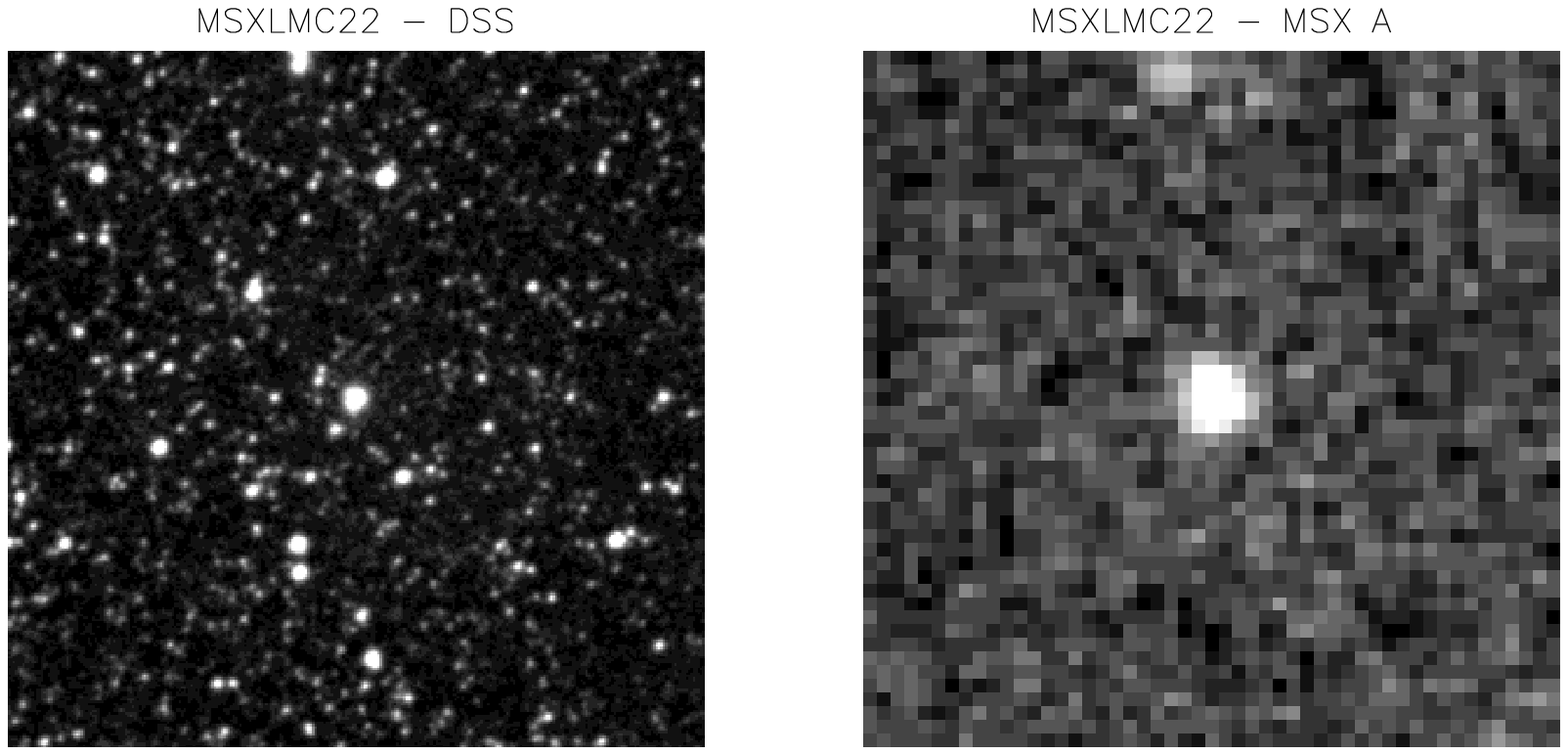}
\protect\vspace*{3mm}
\plottwo{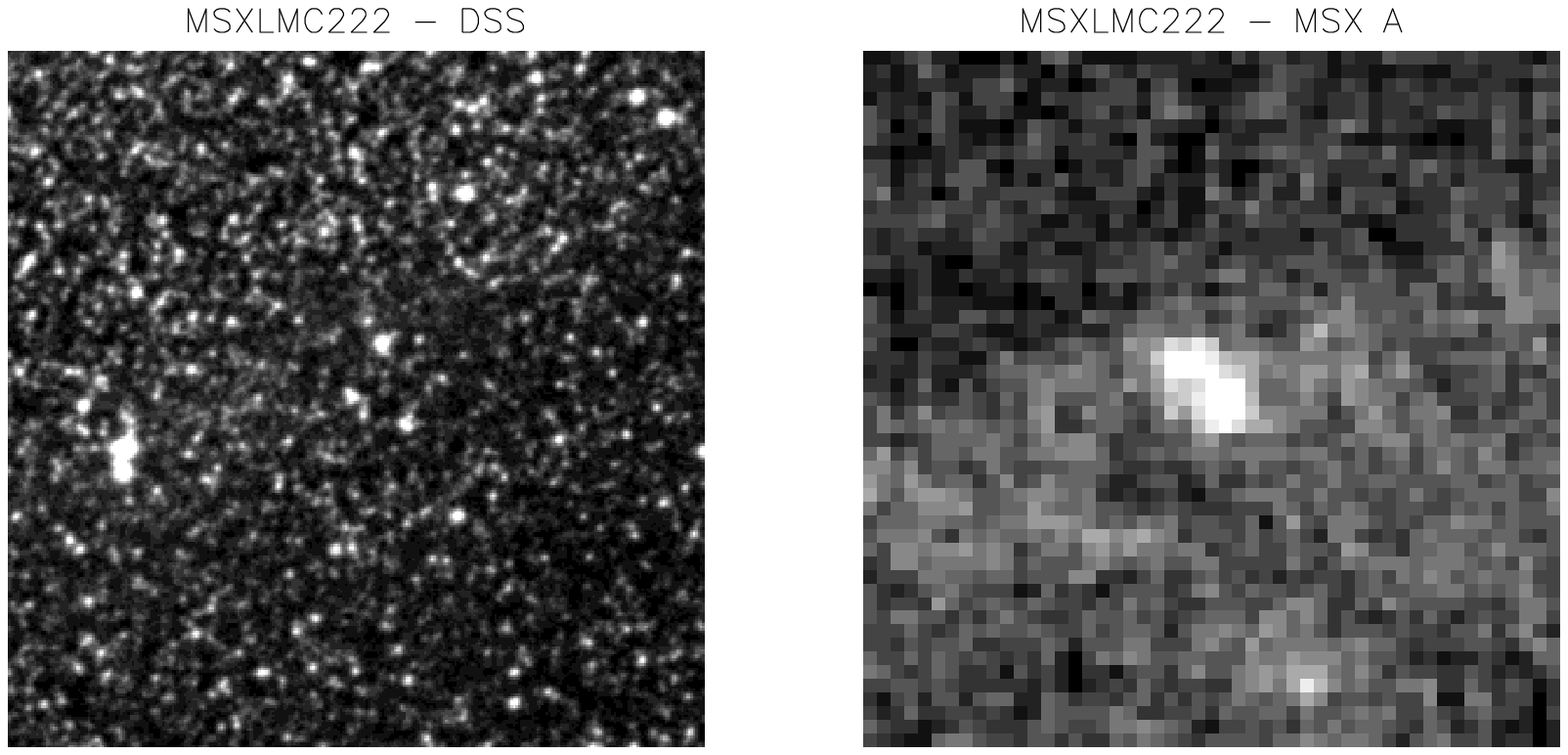}{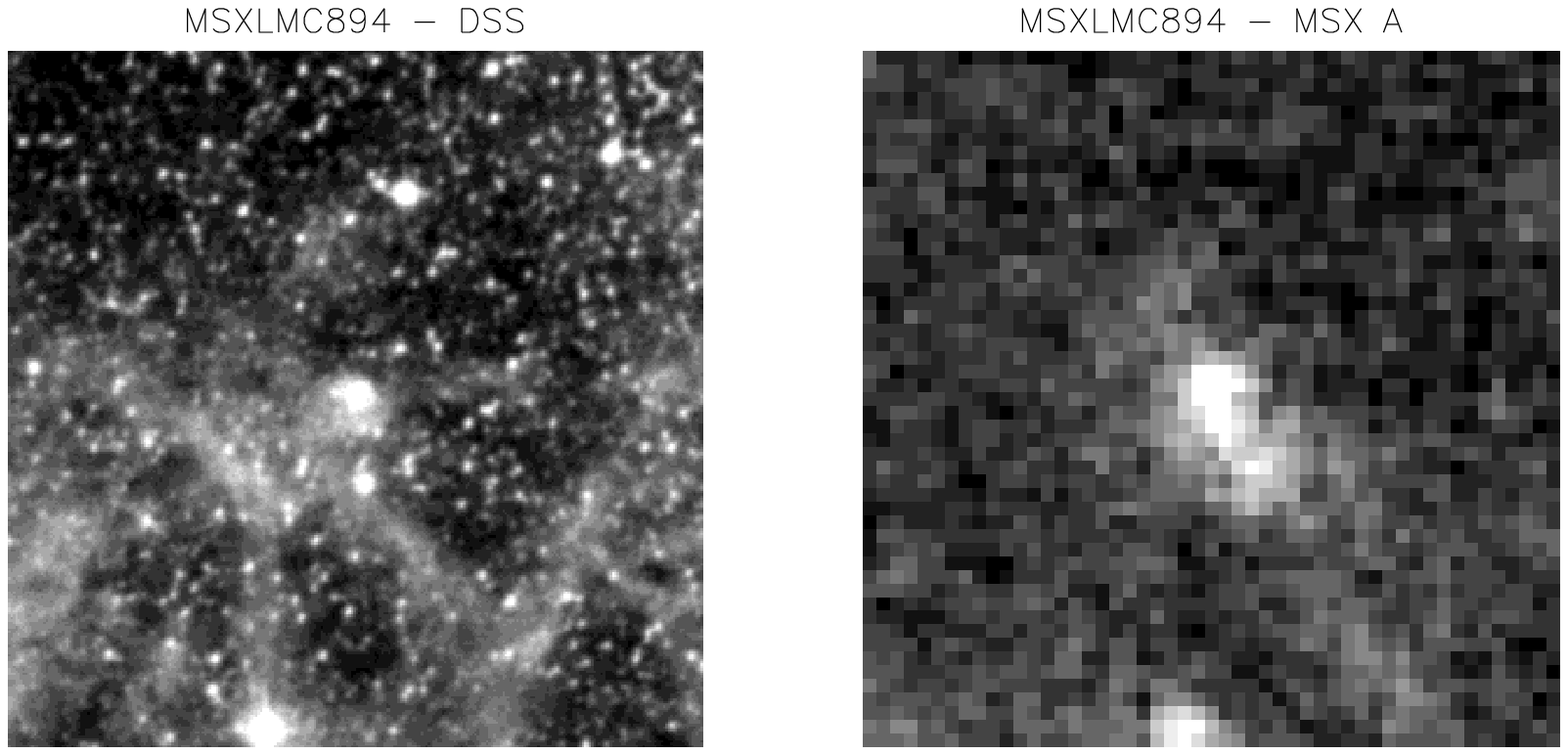}
\protect\vspace*{3mm}
\plottwo{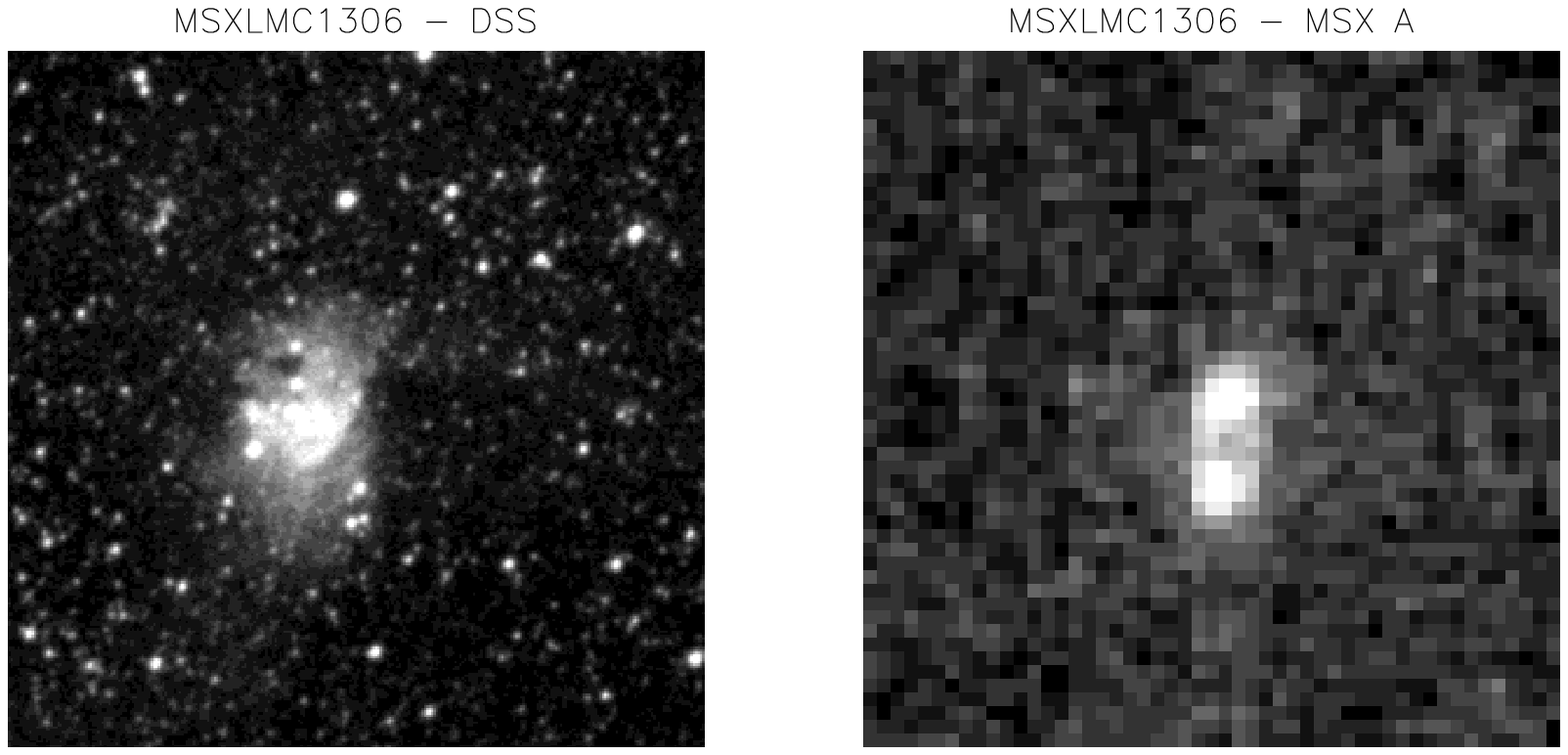}{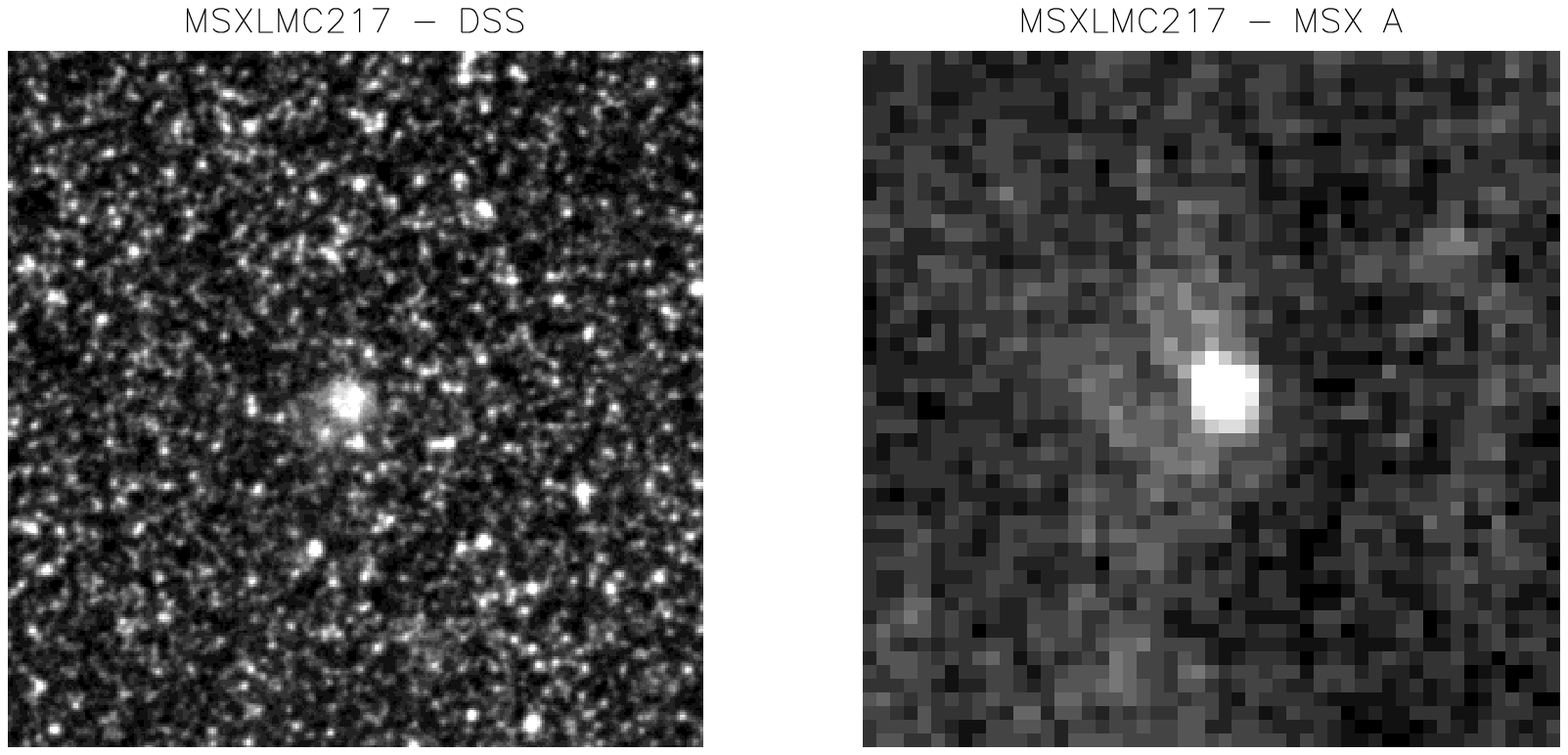}
\protect\vspace*{3mm}
\plottwo{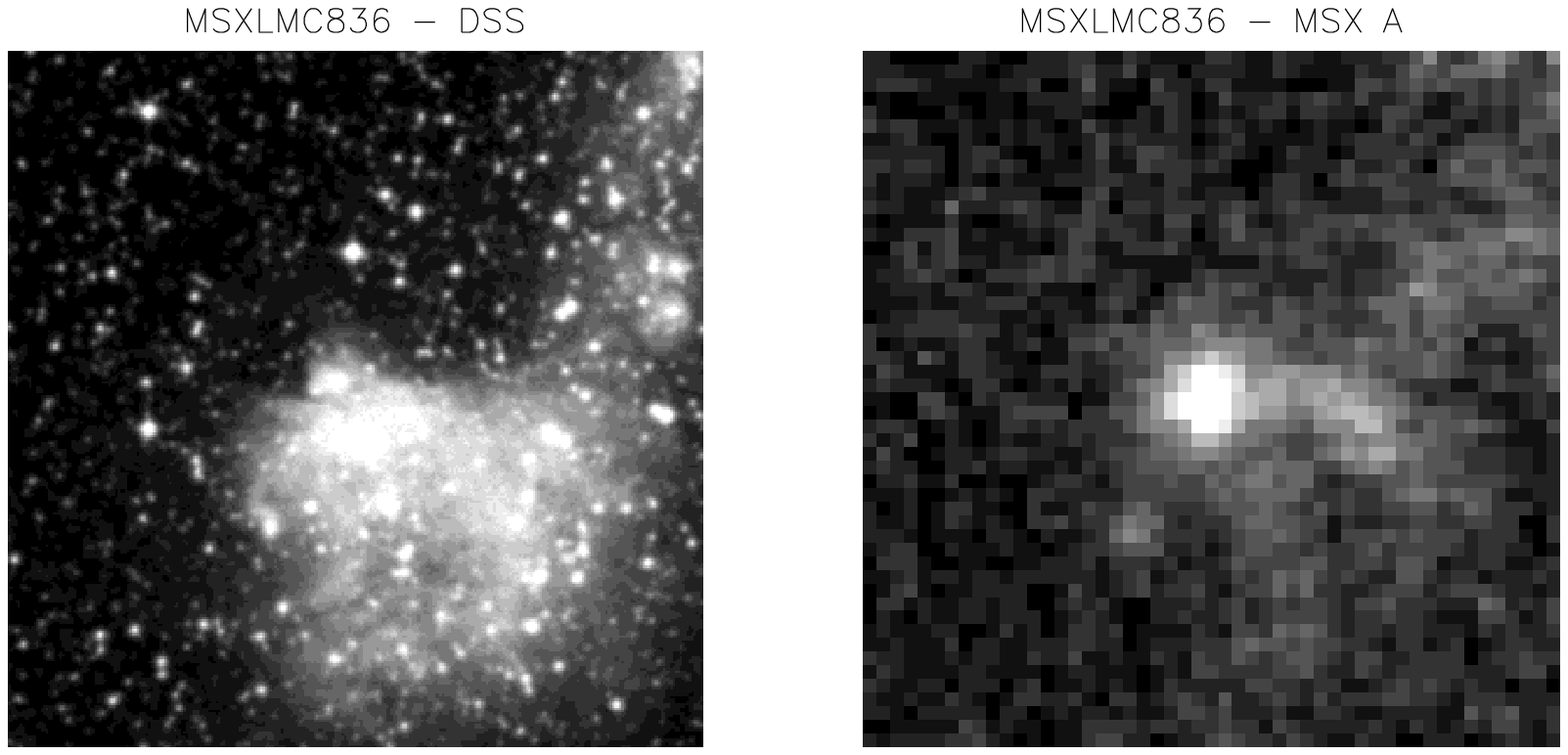}{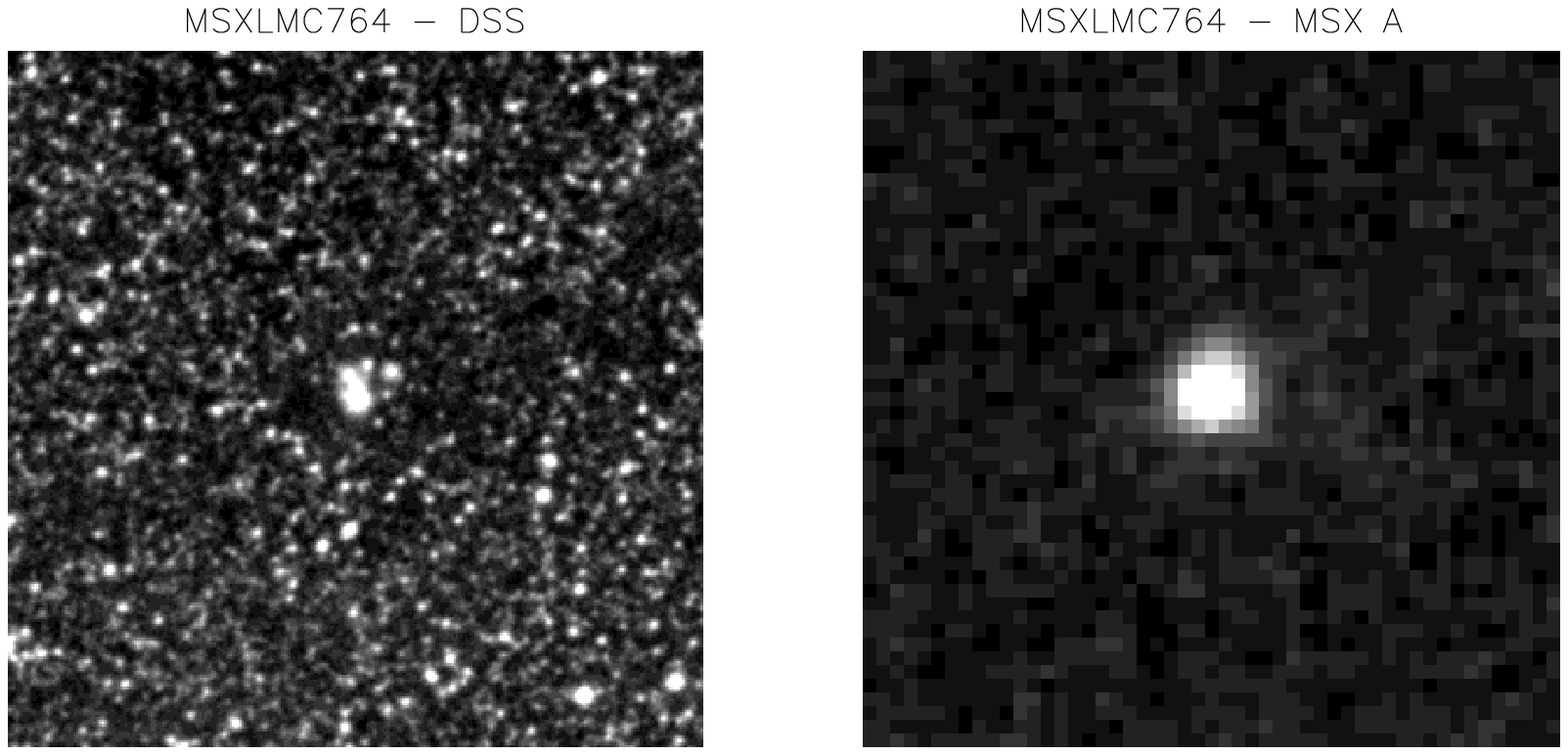}
\protect\vspace*{3mm}
\plottwo{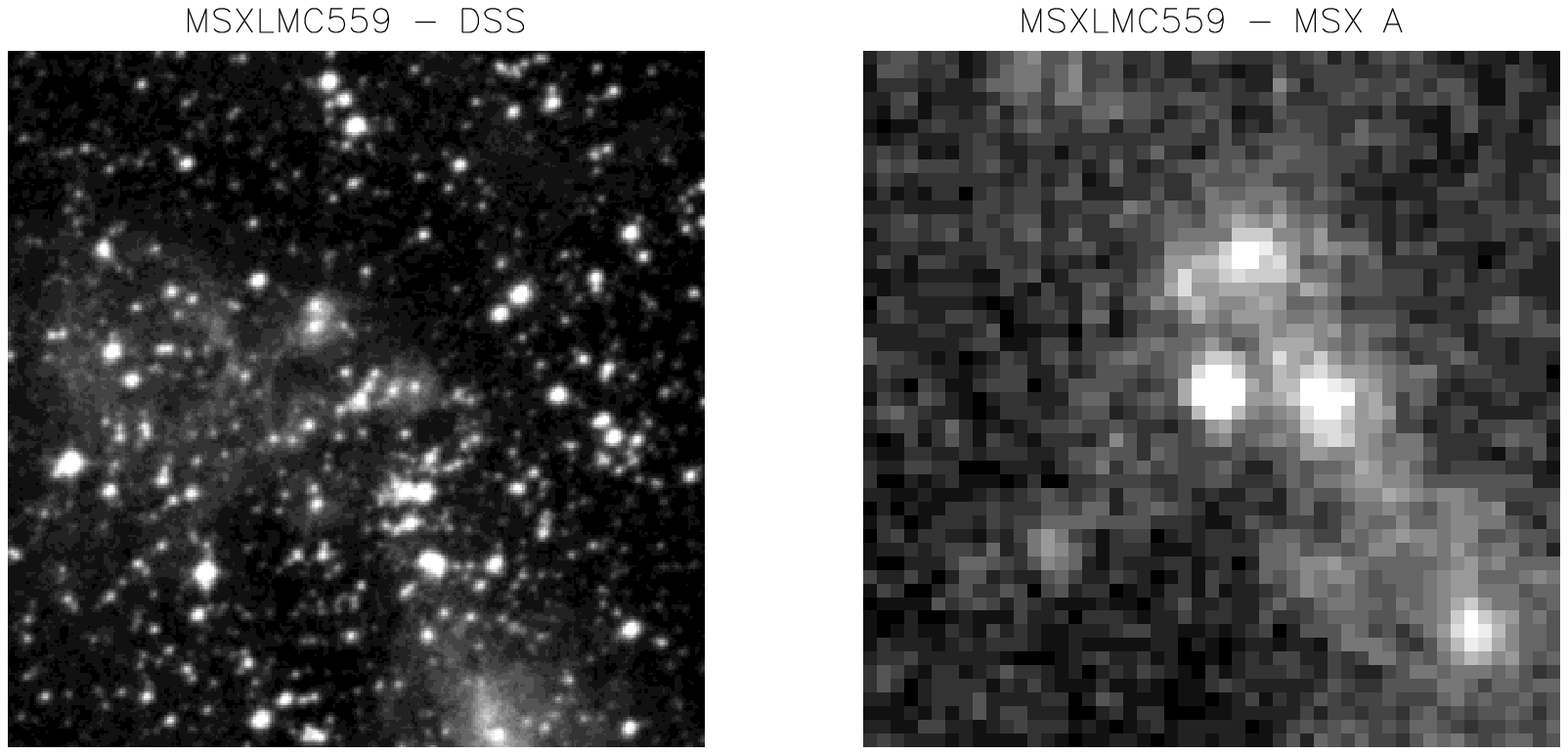}{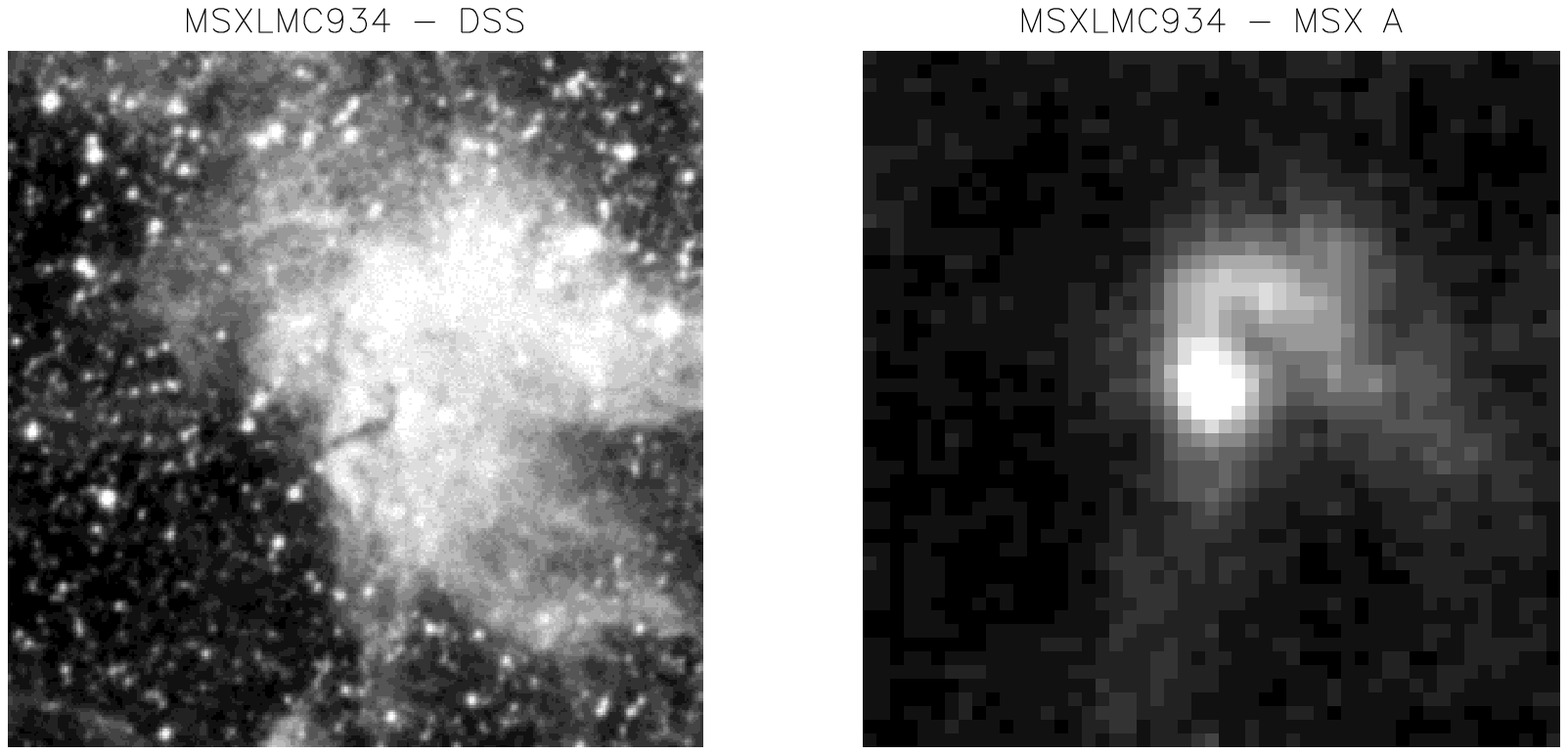}
\protect\vspace*{3mm}
\epsscale{0.37}
\plotone{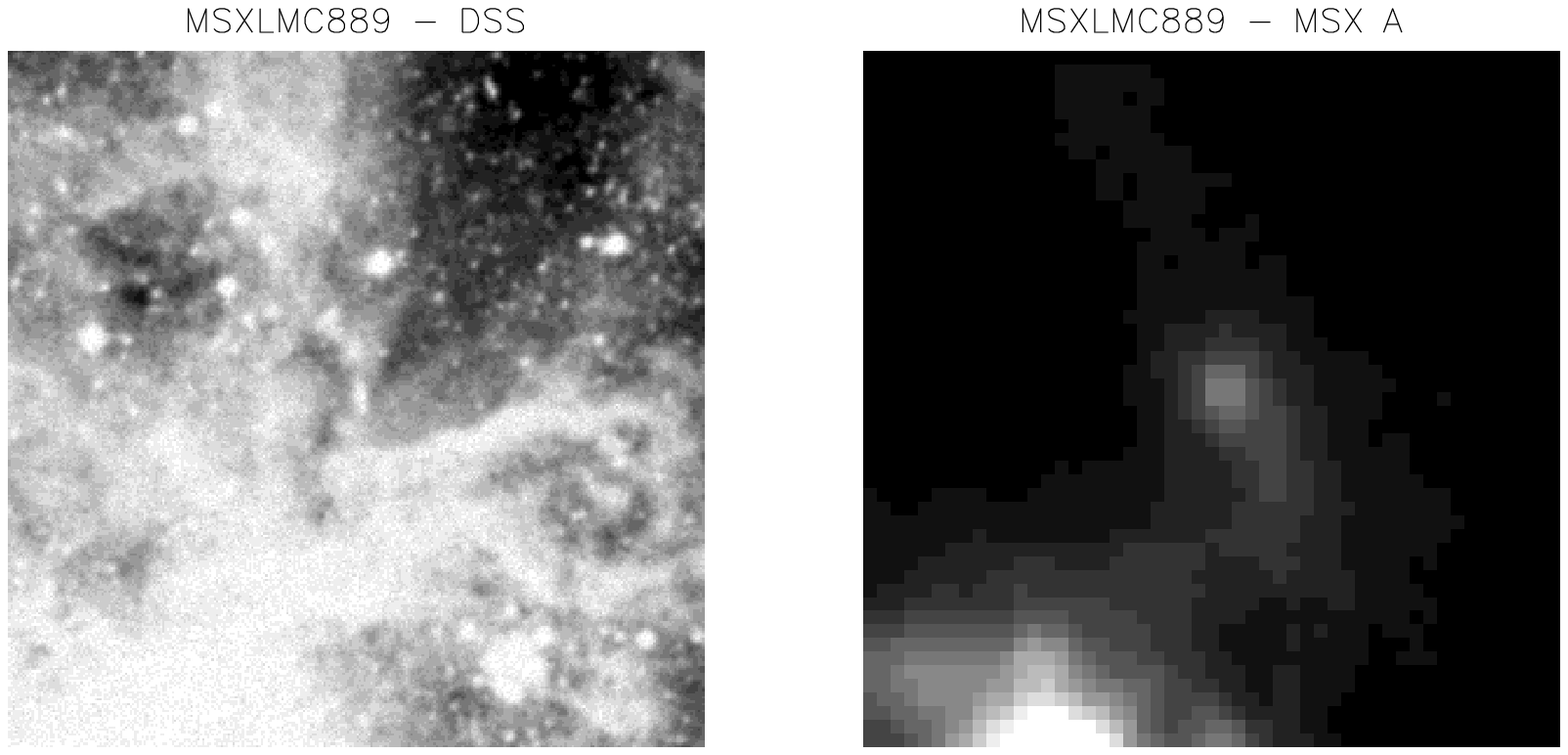}
\caption{Optical (DSS) and infrared (MSX A-band)
images of the emission-line objects in the sample. Each image
is 5\arcmin\ $\times$ 5\arcmin. The extended nebulous emission,
particularly in the MSX 8.3~\micron\ images, argues that these objects
are \protect\htwo\ regions associated with massive star formation
regions, rather than planetary nebulae. \label{fig:hiimontage}}
\epsscale{1.0}
\end{figure}

\clearpage 

\begin{figure}
\epsscale{0.7}
\plotone{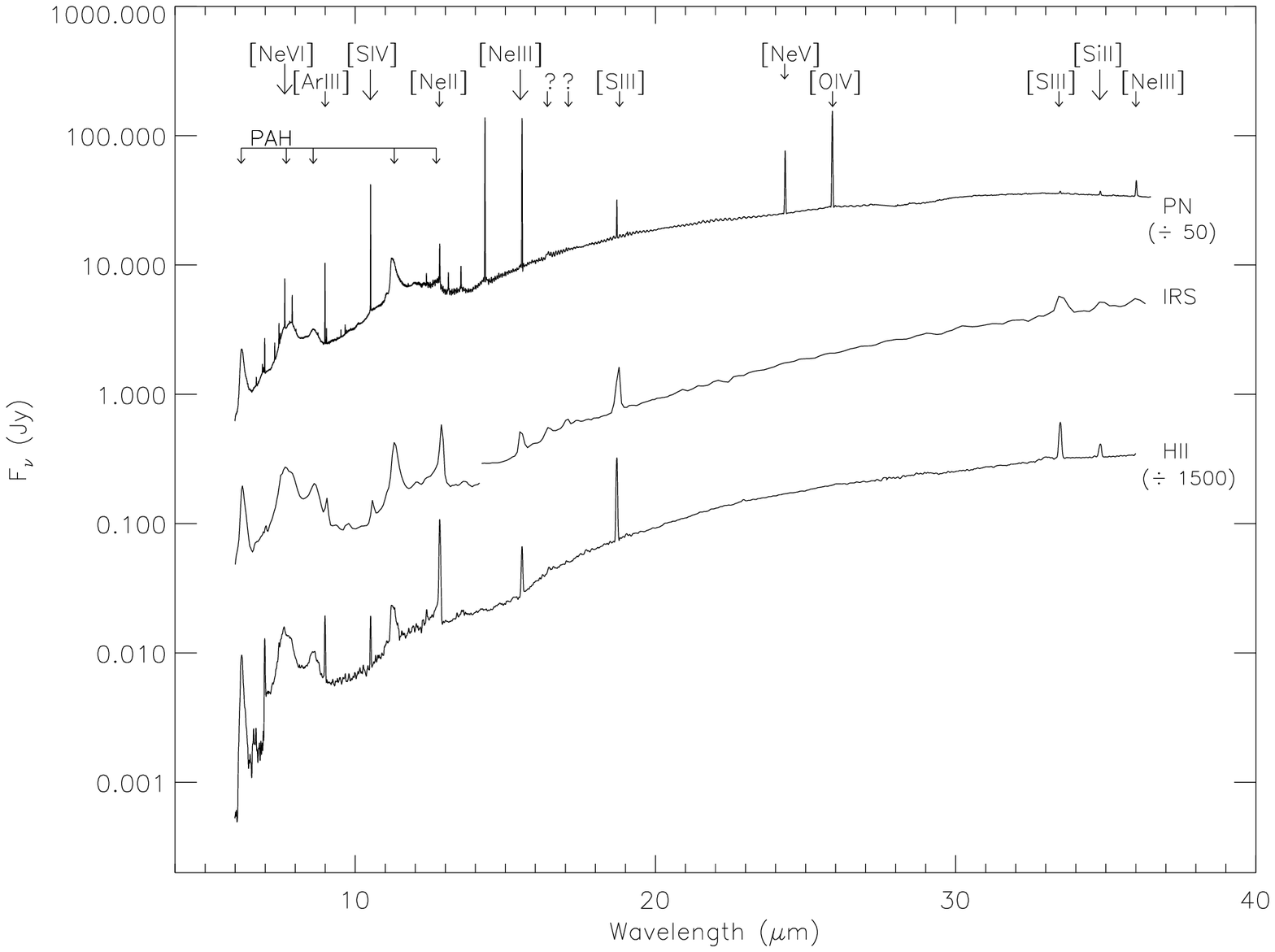}
\caption{Comparison of the \protect\irs\ spectrum of the emission-line source
MSX~LMC~22 {\it (middle)} with the \protect\iso\ spectra of PN NGC7027 {\it
(top)}, which has spectral subclass 4PU in the KSPW system, and H\,{\sc ii}
region IRAS10589-6034 {\it (bottom)}, which has spectral class 5UE. For
clarity, the flux densities of the comparison sources have been scaled by the
values shown. \label{fig:pncomp}} \epsscale{1.0}
\end{figure}

\clearpage

\begin{figure}
\epsscale{0.80}
\plotone{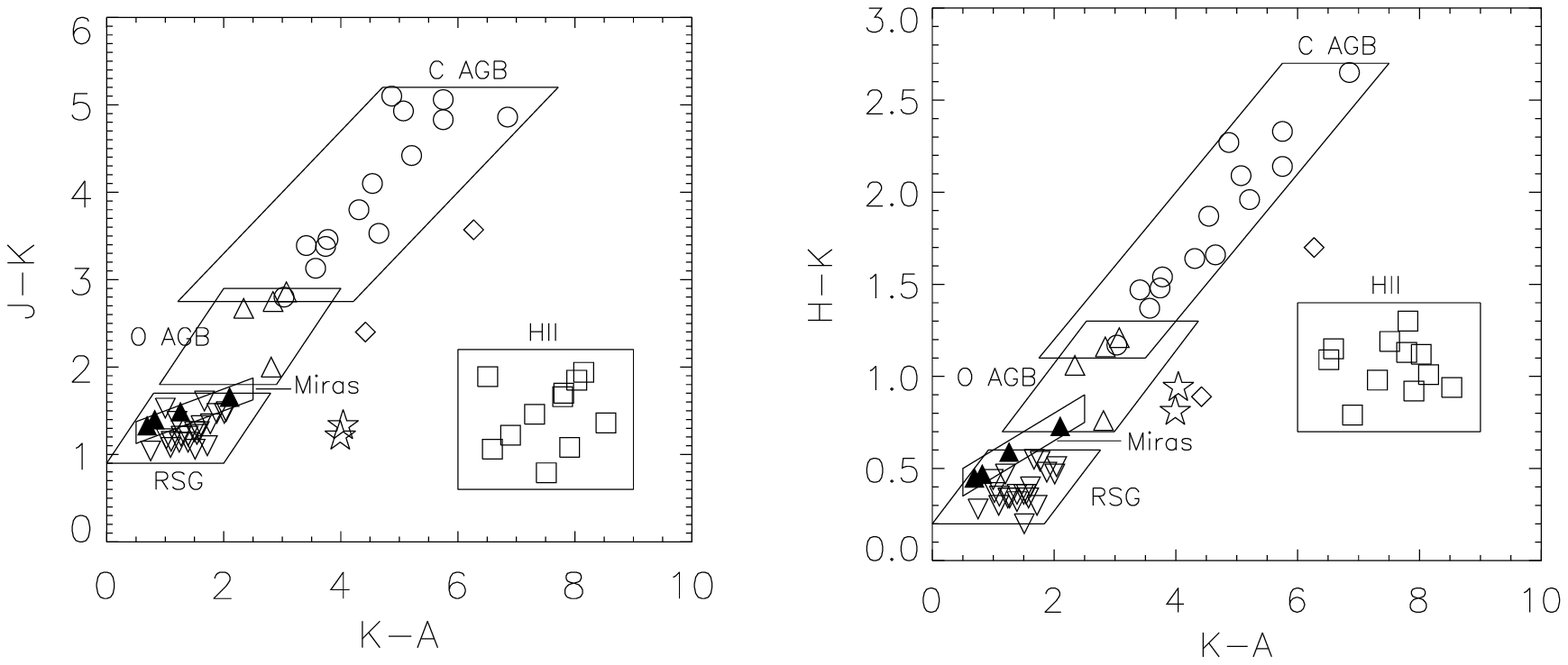}
\caption{\protect\tmass/\protect\msx\ color-color diagram with symbols
  indicating the \protect\irs\ spectral type: RSGs {\it (open, inverted
  triangles)}, O AGB stars {\it (open triangles)}, Galactic Mira variables
  {\it (filled triangles)}, C AGB stars {\it (open circles)}, \protect\htwo\
  regions {\it (open squares)}, B stars {\it (open stars)}, and OH/IR stars
  {\it (open diamonds)}. The boxes indicate the new NIR color criteria to
  classify IR-luminous LMC objects.
  \label{fig:nircols}} \epsscale{1.0}
\end{figure}

\clearpage

\begin{figure}
\epsscale{0.80}
\plotone{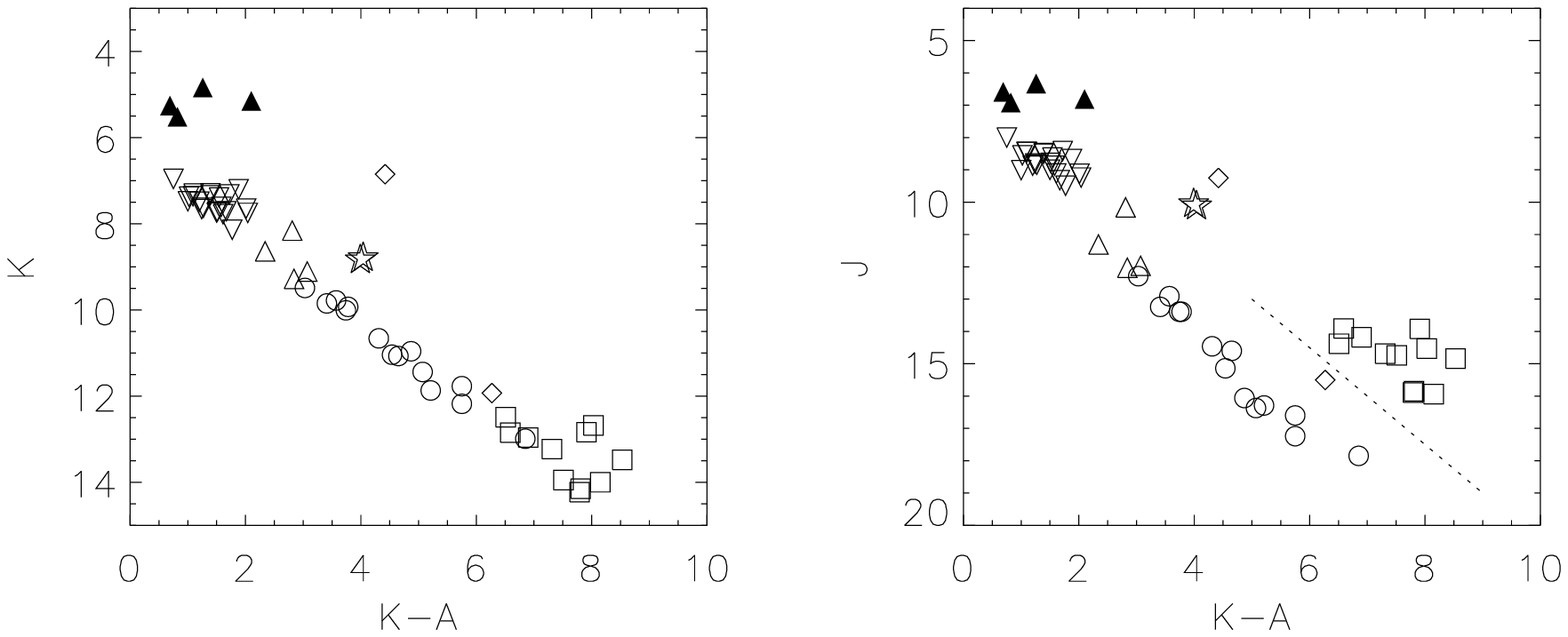}
\caption{Color-magnitude diagrams for our sample derived from
  \protect\tmass/\protect\msx\ photometry.  The symbols are the same as Figure
  \ref{fig:nircols} and indicate the \protect\irs\ spectral type (listed in
  Table \ref{tab:data}).  The filled symbols indicate Galactic objects. The
  dotted line in the right panel indicates the criterion to separate C~AGB
  stars and \protect\htwo\ regions.
  \label{fig:colmag}} \epsscale{1.0}
\end{figure}

\clearpage

\begin{figure}
\epsscale{0.80}
\plotone{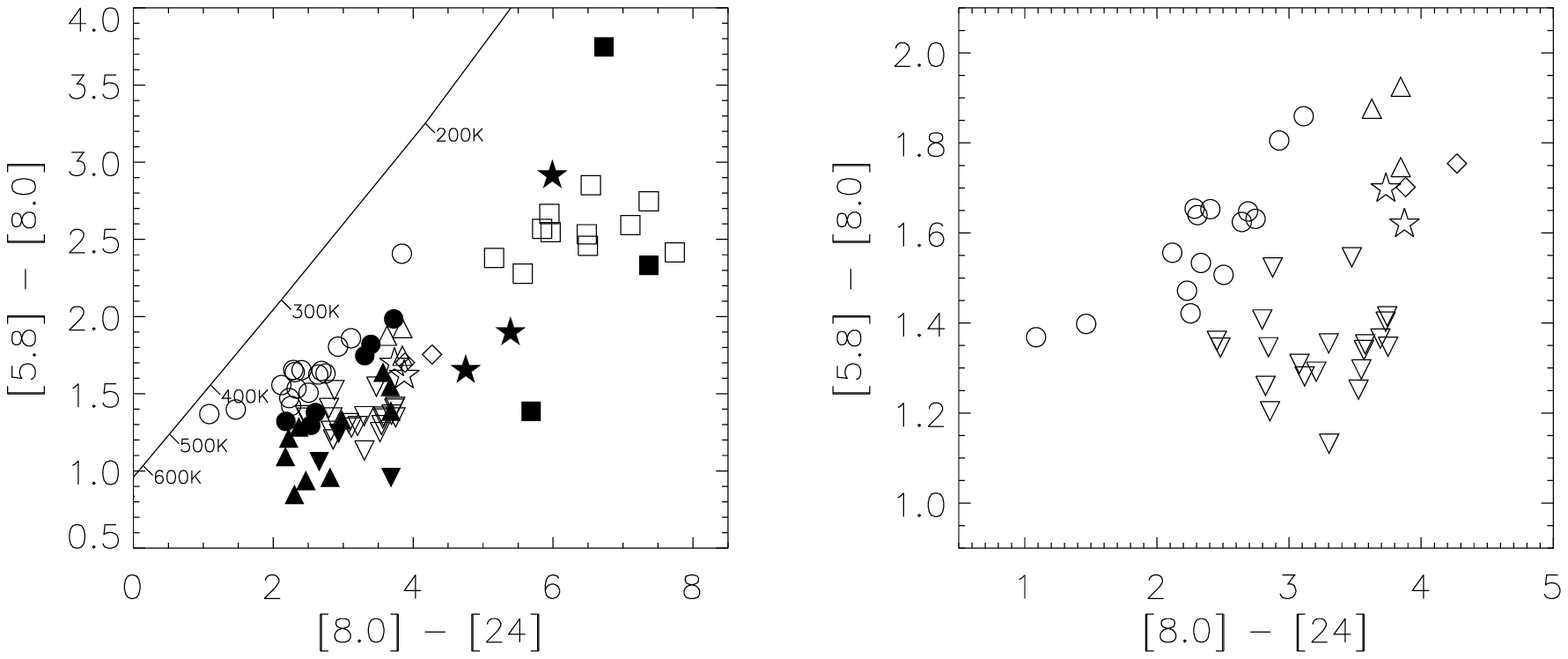}
\caption{{\it Left:} \protect\spitzer\ synthetic \protect\irac/\protect\mips\
  color-color diagram for the program objects, and some additional Galactic
  sources. The synthetic colors for the program objects were derived from the
  \protect\irs\ spectra and the instrument spectral response functions. The
  symbols are the same as in Figure \ref{fig:nircols} and indicate the
  \protect\irs\ spectral type (listed in Table \ref{tab:data}). Additional
  (filled) symbols are colors derived from \protect\iso\ spectra for Galactic
  O-rich {\it (filled triangles)} and C-rich {\it (filled circles)} stars,
  \htwo\ regions {\it (filled squares)}, and PNe {\it (filled stars)} for
  comparison. Also included is a line representing black bodies of different
  temperatures. {\it Right:} An expanded region of the left panel, showing a
  more limited color range and excluding the Galactic objects, to distinguish
  the regions occupied by the O-rich RSGs and C-rich AGB stars more clearly.
  \label{fig:sstcols}} \epsscale{1.0}
\end{figure}

\clearpage

\begin{figure}
\epsscale{0.80}
\plotone{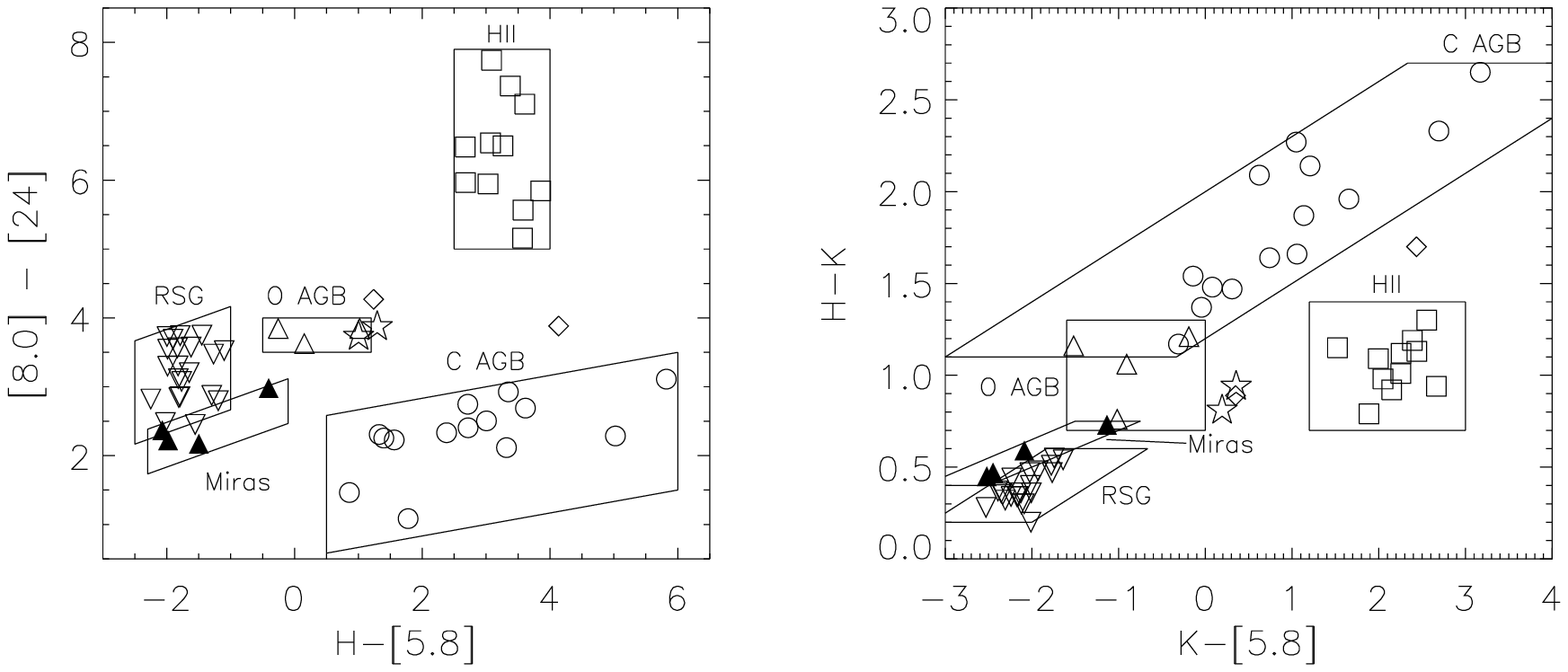}
\caption{\protect\spitzer/\protect\tmass\ color-color diagrams showing the
  proposed criteria to classify IR-luminous LMC objects. Synthetic
  \protect\irac\ and \protect\mips\ fluxes were derived from the \protect\irs\
  spectra and the instrument spectral response functions. The symbols are the
  same as figure \ref{fig:nircols} and indicate the \protect\irs\ spectral
  type (listed in Table \ref{tab:data}).
  \label{fig:sstcrit}} \epsscale{1.0}
\end{figure}

\clearpage

\begin{figure}
\epsscale{0.80} 
\plotone{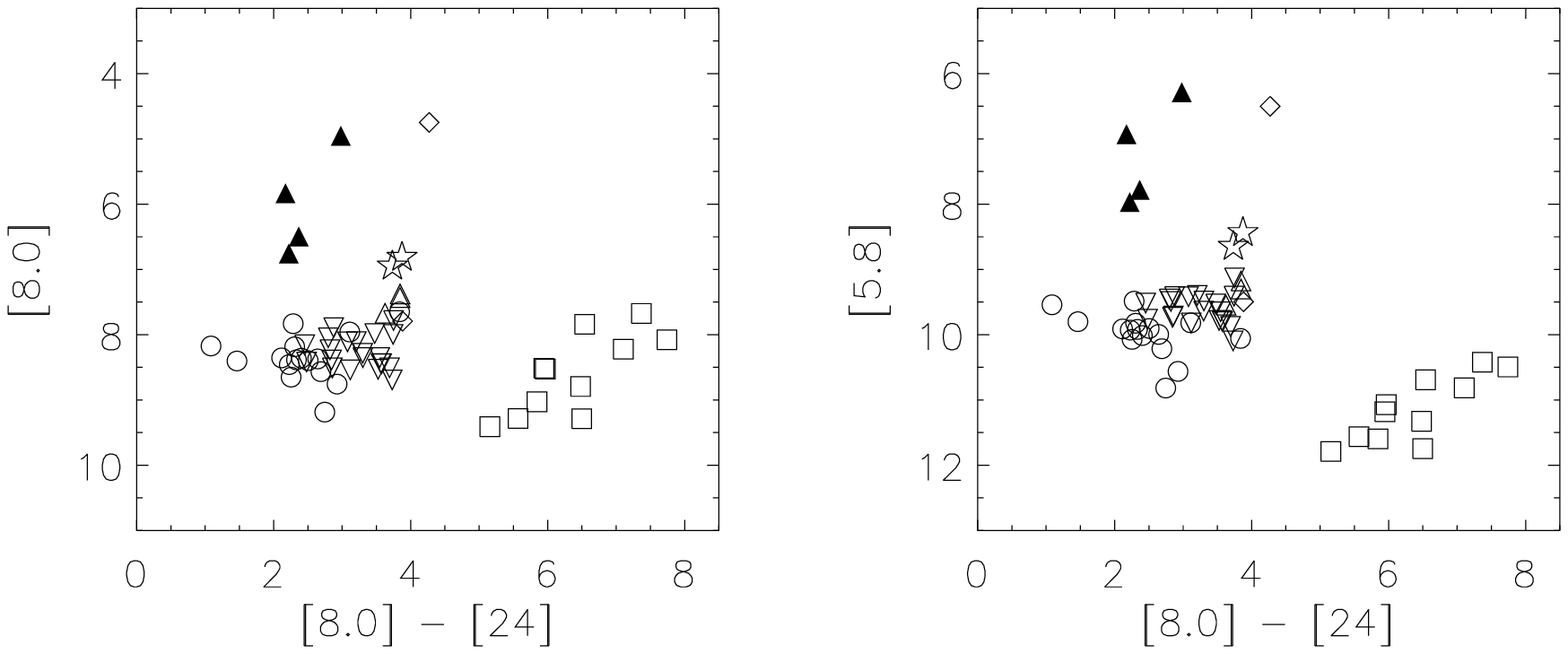}
\caption{Color-magnitude diagrams derived from synthetic
  \protect\spitzer\ photometry.  The symbols are the same as Figure
  \ref{fig:nircols}.
  \label{fig:sstcolmag}} \epsscale{1.0}
\end{figure}

\begin{figure}
\epsscale{0.8}
\plotone{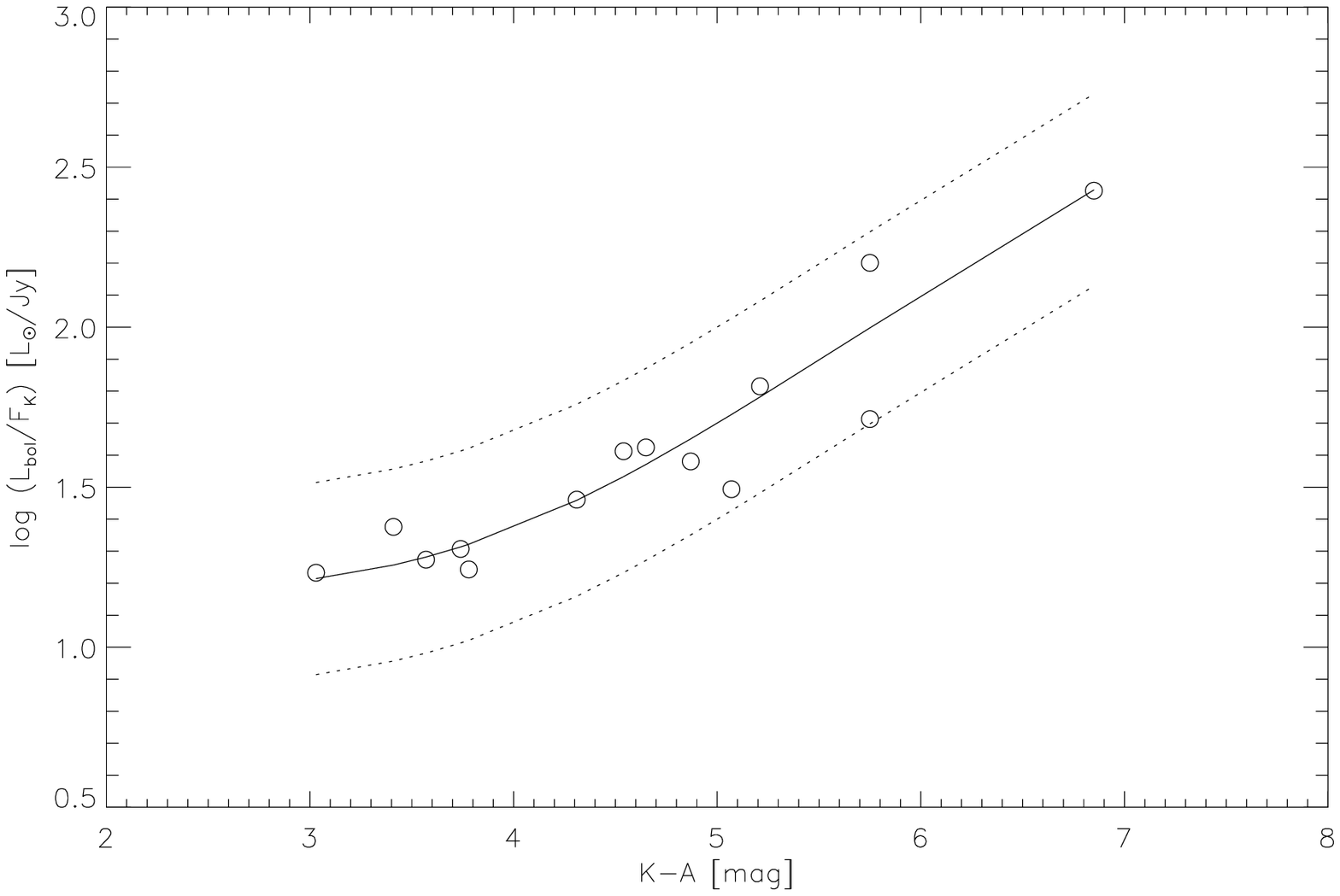}
\caption{The ``bolometric correction'' for the C-rich AGB stars in the sample,
  derived from the luminosity and K-band flux density, compared with the K-A
  color. The solid line shows a least-squares fit to the data points, and the
  dotted lines indicate the maximum error in the predicted luminosity (0.3
  dex). \label{fig:cbolcor}} \epsscale{1.0}
\end{figure}

\begin{figure}
\epsscale{0.8}
\plotone{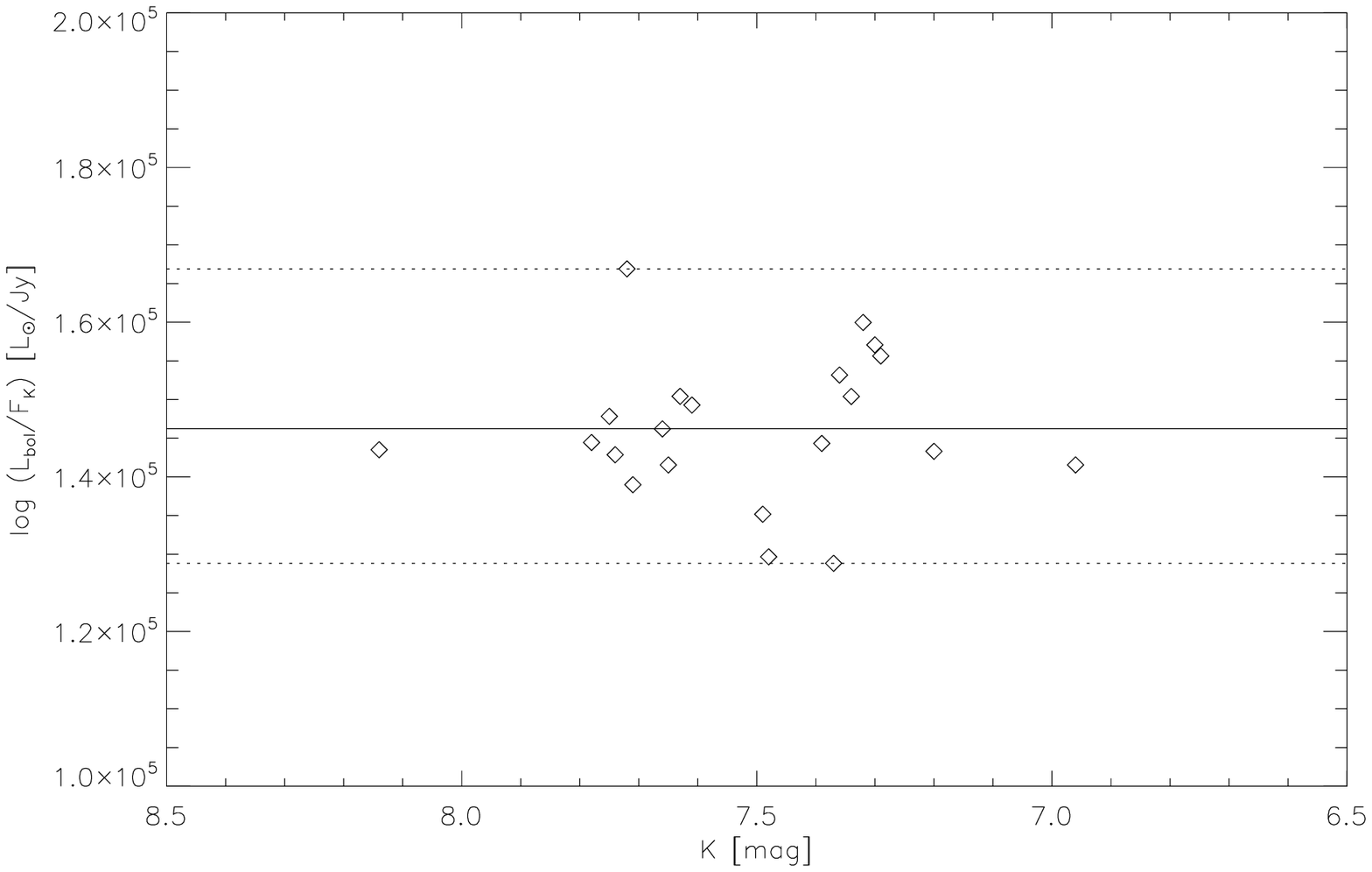}
\caption{The ``bolometric correction'' for the RSGs in the sample, derived
  from the luminosity and K-band flux density, compared with the K
  magnitude. The solid line shows the mean bolometric correction, and the
  dotted lines indicate the uncertainties determined from the sources with the
  largest distance from the mean. \label{fig:rsgbolcor}} \epsscale{1.0}
\end{figure}


\clearpage

\begin{thebibliography}{}
\bibitem[Aoki et al.(1999)]{aok99} Aoki, W., Tsuji, T., \& Ohnaka, K.\ 1999,
\aap, 350, 945
\bibitem[Barlow(1997)]{bar97} Barlow, M.~J.\ 1997, \apss, 251, 15
\bibitem[Barlow(1999)]{bar99} Barlow, M.~J.\ 1999, in IAU Symp.~191:
Asymptotic Giant Branch Stars, ed. T.\ Le Betre, A.\ L\`{e}bre, \& C.\
Waelkens (San Francisco: ASP), 353
\bibitem[Barnbaum et al.(1991)]{bar91} Barnbaum, C., Zuckerman, B., \&
Kastner, J.~H.\ 1991, \aj, 102, 289
\bibitem[Bernard-Salas et al.(2004)]{ber04} Bernard-Salas, J., Houck, J.~R.,
Morris, P.~W., Sloan, G.~C., Pottasch, S.~R., \& Barry, D.~J.\ 2004, \apjs,
154, 271
\bibitem[Bl\"{o}cker \& Schoenberner(1991)]{blo91} Bl\"{o}cker, T., \&
Schoenberner, D.\ 1991, \aap, 244, L43
\bibitem[Chen \& Kwok(1993)]{che93} Chen, P.~S., \& Kwok, S.\ 1993, \apj, 416,
769
\bibitem[Clayton(1996)]{cla96} Clayton, G.~C.\ 1996, \pasp, 108, 225
\bibitem[Draine \& Lee(1984)]{dra84} Draine, B.~T., \& Lee, H.~M.\ 1984, \apj,
285, 89
\bibitem[Draine \& Li(2001)]{dra01} Draine, B.~T., \& Li, A.\ 2001, \apj, 551,
807
\bibitem[Egan, Van Dyk \& Price(2001)]{ega01} Egan, M.~P., Van Dyk, S.~D., \&
Price, S.~D.\ 2001, \aj, 122, 1844
\bibitem[Fazio et al.(2004)]{faz04} Fazio, G.~G., et al.\ 2004, \apjs, 154, 10
\bibitem[Forrest et al.(1981)]{for81} Forrest, W.~J., Houck, J.~R., \&
McCarthy, J.~F.\ 1981, \apj, 248, 195
\bibitem[Freedman et al.(2001)]{fre01} Freedman, W.~L., et 
al.\ 2001, \apj, 553, 47 
\bibitem[Giveon et al.(2002)]{giv02} Giveon, U., Sternberg, A., Lutz, D.,
Feuchtgruber, H., \& Pauldrach, A.~W.~A.\ 2002, \apj, 566, 880
\bibitem[Goebel \& Moseley(1985)]{goe85} Goebel, J.~H., \& Moseley, S.~H.\
1985, \apjl, 290, L35
\bibitem[Groenewegen(2006)]{gro06} Groenewegen, M.~A.~T.\ 2006, \aap, 448, 181
\bibitem[Groenewegen(1999)]{gro99} Groenewegen, M.~A.~T.\ 1999, in IAU
Symp.~191: Asymptotic Giant Branch Stars, ed. T.\ Le Betre, A.\ L\`{e}bre, \&
C.\ Waelkens (San Francisco: ASP), 535
\bibitem[Groenewegen \& Blommaert(1998)]{gro98} Groenewegen, M.~A.~T., \&
Blommaert, J.~A.~D.~L.\ 1998, \aap, 332, 25
\bibitem[Groenewegen et al.(1995)]{gro95} Groenewegen, M.~A.~T., Smith, C.~H.,
Wood, P.~R., Omont, A., \& Fujiyoshi, T.\ 1995, \apjl, 449, L119
\bibitem[Herwig(2005)]{her05} Herwig, F.\ 2005, \araa, 43, 435
\bibitem[Higdon et al.(2004)]{hig04} Higdon, S.\ J.\ U.\ et al.\ 2004, \pasp,
116, 975
\bibitem[Hony et al.(2002)]{hon02} Hony, S., Waters, L.~B.~F.~M., \& Tielens,
A.~G.~G.~M.\ 2002, \aap, 390, 533
\bibitem[Houck et al.(2004)]{hou04} Houck, J.~R., et al.\ 2004, \apjs, 154, 18
\bibitem[Hrivnak et al.(2000)]{hri00} Hrivnak, B.~J., Volk, K., \&
Kwok, S.\ 2000, \apj, 535, 275
\bibitem[Iben \& Renzini(1983)]{ibe83} Iben, I., \& Renzini, A.\ 1983, \araa,
21, 271
\bibitem[Jura(1988)]{jur88} Jura, M.\ 1988, \apjs, 66, 33
\bibitem[Jura \& Kleinmann(1989)]{jur89} Jura, M., \& Kleinmann, S.~G.\ 1989,
\apj, 341, 359
\bibitem[Kastner et al.(2006)]{kas06} Kastner, J.~H., Buchanan, C.~L.,
Sargent, B., \& Forrest, W.~J.\ 2006, \apjl, 638, L29
\bibitem[Kastner et al.(1993)]{kas93} Kastner, J.~H., Forveille, T.,
Zuckerman, B., \& Omont, A.\ 1993, \aap, 275, 163
\bibitem[Kastner \& Weintraub(1998)]{kas98} Kastner, J.~H., \& Weintraub,
D.~A.\ 1998, \aj, 115, 1592
\bibitem[Kraemer et al.(2002)]{kra02} Kraemer, K.~E., Sloan, G.~C., Price,
S.~D., \& Walker, H.~J.\ 2002, \apjs, 140, 389
\bibitem[Kraemer et al.(2005)]{kra05} Kraemer, K.~E., Sloan, G.~C., Wood,
P.~R., Price, S.~D., \& Egan, M.~P.\ 2005, \apjl, 631, L147
\bibitem[Kwok, Volk, \& Bidelman(1997)]{kwo97} Kwok, S., Volk, M, \& Bidelman,
W.P, 1997, \apjs, 112, 557
\bibitem[Lamers et al.(1996)]{lam96} Lamers, H.~J.~G.~L.~M., et al.\ 1996,
\aap, 315, L225
\bibitem[Le Bertre et al.(2001)]{leb01} Le Bertre, T., Matsuura, M., Winters,
J.~M., Murakami, H., Yamamura, I., Freund, M., \& Tanaka, M.\ 2001, \aap, 376,
997
\bibitem[Li \& Draine(2001)]{li01} Li, A., \& Draine, B.~T.\ 2001, \apj, 554,
778
\bibitem[Loup et al.(1997)]{lou97} Loup, C., Zijlstra, A.~A., Waters,
L.~B.~F.~M., \& Groenewegen, M.~A.~T.\ 1997, \aaps, 125, 419
\bibitem[Marigo et al.(1999)]{mar99} Marigo, P., Girardi, L., \&
Bressan, A.\ 1999, \aap, 344, 123
\bibitem[Olnon et al.(1986)]{oln86} Olnon, F.~M., et al.\ 1986, \aaps,
65, 607
\bibitem[Ramos-Larios \& Phillips(2005)]{ram05} Ramos-Larios, G., \& Phillips,
J.~P.\ 2005, \mnras, 357, 732
\bibitem[Rieke et al.(2004)]{rie04} Rieke, G.~H., et al.\ 2004, \apjs, 154, 25
\bibitem[Roche et al.(1993)]{roc93} Roche, P.~F., Aitken, D.~K., \& Smith,
C.~H.\ 1993, \mnras, 262, 301
\bibitem[Schwering(1989)]{sch89} Schwering, P.~B.~W.\ 1989, \aaps, 79,
  105
\bibitem[Sloan et al.(2003)]{slo03} Sloan, G.~C., Kraemer, K.~E., Goebel,
J.~H., \& Price, S.~D.\ 2003, \apj, 594, 483
\bibitem[Sloan et al.(1998)]{slo98} Sloan, G.~C., Little-Marenin, I.~R., \&
Price, S.~D.\ 1998, \aj, 115, 809
\bibitem[Trams et al.(1999)]{tra99} Trams, N.~R., et al.\ 1999, \aap, 346, 843
\bibitem[van Loon et al.(1998)]{van98} van Loon, J.~T., et al.\ 1998, \aap,
329, 169
\bibitem[van Loon et al.(1997)]{van97} van Loon, J.~T., Zijlstra, A.~A.,
Whitelock, P.~A., Waters, L.~B.~F.~M., Loup, C., \& Trams, N.~R.\ 1997, \aap,
325, 585
\bibitem[van Winckel(2003)]{van03} van Winckel, H.\ 2003, \araa, 41, 391
\bibitem[Volk \& Cohen(1989)]{vol89} Volk, K. \& Cohen, M. 1989, \aj, 98, 931
\bibitem[Volk et al.(2002)]{vol02} Volk, K., Kwok, S., Hrivnak, B.~J.,
\& Szczerba, R.\ 2002, \apj, 567, 412
\bibitem[Volk et al.(1991)]{vol91} Volk, K., Kwok, S., Stencil, R., \& Brugel,
E. 1991, \apjs, 77, 607
\bibitem[Wainscoat et al.(1992)]{wai92} Wainscoat, R.~J., Cohen, M., Volk, K.,
Walker, H.~J., \& Schwartz, D.~E.\ 1992, \apjs, 83, 111
\bibitem[Waters et al.(1999)]{wat99a} Waters, L.~B.~F.~M., et al.\ 1999, ESA
SP-427: The Universe as Seen by ISO, 219
\bibitem[Waters \& Molster(1999)]{wat99b} Waters, L.~B.~F.~M., \& Molster,
F.~G.\ 1999, in IAU Symp.~191: Asymptotic Giant Branch Stars, ed. T.\ Le
Betre, A.\ L\`{e}bre, \& C.\ Waelkens (San Francisco: ASP), 209
\bibitem[Werner et al.(2004)]{wer04} Werner, M.~W., et al.\ 2004,
\apjs, 154, 1
\bibitem[Whitelock(1985)]{whi85} Whitelock, P.~A.\ 1985, \mnras, 213, 59
\bibitem[Whitelock et al.(1994)]{whi94} Whitelock, P., Menzies, J., Feast, M.,
Marang, F., Carter, B., Roberts, G., Catchpole, R., \& Chapman, J.\ 1994,
\mnras, 267, 711
\bibitem[Wood et al.(1986)]{woo86} Wood, P.~R., Bessell, M.~S., \& Whiteoak,
J.~B.\ 1986, \apjl, 306, L81
\bibitem[Zijlstra et al.(1996)]{zij96} Zijlstra, A.~A., Loup, C., Waters,
L.~B.~F.~M., Whitelock, P.~A., van Loon, J.~T., \& Guglielmo, F.\ 1996,
\mnras, 279, 32
\bibitem[Zijlstra et al.(2006)]{zij06} Zijlstra, A.~A., et al.\ 2006, ArXiv
Astrophysics e-prints, arXiv:astro-ph/0602531
\end{thebibliography}
\end{document}